\documentclass{aa}

\usepackage[varg]{txfonts}
\usepackage{amssymb,color,verbatim}
\usepackage{graphicx}
\usepackage{natbib}
\bibpunct{(}{)}{;}{a}{}{,}
   
\newcommand{\amm}{NH$_3$}
\newcommand{\wat}{H$_2$O}
\newcommand{\met}{CH$_3$OH}

\newcommand{\kms}{km~s$^{-1}$}
\newcommand{\cmc}{cm$^{-3}$}

\newcommand{\ms}{$M_{\odot}$}
\newcommand{\ls}{$L_{\odot}$}

\newcommand{\msyr}{$M_{\odot}$~yr$^{-1}$}

\def\vlsr{V$_{\rm LSR}$}
\newcommand{\pas}{$\rlap{.}^{\prime\prime}$}
\newcommand{\degree}{$^{\circ}$}

\newcommand{\trot}{$T_{rot}$}
\newcommand{\tex}{$T_{ex}$}
\newcommand{\ncol}{N$_{\rm col}$}

\begin{document}


\title{Hot ammonia around young O-type stars}

\subtitle{III. High-mass star formation and hot core activity in W51~Main}

 \titlerunning{Hot NH$_3$ around Young O-type Stars: III. W51~Main}

\author{ C. Goddi \inst{1,2}
\and A. Ginsburg \inst{3}
\and Q. Zhang \inst{4}
}

\offprints{C. Goddi,\\\email{c.goddi@astro.ru.nl}}

\institute{Department of Astrophysics/IMAPP, Radboud University Nijmegen, PO Box 9010, 6500 GL Nijmegen, the Netherlands
\and  ALLEGRO/Leiden Observatory, Leiden University, PO Box 9513, NL-2300 RA Leiden, the Netherlands  
\and ESO, Karl-Schwarzschild-Strasse 2, D-85748 Garching bei M{\" u}nchen
\and Harvard-Smithsonian Center for Astrophysics, 60 Garden Street, Cambridge, MA 02138
}



\abstract
{This paper is the third in a series of \amm\,multilevel imaging studies in well-known high-mass star forming regions.
}
{We want to map at sub-arcsecond resolution  highly-excited inversion lines of \amm\,in the high-mass star forming region W51~Main   (distance=5.4~kpc).  }
{Using the Karl Jansky Very Large Array (JVLA), we have mapped the hot and dense molecular gas in W51~Main, with $\sim$0\pas2--0\pas3 angular resolution, 
in five metastable (J=K) inversion transitions of ammonia (NH$_3$): (J,K)=(6,6), (7,7), (9,9), (10,10), and (13,13). 
}
{We have identified and characterized two main centers of high-mass star formation in W51-Main: 
the W51e2 complex and the W51e8 core ($\sim$6\arcsec\,southward of W51e2). 
The former breaks down into three further sub-cores:  
W51e2-W, which surrounds the well known hypercompact (HC) HII region, where hot \amm\,is observed in absorption, 
and two additional dusty cores, W51e2-E ($\sim$0\pas8 to the East) and W51e2-NW ($\sim$1\arcsec\,to the North), where hot \amm\,is observed in emission. 
 The velocity maps towards the HC HII region show a clear velocity
gradient along east-west in all lines.  The gradient may indicate
rotation, though any Keplerian motion must be on smaller 
scales ($<$1000 AU) as we do not directly observe a Keplerian velocity profile. 
 The absence of outflow and/or maser activity and the low amount of molecular gas available for accretion ($\sim$5~\ms, assuming [\amm]/[H$_2$]=10$^{-7}$) with respect to the  mass of the central YSO estimated from radio luminosity ($>$20~\ms), both indicate that the central YSO has already accreted most of its final mass. 
 On the other hand, the nearby W51e2-E, while not displaying evidence for rotation, shows signatures of infall in a hot dense core ($T\sim 170$~K, $n_{H_2} \sim 5 \times 10^7$~\cmc), based on asymmetric spectral profiles (skewed towards the blueshifted component) in optically thick emission lines of \amm.  The relatively large amount of hot molecular gas available for accretion ($\sim$20~\ms\, within about half an arcsecond or 2500~AU), along with strong outflow and maser activity, indicates that the main accretion center in the W51e2 complex is W51e2-E rather than W51e2-W. 
 Finally, W51e2-NW and W51e8, although less dense ($n_{H_2} \sim 2 \times 10^7$~\cmc\,and $\sim 3 \times 10^6$~\cmc), are also hot cores ($T_{\rm gas}\sim$140 and 200~K) and contain a significant amount of  molecular gas ($M_{\rm gas}\sim$30~\ms\,and $\sim$70~\ms, respectively). We speculate that they may host high-mass YSOs either  at a previous evolutionary stage to or with lower mass than W51e2-E and W51e2-W.
}
{Using high-angular resolution multi-level imaging of highly-excited \amm\,metastable lines, we characterized the physical and dynamical properties of four individual high-mass young stars forming in the W51~Main clump.  
}

\keywords{ISM: individual objects (W51~Main) --- ISM: molecules --- ISM: abundances  }
 
\maketitle

\section{Introduction}
\label{intro}

W51 (at 5.4 kpc; \citealt{Xu09}) is one of the most luminous high-mass star forming  regions (HMSFRs) in the Galaxy, with a luminosity of about $\sim 2\times 10^7$ \ls\,(within a 2 pc radius), implying a total stellar mass of about 7000~\ms, corresponding to at least 20  O-type stars (i.e greater than 20~\ms; Ginsburg et al. 2016, in prep.). These massive stars are associated with three main regions: 
W51 IRS1, W51 IRS2 (30\arcsec\,northwest or NW of IRS1), and W51 Main (30\arcsec\,southeast or SE of IRS1).  
While  W51 IRS1 is associated with an evolved HII region (size$\sim$1 pc) and is deprived of molecular gas or dust, W51 Main  and W51 IRS2 are sites of active high-mass star formation (HMSF).
This paper is the third in a series of \amm\,multilevel imaging studies in luminous HMSFRs associated with strong hot core activity, presumably hosting O-type  stars forming at their centers (for a general description, see \citealt{Goddi15a}; hereafter Paper I).    
A study on W51 IRS2 was reported in the second paper of the series (\citealt{Goddi15b}; hereafter Paper II).  
This paper focuses on W51~Main. 

W51 Main is a very active region of HMSF and contains a cluster of ultracompact (UC) and hypercompact (HC) HII regions, labelled as e1 to e8 \citep{Gaume93,Mehringer94,ZhangHo97}. 
The region shows widespread \amm~thermal emission \citep{ZhangHo97} as well as strong maser emission from OH, \met, \wat, and \amm\, \citep{ZhangHo95,Imai02,FishReid07,Etoka12,Surcis12}.  

W51e2 is the strongest and best studied HC HII region in the W51~Main cluster, and it is believed to be powered by an O8-type young star \citep[e.g.,][]{Shi10a}. 
A number of  interferometric studies conducted with varying angular resolutions,  at centimetre (cm) and (sub)millimetre (mm) bands, identified molecular and ionized gas undergoing infall and rotation towards W51e2. 
VLA observations of the \amm~inversion lines (1,1) and (2,2) seen in absorption (1\pas1 beamsize) revealed radial infall on scales larger than 5000 AU  towards the W51e2 core \citep{ZhangHo97}. 
 Higher-angular resolution observations of the (3,3) \amm\,absorption line (0\pas3 beamsize)  showed signatures of rotation within 2000 AU based on a position-velocity (pv) diagram \citep{ZhangHo97}.  
\citet{Zhang98} identified a velocity gradient in a CH$_3$CN transition at 2 mm, deriving  a position angle (P.A.) of $ 20 \pm 20$\degree.  
\citet{KetoKlaassen08} imaged the H53$\alpha$ radio recombination line (RL) with the VLA (0\pas45 beamsize) and 
they claimed rotation in the ionized gas along the axis of a molecular bipolar outflow (oriented NW-SE) imaged with the SMA in the CO (2-1) line (1\arcsec~beamsize), suggesting a simple  inflow/outflow picture in a single high-mass YSO.
However, higher resolution observations, using the SMA at the wavelengths of 0.85 mm (0\pas3 beamsize) and 1.3 mm (0\pas7 beamsize), revealed a more complex picture, by resolving  W51e2 into three sub-cores \citep{Shi10a}:  W51e2-W, corresponding to the HC HII region,  W51e2-E,  located about 1\arcsec\,east of the HC HII region and corresponding to the brightest dust continuum source, and  
W51e2-NW, the weakest continuum component,  located about 1\arcsec\, NW of the HC HII region. 
 \citet{Shi10b} imaged the CO (3-2) line (with a 0\pas7 beamsize) and established that  the driving source of the powerful molecular outflow in this region is the protostellar core W51e2-E, and not the HC HII region W51e2-W, challenging the scenario proposed by \citet{KetoKlaassen08}. 
\citet{Etoka12} used  MERLIN to image the Class II 6.7 GHz \met\,masers (typical signpost of HMSF), and found that the bulk of maser emission is indeed concentrated towards W51e2-E, and not the HC HII region W51e2-W. This further supports the scenario proposed by \citet{Shi10a} where the ongoing star formation activity in the region is not concentrated on the HC HII region but towards its companion 1\arcsec\,to the East.  

While the subarcsecond SMA study was successful in resolving multiple components,   a multi-level imaging study of the same molecule at subarcsecond resolution is required to study the kinematics and the physical conditions of hot molecular  gas  surrounding individual  high-mass YSO(s), and thus characterize their physical properties.
With this in mind,  we imaged  five \amm~inversion lines with energy levels high above  the ground state (equivalent to 400-1700~K), 
at an angular resolution of about 0\pas2, towards W51~Main.

The current paper is structured as follows.
The  observational setup and data calibration procedures are described in \S 2.
Maps and spectral profiles of different maser transitions are presented in \S 3. 
In \S 4 we present our analysis on the physical conditions of the molecular gas based on the \amm\,measurements. In \S 5, we discuss the  star formation activity in W51~Main. 
 Finally,  our conclusions are drawn in \S 6.

\begin{table*}
\caption{Parameters of JVLA observations toward W51.}             
\label{obs}      
\centering                        
\begin{tabular}{ccclccc} 
\hline\hline                 
\noalign{\smallskip}
\multicolumn{1}{c}{Transition$^{a}$} & \multicolumn{1}{c}{$\nu_{\rm rest}$} & \multicolumn{1}{c}{$E_u/k^{b}$} & \multicolumn{1}{c}{Date}  & \multicolumn{1}{c}{JVLA} & \multicolumn{1}{c}{Beamwidth$^{c}$}  & \multicolumn{1}{c}{RMS$^{d}$} \\ 
\multicolumn{1}{c}{(J,K)} & \multicolumn{1}{c}{(MHz)} & \multicolumn{1}{c}{(K)} & \multicolumn{1}{c}{(yyyy-mmm-dd)} & \multicolumn{1}{c}{Receiver}  & \multicolumn{1}{c}{$\theta_M('') \times \theta_m(''); \ P.A.(^{\circ})$} &  \multicolumn{1}{c}{(mJy/beam)}   \\
\noalign{\smallskip}
\hline
\noalign{\smallskip}
\multicolumn{7}{l}{\amm}\\
\noalign{\smallskip}
(6,6)   & 25055.96    &  408  & 2012-May-31 & K  & $0.29 \times 0.23;\ -17$  & 1.5 \\
(7,7)   & 25715.44    &  539  & 2012-May-31 & K  & $0.28 \times 0.23;\ -2 \ \ $  & 1.6 \\
(9,9)   & 27477.94    &  853  & 2012-Jun-21 & Ka & $0.24 \times 0.22;\ +53$  & 1.3 \\
(10,10) & 28604.75    & 1035  & 2012-Aug-07 & Ka & $0.23 \times 0.21;\ +61$  & 2.9 \\
(13,13) & 33156.84    & 1691  & 2012-Jun-21 & Ka & $0.20 \times 0.18;\ +52$  & 1.9 \\
\noalign{\bigskip}
\multicolumn{7}{l}{Other molecular transitions}\\
\noalign{\smallskip}
\met$^{e}$ &27472.53&234&2012-Jun-21&Ka&$0.24 \times 0.22;\ +53$  & 1.3 \\
CH$_3$CN$^{f}$ &36793.71&10&2012-Aug-07&Ka& $0.18 \times 0.16;\ +67 $ & 3.7\\
\noalign{\smallskip}
\hline   
\end{tabular}
\tablefoot{\\
(a) Transitions include ortho-\amm~($K=3n$) and para-\amm~($K\neq3n$).  \\
(b) Energy above the ground  reported from the JPL database.  \\ 
(c) Synthesized beams  in images made with the CASA task CLEAN with a robust parameter set to 0.5. \\
(d) RMS noise in a 0.4~\kms~channel without primary beam correction. 
After primary beam correction,  the noise level increases by up to 25\%.  \\
(e)  The $J_K$= 13$_2$-13$_1$ line of CH$_3$OH was detected in the same baseband as the \amm\,(9,9) line.\\
(f)  The CH$_3$CN (2-1) line was observed in a separated baseband paired with the \amm\,(10,10) line, but was not detected.
}
\end{table*}

\section{Observations and data reduction}
\label{obser}
Observations of NH$_3$ towards the W51 complex were conducted using the Karl G. Jansky Very Large Array (JVLA) of the National
Radio Astronomy
Observatory (NRAO)\footnote{NRAO is a facility of the National Science Foundation operated under cooperative agreement by Associated Universities, Inc.} in the B configuration. 
The observing setup and data reduction procedures were already described in detail in Paper I and II; but we  summarize them here as well. 
By using the broadband JVLA K- and Ka-band receivers, we
observed  a total of five metastable inversion transitions of NH$_3$: 
($J,K$)=(6,6), (7,7), (9,9), (10,10), and (13,13)  at  the 1~cm band, with frequencies ranging from \ $\approx 25$~GHz for the (6,6) line to $\approx 33$~GHz for the (13,13) line.    
Transitions were observed in pairs of independently tunable basebands 
during 6h tracks (two targets per track:  W51 -- this paper; NGC7538~IRS1 -- Paper I) on three different dates in 2012: 
the (6,6) and (7,7) lines on May 31 at K-band, the (9,9) and (13,13) lines on June 21, 
and the (10,10) transition on August 7, both  at Ka-band.  
Each baseband had eight sub-bands with a 4~MHz bandwidth   ($\approx$40~\kms\ at 30~GHz), providing a total coverage of 32~MHz ($\approx$320~\kms\ at 30~GHz). 
Each sub-band consisted of 128 channels with a separation of 31.25~kHz ($\approx$0.3~\kms\ at 30~GHz). 
The typical on-source integration time was about 80 min. Each transition was observed with 
fast switching, where 80s scans on-target were alternated with 40s
scans  on the nearby (1.2$^{\circ}$ on the sky) QSO J1924+1540  (measured flux density 0.6--0.7~Jy, depending on frequency). 
We derived  absolute flux calibration from observations of 3C~48 ($S_{\nu}$ = 0.5--0.7~Jy, depending on frequency), 
and bandpass calibration from observations of 3C~84 ($S_{\nu}$ = 27--29~Jy, depending on frequency).

The data were edited, calibrated, and imaged in a standard fashion using the Common Astronomy Software Applications (CASA) package. 
We fitted and subtracted continuum emission  from the spectral line data in the uv plane  
using CASA task UVCONTSUB, combining the continuum (line-free) signal from all eight sub-bands around the \amm\ lines. 
Before imaging, we performed self-calibration on the strong  (6,6) \amm\,maser detected in W51-North (velocity of 47.6 \kms, and peak flux density $\sim$5 Jy; see Paper II). 
We then applied the self-calibration solutions from the reference channel with the maser to the dataset containing the lines (6,6) and (7,7)\footnote{ While the self-calibration solutions had a significant effect on the image quality of the maser line, improving  its dynamic range by a factor of 4,  they had  a more limited effect on the images of the continuum and the \amm\,thermal line emission (the latter occurs at a different velocity with respect to the maser, $\sim$57~\kms\,vs. $\sim$47~\kms, respectively; see Paper II). The quality of the final maps obtained applying the self-calibration solutions was nevertheless better also in the case of  thermal emission.}. 
Since the (9,9) maser line  was much weaker than the (6.6) line (peak flux density $\sim$0.4 Jy), 
we did not perform self-calibration on the dataset containing the (9,9) and (13,13) lines (nor  the 10,10 transition). 
Using the CASA task CLEAN, we imaged the W51~Main region with a cell size of 0\farcs04, covering a 
20\arcsec~field around the position  $\alpha(J2000) = 19^h 23^m 43^s.90$,   $\delta(J2000) = +14^{\circ} 30' 34$\pas6. 
We adopted Briggs weighting with a ROBUST parameter set to 0.5 and smoothed the velocity resolution to 0.4~\kms, for all transitions. 
The resulting synthesized clean beam FWHM were \ 0\farcs19--0\farcs26 (depending on frequency)  
and the typical RMS noise level per  channel was \ $\approx$1.5~mJy~beam$^{-1}$ (except for the dataset containing the 10, 10 doublet, which was noisier due to bad atmospheric conditions and other issues). 
Since the observations were conducted pointing the telescopes at W51-IRS1 
(with a sky position of $\alpha(J2000) = 19^h 23^m 42^s.00$,   $\delta(J2000) = +14^{\circ} 30' 50$\pas0),  
to include both W51-IRS2 and W51 Main in the JVLA antennae's primary beam,  
we applied primary beam corrections during cleaning (on the order of 15-25\%, depending on transition). 
Table~\ref{obs} summarizes the observations.

\begin{table*}
\caption{Parameters of the \amm\, inversion lines observed around W51~Main}             
\label{lines}      
\centering                        
\begin{tabular}{ccccccccclcccl} 
\hline\hline                 
\noalign{\smallskip}
\multicolumn{1}{c}{Line} &  & \multicolumn{1}{c}{F$_{\rm peak}$} &  \multicolumn{1}{c}{V$_{c}$} & \multicolumn{1}{c}{$\Delta V_{1/2}$}&\multicolumn{1}{c}{F$_{\rm int}$} &\multicolumn{1}{c}{F$_{\rm cont}$}\\ 
\multicolumn{1}{c}{(J,K)} &  &  \multicolumn{1}{c}{(Jy)}  & \multicolumn{1}{c}{(km/s)}     & \multicolumn{1}{c}{(km/s)}&  \multicolumn{1}{c}{(Jy~km/s)} &  \multicolumn{1}{c}{(Jy)}\\
\noalign{\smallskip}
\hline
\noalign{\bigskip}
 \multicolumn{6}{c}{W51e2-W (HC HII)} \\
 \multicolumn{6}{c}{J2000 19:23:43.9054 +014.30.34.487, $0.8\arcsec\times0.9\arcsec$, PA$=0\deg$} \\
\noalign{\smallskip}
(6,6)   &     &  $-0.336\pm0.001$ & $57.14\pm0.01$ & $7.29\pm0.03$ & $-2.61\pm0.01$  \ \   & { 0.362}\\
(7,7)   &     &  $-0.310\pm0.001$ & $57.09\pm0.01$ & $5.93\pm0.02$ & $-1.96\pm0.01$  \ \   & { 0.362} \\
(9,9)   &     &  $-0.227\pm0.001$ & $57.22\pm0.01$ & $5.54\pm0.03$ & $-1.34\pm0.01$  \ \   & { 0.346} \\
(10,10) &   &  $-0.150\pm0.004$ & $57.45\pm0.06$ & $4.38\pm0.14$ & $-0.70\pm0.03$  \ \   & { 0.258} \\
(13,13) &   &  $-0.072\pm0.004$ & $57.53\pm0.12$ & $3.87\pm0.28$ & $-0.30\pm0.03$  \ \   & { 0.369} \\
\noalign{\smallskip}
\met    &     &  $-0.287\pm0.006$ & $57.72\pm0.04$ & $4.12\pm0.09$ & $-1.30\pm0.03$  \ \   & { 0.346} \\
\noalign{\smallskip}
\noalign{\bigskip}
 \noalign{\bigskip}
 \multicolumn{6}{c}{W51e2-E (Protostar)} \\
 \multicolumn{6}{c}{J2000 19:23:43.9600 +014.30.34.500, $0.9\arcsec\times 0.9\arcsec$, PA$=0\deg$} \\
 
\noalign{\smallskip}
(6,6)   &     &  $0.044\pm0.001$ & $57.00\pm0.12$ & $14.5\pm0.3$ & $0.68\pm0.02$  \ \   & --\\
(7,7)   &     &  $0.046\pm0.001$ & $56.30\pm0.10$ & $10.9\pm0.2$ & $0.53\pm0.02$  \ \   & -- \\
(9,9)   &     &  $0.023\pm0.001$ & $56.24\pm0.19$ & $11.6\pm0.5$ & $0.28\pm0.02$  \ \   & -- \\
(10,10) &   &  $0.020\pm0.003$ & $54.81\pm0.70$ & $8.74\pm1.65$ & $0.18\pm0.05$  \ \   & --\\
(13,13) &   &  $0.036\pm0.004$ & $55.53\pm0.27$ & $4.63\pm0.64$ & $0.18\pm0.03$  \ \   & --\\
\noalign{\smallskip}
\met    &     &  $0.040\pm0.003$ & $56.48\pm0.27$ & $8.80\pm0.65$ & $0.37\pm0.03$  \ \   & -- \\
\noalign{\smallskip}
\noalign{\bigskip}
 \noalign{\bigskip}
\multicolumn{6}{c}{W51e2-NW (Protostar)} \\
 \multicolumn{6}{c}{J2000 19:23:43.900 +014.30.35.980, $1.3\arcsec \times 2\arcsec$, PA$=0\deg$} \\
\noalign{\smallskip}
(6,6)   &     &  $0.099\pm0.002$ & $55.72\pm0.08$ & $10.7\pm0.2$ & $1.12\pm0.03$  \ \   & --\\
(7,7)   &     &  $0.075\pm0.002$ & $55.24\pm0.09$ & $7.72\pm0.22$ & $0.62\pm0.02$  \ \   & -- \\
(9,9)   &     &  $0.028\pm0.002$ & $55.64\pm0.30$ & $7.36\pm0.70$ & $0.22\pm0.03$  \ \   & -- \\
(10,10) &   &  $--$ & $--$ & $--$ & $--$  \ \   & --\\
(13,13) &   &  $--$ & $--$ & $--$ & $--$  \ \   & --\\
\noalign{\smallskip}
\met    &     &  $0.084\pm0.004$ & $55.46\pm0.16$ & $6.8\pm0.4$ & $0.61\pm0.03$  \ \   & -- \\
\noalign{\smallskip}
\noalign{\bigskip}
 \noalign{\bigskip}
  \multicolumn{6}{c}{W51e2-E+NW (Entire core)} \\
\noalign{\smallskip}
(6,6)   &     &  $0.432\pm0.003$ & $55.38\pm0.04$ & $10.7\pm0.2$ & $4.90\pm0.06$  \ \   & --\\
(7,7)   &     &  $0.310\pm0.004$ & $55.53\pm0.05$ & $8.57\pm0.1$ & $2.82\pm0.05$  \ \   & -- \\
(9,9)   &     &  $0.099\pm0.004$ & $55.90\pm0.20$ & $8.98\pm0.5$ & $0.94\pm0.07$  \ \   & -- \\
(10,10) &   &  $0.059\pm0.007$ & $54.32\pm0.8$ & $13.0\pm2$ & $0.82\pm0.16$  \ \   & --\\
(13,13) &   &  $0.069\pm0.012$ & $55.77\pm0.43$ & $5.1\pm1.0$ & $0.38\pm0.10$  \ \   & --\\
\noalign{\smallskip}
\met    &     &  $0.444\pm0.010$ & $55.6\pm0.1$ & $8.3\pm0.2$ & $3.92\pm0.09$  \ \   & -- \\
\noalign{\smallskip}
\noalign{\bigskip}
 \noalign{\bigskip}
 \multicolumn{6}{c}{W51e8} \\
 \multicolumn{6}{c}{J2000 19:23:43.9076 +014.30.28.068, $3.1\arcsec \times 4.3\arcsec$, PA=$0\deg$} \\
\noalign{\smallskip}
(6,6)   &     &  $0.279\pm0.003$ & $59.06\pm0.07$ & $12.39\pm0.17$ & $3.67\pm0.07$  \ \   & --\\
(7,7)   &     &  $0.190\pm0.004$ & $59.73\pm0.11$ & $11.20\pm0.26$ & $2.27\pm0.07$  \ \   & -- \\
(9,9)   &     &  $0.083\pm0.005$ & $59.65\pm0.31$ & $11.20\pm0.73$ & $0.99\pm0.09$  \ \   &-- \\
(10,10) &   &  $0.059\pm0.009$ & $60.60\pm0.92$ & $12.36\pm2.17$ & $0.78\pm0.18$  \ \   & -- \\
(13,13) &   &  $0.063\pm0.012$ & $59.82\pm1.24$ & $13.28\pm2.93$ & $0.88\pm0.26$  \ \   & --\\
\noalign{\smallskip}
\met    &     &  $0.260\pm0.009$ & $58.43\pm0.14$ & $7.82\pm0.32$ & $2.16\pm0.08$  \ \   &-- \\
\noalign{\smallskip}
\hline   
\end{tabular}
\tablefoot{
The peak fluxes (F$_{\rm peak}$, col. 2), the central velocities (V$_{c}$; col. 3), the FWHM line-width ($\Delta V_{1/2}$; col. 4), and the velocity-integrated flux (F$_{\rm int}$; col. 5) are estimated from single-Gaussian fits to the spectral profiles of the main lines.
 The table reports also the continuum emission flux density at the corresponding frequency of the inversion lines seen in absorption (F$_{\rm cont}$; col. 6). 
 }
\label{nh3_lines}
\end{table*}
\begin{figure}
\centering
\includegraphics[width=0.5\textwidth]{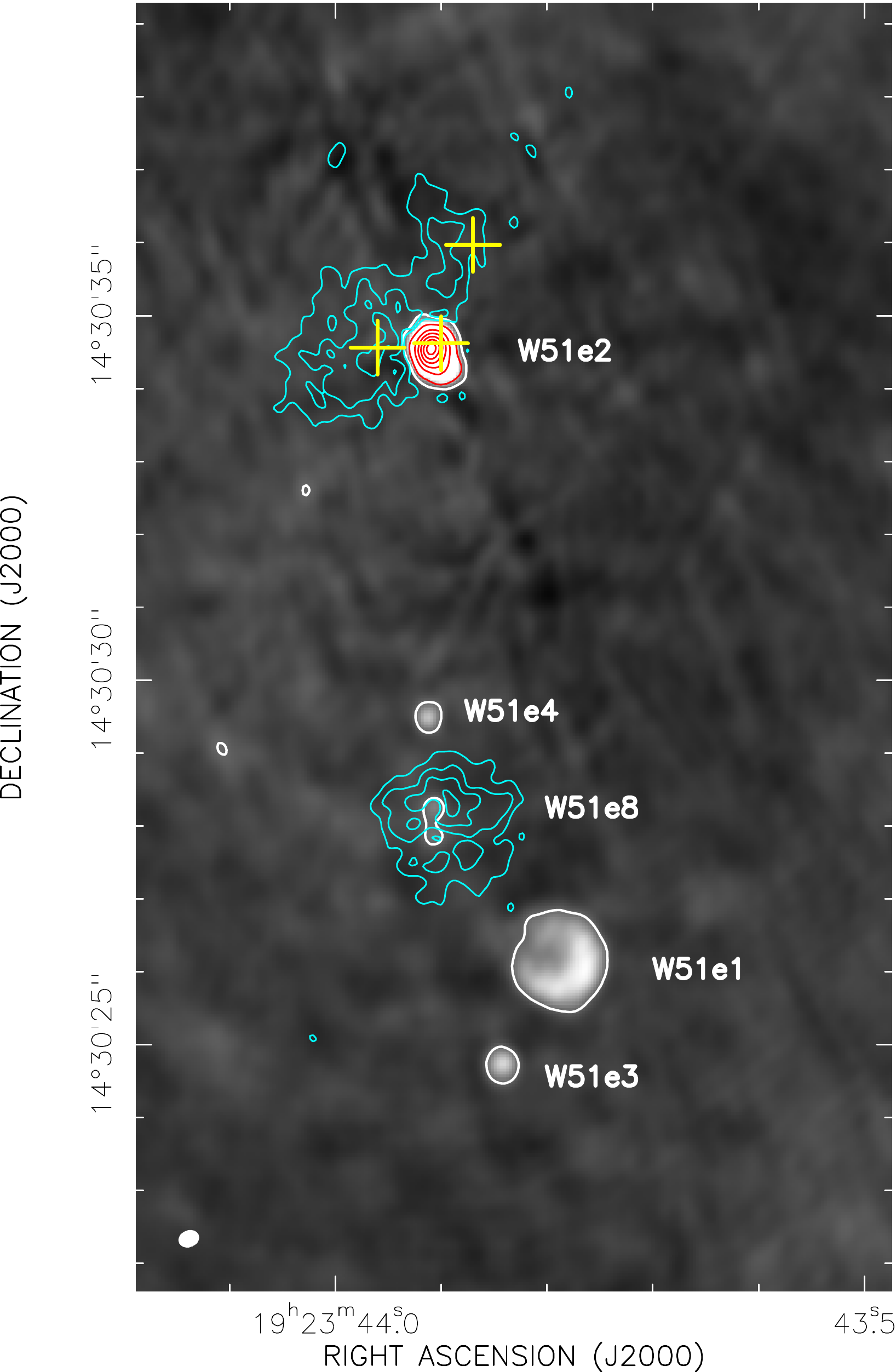}
\caption{Distribution of ionised and hot molecular gas in W51~Main as observed with the JVLA-B array.  
The 25 GHz continuum emission (gray scale and white contours) is produced by the cluster of HII regions W51e1, e2, e3, e4, and e8 (labelled in the plot). The 25 GHz continuum (white) contours indicate 2 mJy flux levels per beam   
 (corresponding to $3\sigma$ where $\sigma \sim0.7$  mJy beam$^{-1}$). 
The total intensity (0th moment) map of the (6,6) inversion transition of  NH$_3$ is overlaid on the continuum, showing hot and dense molecular gas towards W51e2 and W51e8. 
The \amm\,emission is displayed with cyan contours, representing 30\% to 90\% with steps of 20\% of the line peak for the (6,6) line, 107 mJy beam$^{-1}$ km s$^{-1}$).  
 The \amm\,absorption is displayed with red contours, representing factors 1, 5, 9, 13,.... of --50~mJy~beam$^{-1}$~\kms, for all transitions.  
 No flux cutoff was applied.  
The images were integrated over the velocity range 48~\kms~to 70~\kms, while the velocity resolution was smoothed to 0.4~\kms. 
The synthesized beam (0\pas29$\times$0\pas23)  is shown in the lower left corner of the panel. 
The images were constructed with 0\pas04 pixels for all transitions. 
The yellow crosses around W51e2 mark the positions of the 0.8mm  continuum peaks identified by \citet{Shi10a} and corresponding to sources W51e2-E, W51e2-W, and W51e2-NW (from left to right), claimed to be individual high-mass YSOs. 
}
\label{w51main}
\end{figure}

\section{Results}
\label{res}
We have mapped the hot \amm~gas in  five highly excited metastable inversion transitions  of NH$_3$ (J,K)=(6,6), (7,7), (9,9), (10,10), and (13,13) with 0\pas26-0\pas19 resolution towards W51~Main.  
Besides \amm, we mapped also  the $J_K$= 13$_2$-13$_1$ line of CH$_3$OH (with a rest frequency of 27.47253 GHz), while the CH$_3$CN (2-1)  line (with a rest frequency of 36.79371 GHz) was included in our bandwidth but not detected. 
The parameters of all the observed transitions and the JVLA observations   are reported in Table~\ref{obs}.  
Using the line-free sub-bands, we also produced images of the radio continuum emission with an RMS noise level of $\sim$0.6--0.8~mJy~beam$^{-1}$, at frequencies 25-36 GHz.  

In the following we will discuss the morphology and velocity field of the  molecular gas mapped in  hot \amm\,and the \met\, line (Sect.~\ref{maps-res}), 
as well as the spectral profiles of  the observed lines (Sect.~\ref{spec}).

\subsection{Distribution and velocity of  hot molecular (\amm) gas}
\label{maps-res}

 Figure~\ref{w51main} shows the distribution of hot and dense molecular gas, as traced by the \amm\,(6,6) line\footnote{The lower excitation line traces best the bulk of the  emission from hot molecular gas.}, with respect  to the known HII regions in the W51~Main complex, as traced by the  25 GHz continuum emission (displayed with white contours). 
\amm\,is observed both in absorption (displayed with black contours)  against the strong HC HII region W51e2-W as well as in emission  (displayed with cyan contours) in hot  gas associated with other dense cores in the region: the dust continuum sources W51e2-E and W51e2-NW (to the east and north of the HC HII region, respectively), and the southern core associated with W51e8. 
There is no dense gas towards the other HII regions in the cluster; in particular, we do not see dense gas towards the cometary HII region W51e1, confirming previous results from low-excitation \amm\,transitions \citep{ZhangHo97}. 

In order to study the morphology as well as the kinematics of  hot molecular gas, 
we performed a moment analysis deriving maps of total intensity (0$^{th}$ moment) and velocity field  (1$^{st}$ moment), 
as well as  pv-diagrams, for each detected transition. 
In the following, we will discuss separately the results obtained towards individual sources.

\subsubsection{The HC HII region W51e2-W}
\label{w51e2-res}

\begin{figure*}[htb]
\centering
\includegraphics[angle=-90,width=\textwidth]{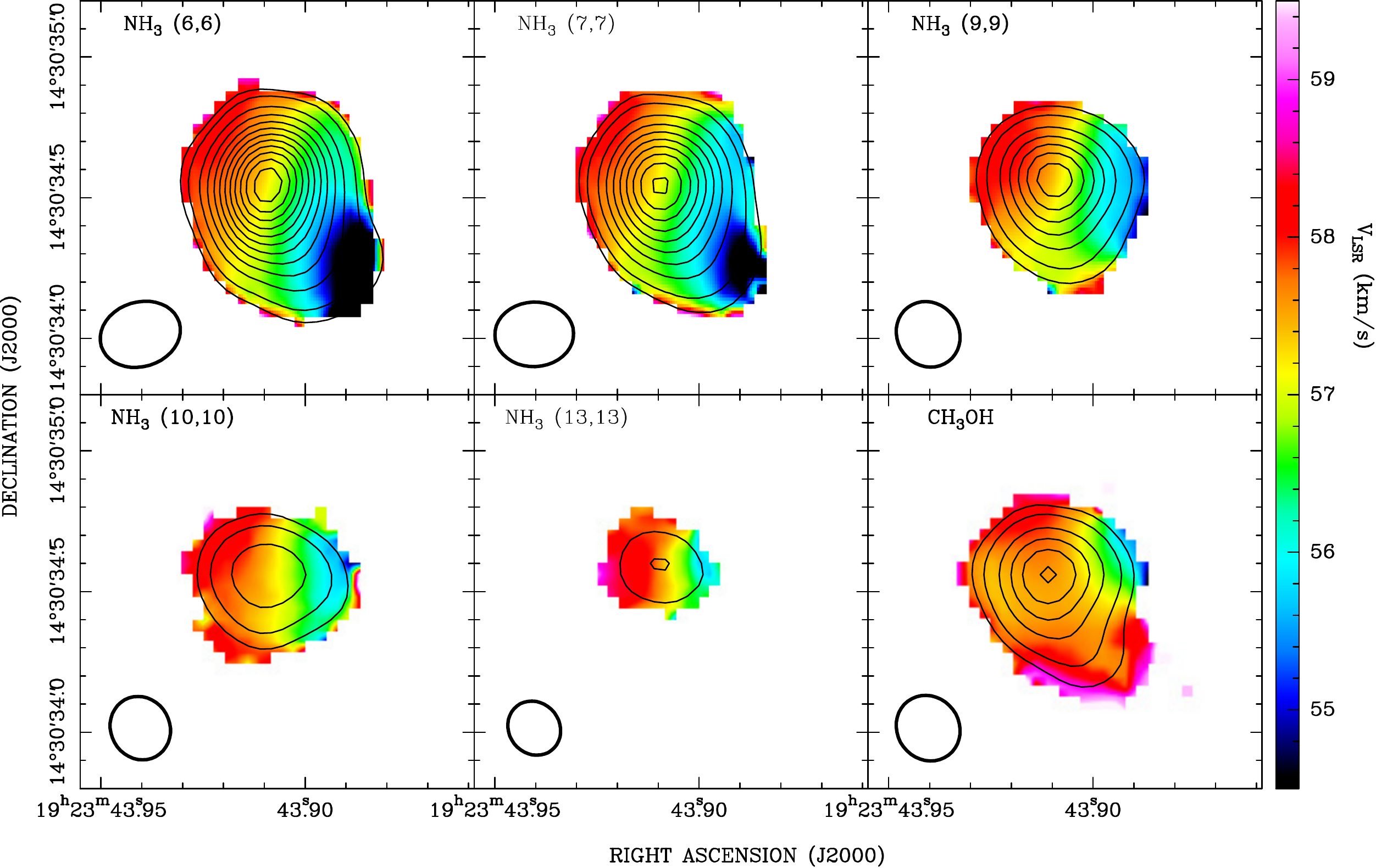}
\caption{Velocity fields of  five inversion transitions of  NH$_3$ as well as the \met\,line, as measured in absorption toward the HC HII region W51e2-W with the JVLA B-array.  
The total intensity 0$^{th}$ moment maps ({\it contours})  are overlaid on the velocity field 1$^{st}$ moment maps ({\it images}). 
Colors indicate V$_{LSR}$ in \kms. The images were constructed with a 0\pas04 pixel for all transitions. 
 The contours represent factors 1, 2, 4, 6,.... of --50~mJy~beam$^{-1}$~\kms, for all transitions.  
A flux cutoff of --5~mJy~beam$^{-1}$ ($\sim3\sigma$) was used to create  1$^{st}$ moment maps for the (6,6), (7,7), and (9, 9)  transitions, 
and a slightly higher  cutoff of --7~mJy~beam$^{-1}$ for the (10,10) and (13,13) transitions. 
Note that while the lower-excitation  transitions (6,6) and (7,7) show an apparent velocity gradient along NE-SW, 
higher-excitation  lines show consistently a gradient along E-W. 
We ascribe this difference to  a weaker component with blue shifted velocity detected towards SW in the lower-excitation lines. 
The synthesized beams (0\pas19--0\pas26)  are shown in the lower left corner of each panel (see Table~\ref{obs}). 
}
\label{mom1}
\end{figure*}

Towards We51e2-W, the \amm\,inversion transitions are observed in absorption and are detected up to  the (13,13) doublet (1700 K above the ground), indicating that there is a significant amount of hot gas presumably still surrounding  the HC HII region.  
Figure~\ref{mom1} shows the total intensity maps (with black contours) superimposed on the intensity-weighted  velocity field maps (shown in colours), for all \amm\,transitions as well as the \met\,line. 

Let us first consider the morphology of the \amm~absorption.  
For lower excitation transitions, \amm\, (6,6) and (7,7), as well as the \met\,line, 
we resolve the molecular gas absorption in a central core and an extra component to the SW. 
A Gaussian fit to the continuum emission from the core provides a deconvolved FWHM size of 0\pas27$\times$0\pas22 (averaged over frequencies which do not display the SW component: 27 to 36 GHz): the core of the HC HII region is therefore not resolved in our maps.  
Our measurements provide an upper limit to the linear size of the HC HII region of about 1300 AU (at the source distance of 5.4 kpc). 
The SW "extension" was previously detected  in both continuum emission at 3.6, 1.3, and 0.7 cm \citep{Gaume93,Shi10a} as well as RLs H26$\alpha$, H53$\alpha$, and H66$\alpha$ \citep{Shi10a}.  In particular, \citet{Gaume93} estimated the spectral indices for both the compact core and the SW extension, finding $\alpha$ of about 2 (optically thick) and 0.4, respectively, and suggested the presence of a one-side (SW) collimated ionized flow emitted from the core. 
\citet{Shi10a} used RLs from the cm to submm wavelengths to constrain electron temperatures and densities, and suggested that W51e2 powers  both a HC HII core, possibly an ionized disk (see also \citealt{KetoKlaassen08}), and a lower density ionised outflow with a single lobe extended to the SW. 
This scenario is not inconsistent with our \amm\,observations. 
The hot and dense material  surrounding  the HC HII region could  in principle be available for accretion onto the central O-type YSO and potentially make-up a molecular disk.     
 The SW extension  has the weakest integrated absorption 
 and it is not detected in the highest-$JK$ transitions, 
 indicating lower temperatures and/or densities than the core, as one would expect for outflowing gas. 
 The presence of dense gas to the W and N of W51e2 (see Fig.~\ref{w51main}), may explain why the outflow extends only towards the SW. 
 In order to establish the nature of the two components we can use the velocity field maps. 
 
\begin{figure}
\centering
\includegraphics[width=0.5\textwidth]{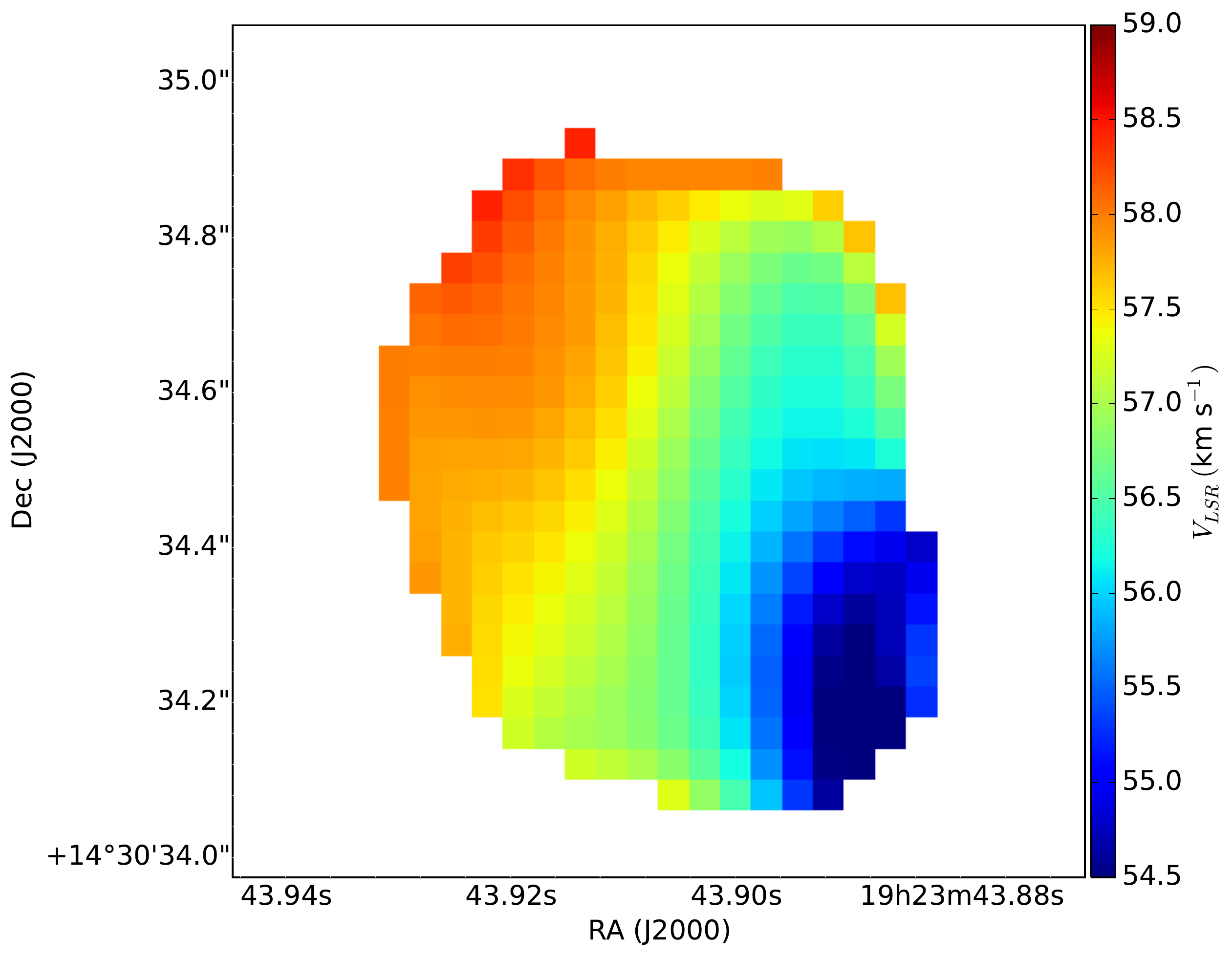}
\includegraphics[width=0.5\textwidth]{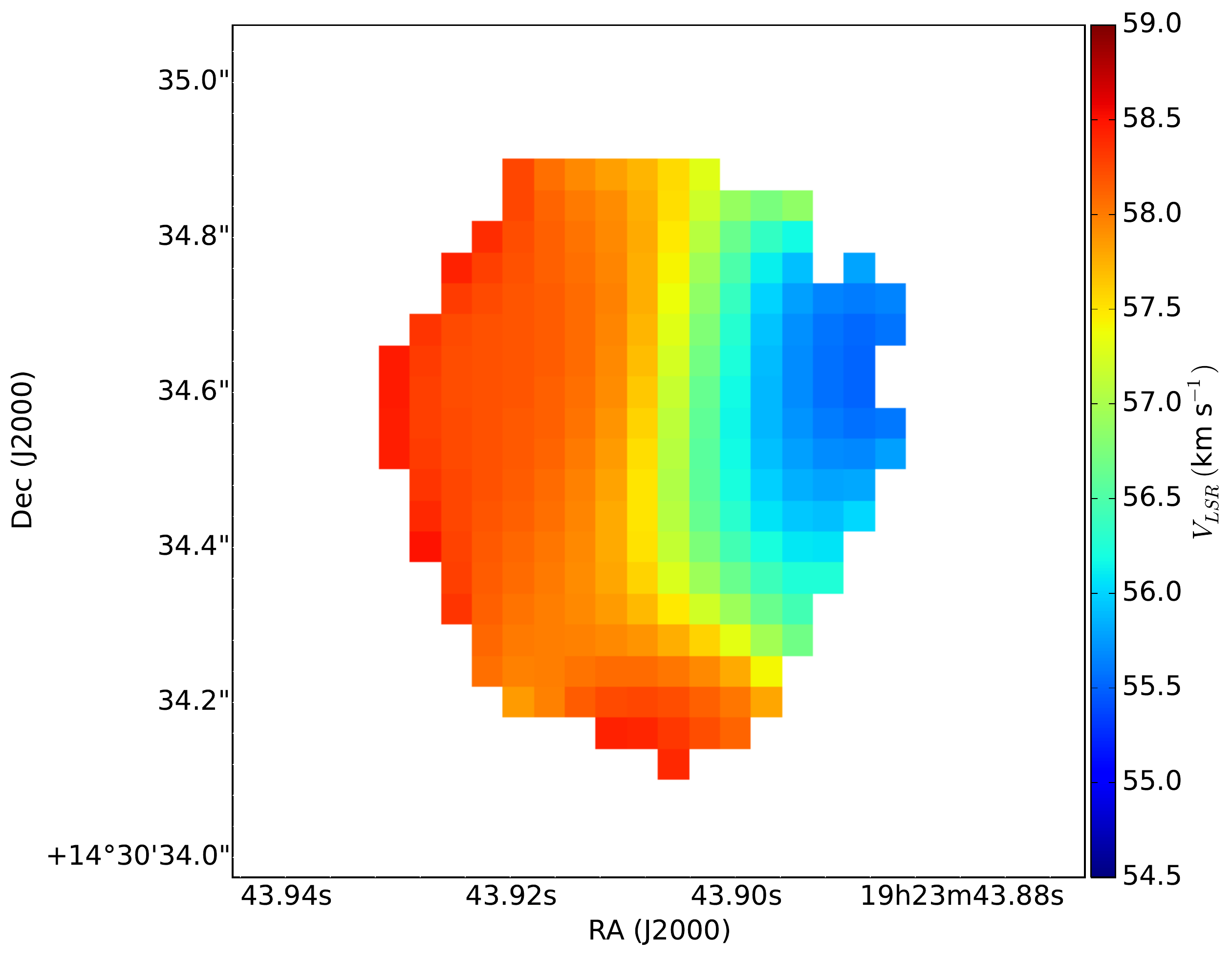}
\caption{Velocity field of the  NH$_3$ (6,6) inversion line for both the main line ({\it upper panel}) and one of the satellite components ({\it lower panel}), as fitted toward the HC HII region W51e2-W with the JVLA B-array.   
The images were constructed with 0\pas04 pixels. 
Colors indicate V$_{LSR}$ in \kms\,(color bar).  
The main component has an apparent velocity gradient along NE-SW, whereas the true velocity gradient displayed by the satellite component is along E-W (see text).
The velocity field maps were created by fitting the data cube (channel-by-channel) using pyspeckit \citep{Ginsburg2011c}. 
}
\label{e2abs-hf-vel}
\end{figure}

Perhaps the most striking feature about Figure~\ref{mom1} is  that the \amm~absorption  shows a well-defined velocity gradient in each line. 
This gradient appears to change orientation with the excitation line, going from NE-SW for the lower-excitation lines, (6,6) and (7,7), to E-W for more highly excited  lines, (9,9), (10,10), and (13,13). 
The velocity field of the lower-excitation lines, however, is affected by the SW extension, which is blue-shifted with respect to the core component and  may potentially hide the true sense of rotation in the compact core. 
Besides the main component, the spectral profile of an \amm\,inversion line displays four  symmetric satellite hyperfine components (see \S~\ref{spec}). 
Since they are more optically thin and do not display the SW extension, they can reveal the true velocity field more reliably than the main component.  
Therefore, we created a velocity field map for  one of the hyperfine components from the (6,6) doublet (which has the highest SNR). 
A comparison between the main and the satellite line velocity fields for the (6,6) doublet is shown in Figure~\ref{e2abs-hf-vel}. 
While the main line shows  a velocity gradient oriented NE-SW, the satellite component 
reveals a velocity gradient oriented E-W (i.e. at a position angle [P.A.] $\sim$90\degree, north to east), consistent with the higher excitation lines. 
We conclude that the true sense of rotation of the molecular gas is E-W. 
This is inconsistent with an outflow along SW (see \S~\ref{discussion}).

%
%
%
 
\begin{figure}
\centering
\includegraphics[angle=-90,width=0.5\textwidth]{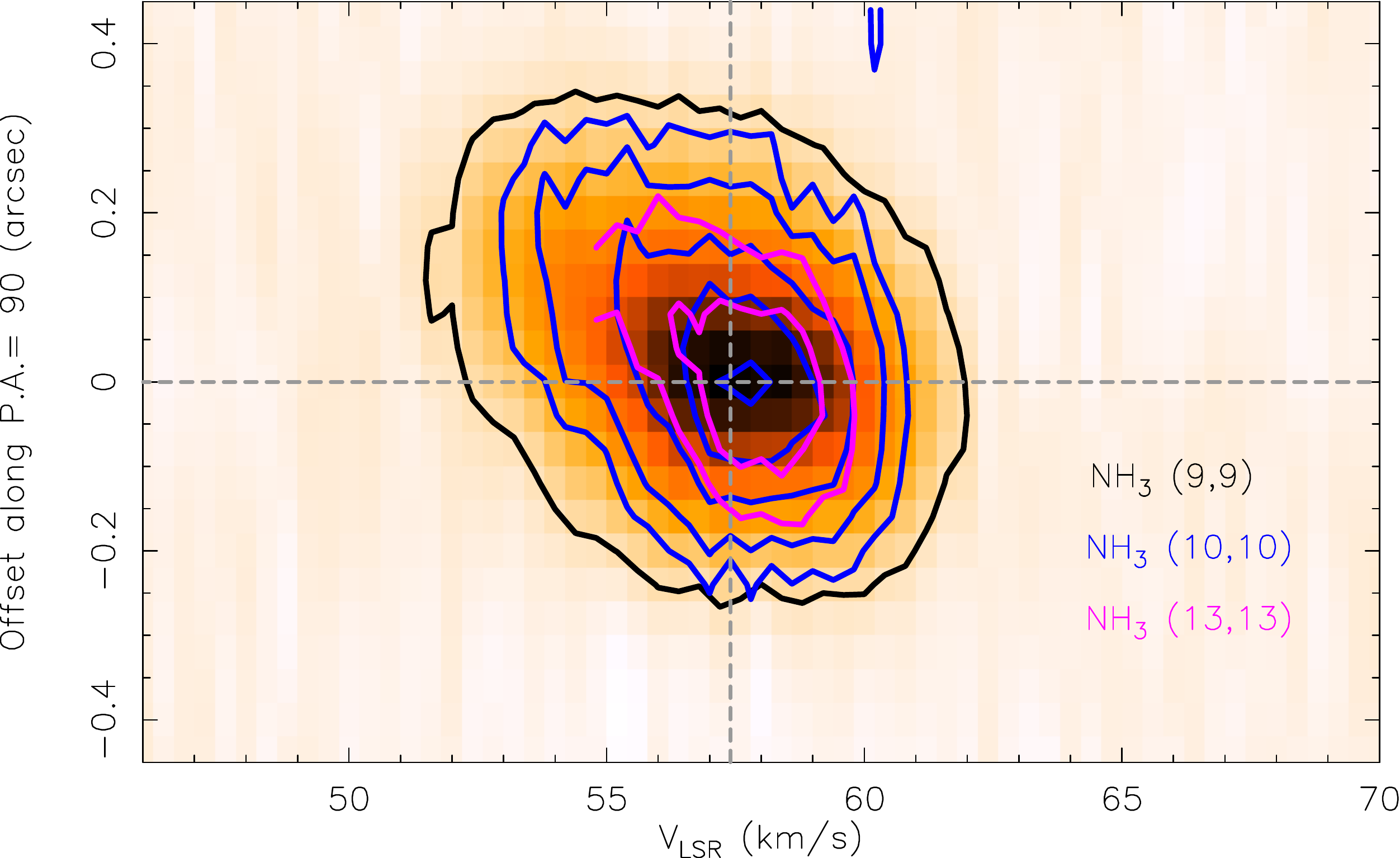}
\caption{Overlay of position-velocity diagrams of the (J,K) = (9,9), (10,10), (13,13) lines observed with the JVLA B-array towards W51e2-W. 
The cut is taken at the peak of the \amm\,core with a P.A. = 90\degree. 
The image shows the pv-diagram for the (9,9) doublet. 
The contours are -0.005 Jy beam$^{-1}$ for the (9,9) transition (in black), 
-0.005, -0.01, -0.02, -0.03, -0.04 Jy beam$^{-1}$ for the   (10,10) transition (in blue), 
-0.005 and -0.01 Jy beam$^{-1}$ for the   (13,13) transition (in cyan), respectively.  
The vertical dashed gray line indicates the velocity at 57.4~\kms. 
The P.A. is measured from north to east.
}
\label{pv_991013}
\end{figure}

 Besides 1$^{st}$ moment maps,  we also made pv-plots of the \amm\, inversion lines.   
Figure~\ref{pv_991013} shows an overlay of pv-diagrams for the most highly-excited lines\footnote{We excluded the lower-excitation lines because their velocity fields are affected by the SW component.}: (J,K) = (9,9), (10,10), and (13,13).  
The cuts  are taken at the peak of the \amm\,core along the direction of the main velocity gradient observed in the velocity field maps, P.A. = 90\degree\,(E-W).    
  Although a clear velocity gradient is evident in the pv-diagrams as well, there is no evidence of  steepening of such a gradient with  increasing  excitation energy (as also displayed in the 1$^{st}$ moment maps).  

%

\subsubsection{The W51e2 complex}
\begin{figure*}
\centering
\includegraphics[angle=-90,width=\textwidth]{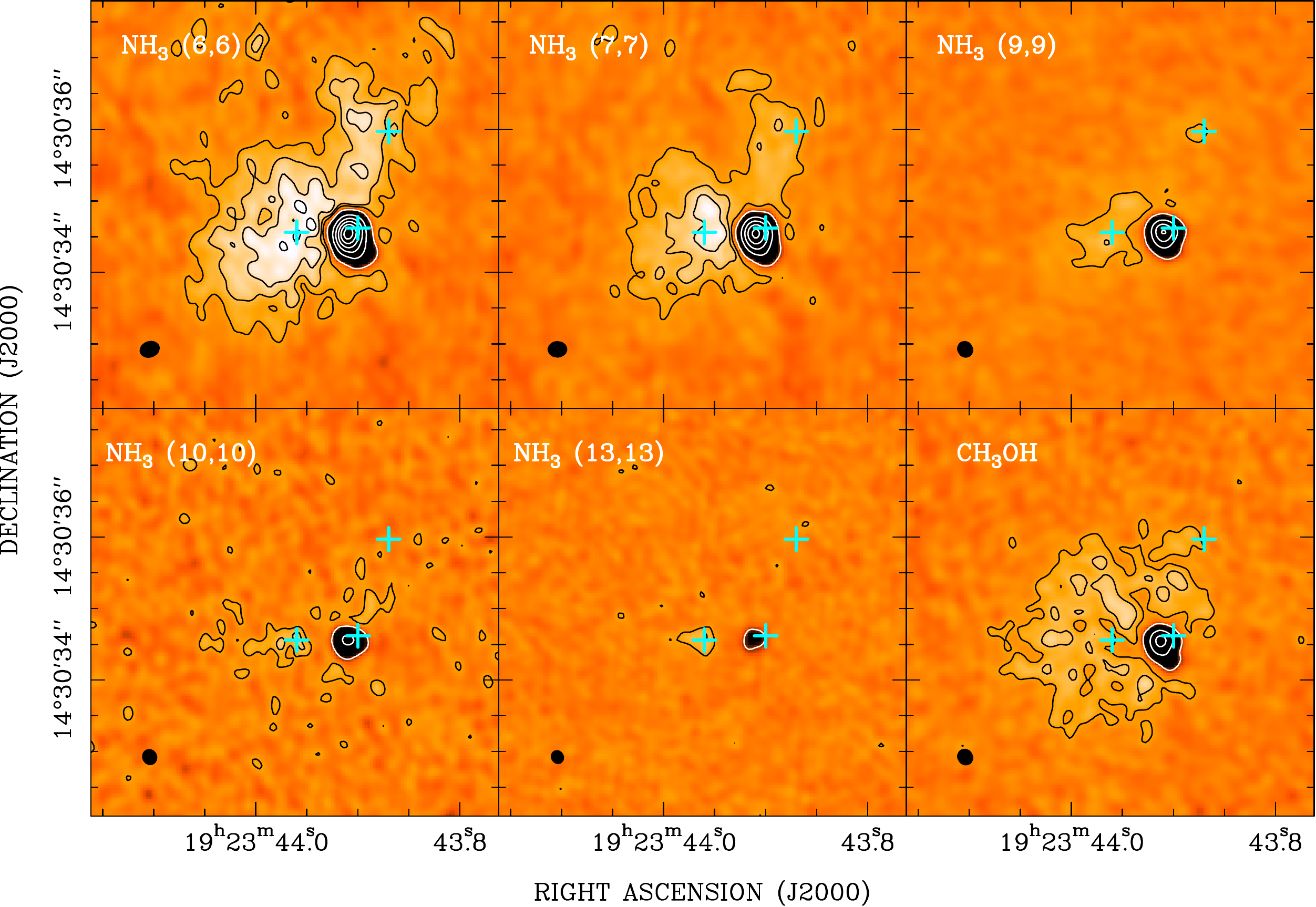}
\caption{Total intensity images of 5 inversion transitions of  NH$_3$ as well as the \met\,line towards W51e2. 
Light colours indicate emission and dark colours absorption. 
Contours of the \amm\,total intensity are also overplotted in the image: 
emission is displayed with black contours, representing factors 1, 2, 3, 4,.... of 20~mJy~beam$^{-1}$~\kms; absorption is displayed with white contours, representing factors 1, 5, 9, 13,.... of --50~mJy~beam$^{-1}$~\kms), for each transition, respectively. 
The images were integrated over the velocity range covering  the main hyperfine component for each transition (from 48~\kms~to 70~\kms).  
The velocity resolution was smoothed to 0.4~\kms\,for all transitions. 
The synthesized beams (0\pas19--0\pas26)  are shown in the lower left corner of each panel (see Table~\ref{obs}). The images were constructed with 0\pas04 pixels for all transitions. 
The cyan crosses mark the positions of the 0.8mm  continuum peaks identified by \citet{Shi10a} and corresponding to sources W51e2-E, W51e2-W, and W51e2-NW (from left to right), claimed to be individual high-mass YSOs. 
}
\label{w51e2em-mom0}
\end{figure*}

Figure~\ref{w51e2em-mom0} shows the total intensity images of various NH$_3$ inversion transitions as well as the \met\,line, integrated over their line-widths, in the surroundings of   W51e2. 
As already pointed out, while  \amm\,is  seen in absorption towards the HC HII region (displayed with white contours), hot molecular gas is  observed in emission (displayed with black contours) to the East and to the North of the HC HII region. 
The structure of the emission from lower-excitation lines (e.g., the 6, 6 and 7, 7 \amm\,doublets or  \met) is fairly consistent with the  dust emission  imaged at 0.8mm by \citet{Shi10a} (see their Fig. 1), 
although  peaks of warm dust (indicated with crosses in our Figure~\ref{w51e2em-mom0}) and warm ammonia emission do not correspond exactly. 
It is interesting to note that even the most highly excited lines, (10,10) and (13,13),  show  emission (although weak) towards  W51e2-E.
This  demonstrates that  W51e2-E is  powering a prominent hot core.   
On the other hand, in the vicinity of the dust peak W51e2-NW, hot \amm\,emission is observed only up to the (9,9) doublet and it is generally weaker than towards W51e2-E, indicating lower gas temperature and density (see \S~\ref{phycon}). 

\begin{figure}
\centering
\includegraphics[width=0.5\textwidth]{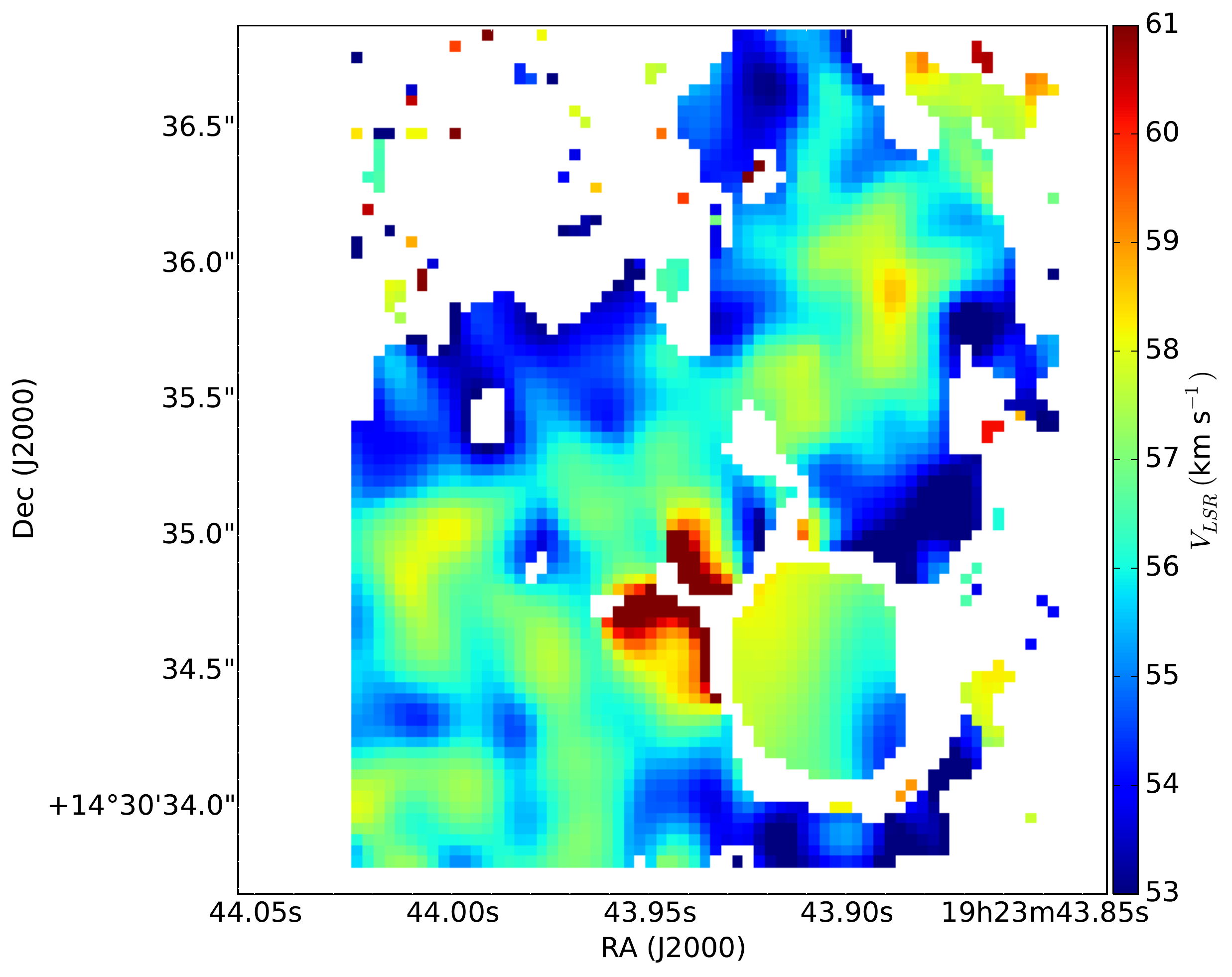}
\includegraphics[width=0.5\textwidth]{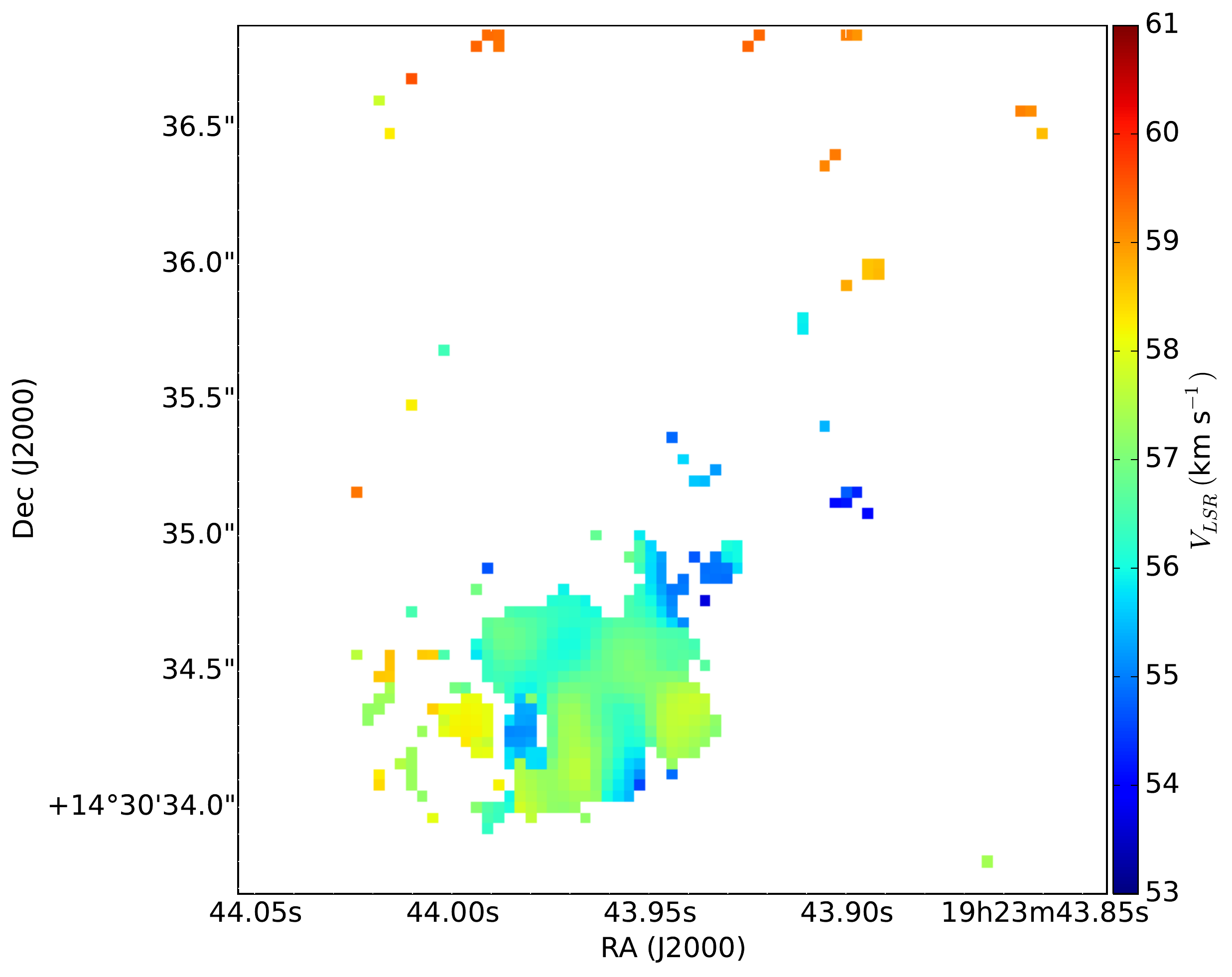}
\caption{Velocity field of the  NH$_3$ (6,6) inversion line for both the main line ({\it upper panel}) and one of the satellite components ({\it lower panel}), as measured in the W51e2 complex.   
Colors indicate V$_{LSR}$ in \kms. The images were constructed with 0\pas04 pixels. 
Colors indicate V$_{LSR}$ in \kms, from 53 to 61~\kms\,by 0.4~\kms\,(color bar).  
}
\label{e2em-vel}
\end{figure}

We also created velocity field maps for the \amm\,emission in the W51e2 core. 
We used the (6,6) doublet which has the highest SNR and made velocity maps in both the main component and one of the hyperfine satellites (Figure~\ref{e2em-vel}). 
Although the velocity field  does not show any clear regular pattern, there is an interesting redshifted component near W51e2-E (Figure~\ref{e2em-vel}, upper panel), which may indicate infalling gas (see discussion in \S~\ref{discussion}).  
Unfortunately this redshifted emission is fairly weak, and we could not detect it in the hyperfine satellites to confirm conclusively its nature (Figure~\ref{e2em-vel}, lower panel).

\begin{figure}
\centering
\includegraphics[angle=-90,width=0.5\textwidth]{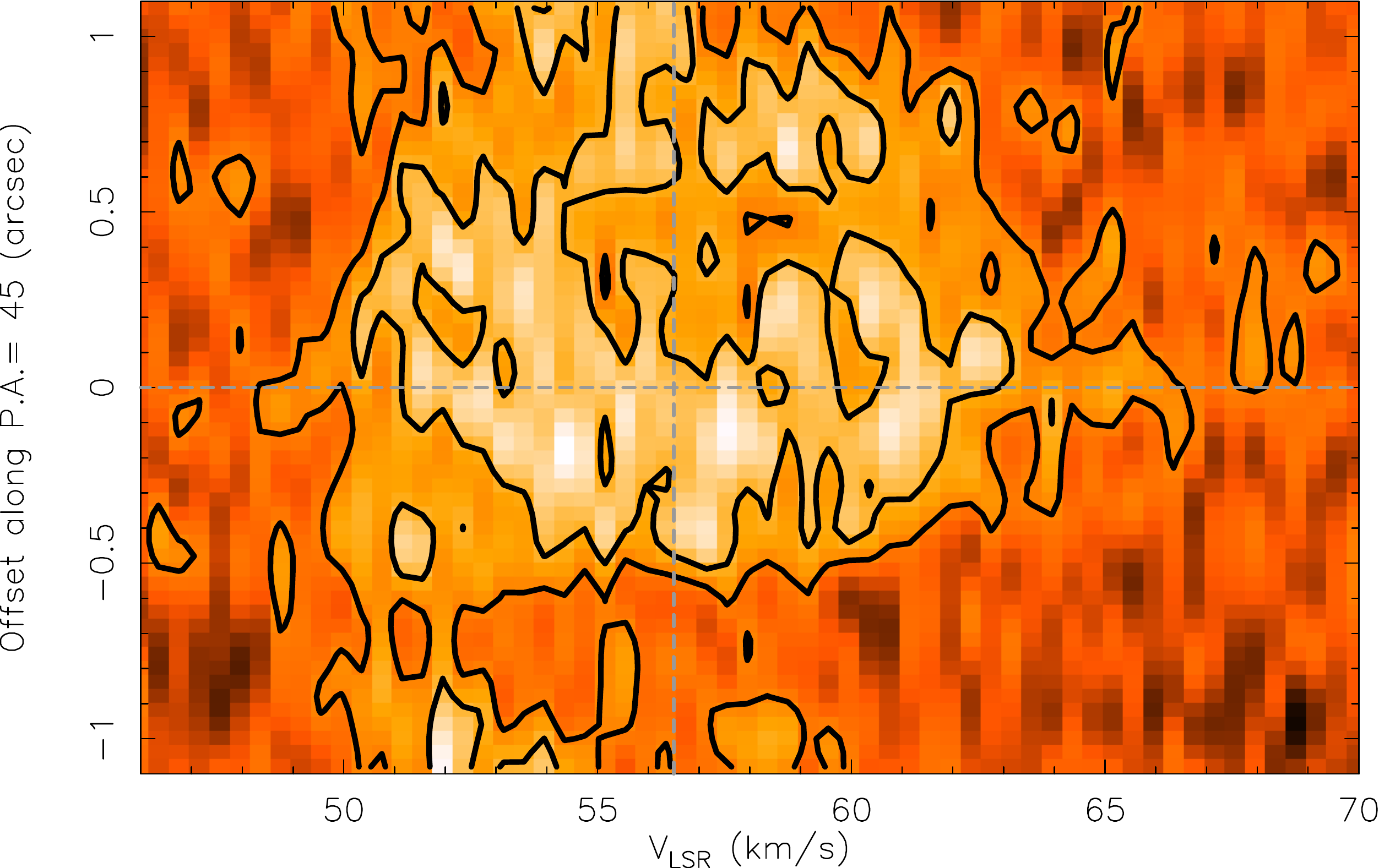}
\caption{Position-velocity diagram of the (J,K) = (6,6) line observed in emission around the protostar W51e2-E. 
The cut is taken at the presumed location of the protostar (see text) with a P.A. = 45\degree\,(north to east). 
The contours are 0.005 and 0.01 Jy beam$^{-1}$.   
The vertical dashed gray line indicates the velocity at 56.4~\kms. 
}
\label{pv_e2e}
\end{figure}

In order to better examine any potentially regular velocity pattern in the molecular gas surrounding W51e2-E, 
we made  a pv-diagram for the (6,6) doublet (which again has the highest SNR),  shown in Figure~\ref{pv_e2e}. 
The cut is taken at the presumed position of the protostar, which we assume is coincident with the peak of the (13,13) \amm\,emission (see also \S~\ref{discussion}), 
and at P.A. = 45\degree, i.e. perpendicular to the axis of the known CO outflow (see Figure~\ref{nh3+co} and \citealt{Shi10b}).    
From the pv-plot, there is no evidence for a velocity gradient that may indicate gas rotation. 
Nevertheless, the pv-diagram shows a C-shaped structure (in the blueshifted side) as well as an o-shaped structure (when the redshifted side is also included; although this feature is less clear). This pattern is expected for a radially infalling core, where the l.o.s. velocity displacement is expected to be maximum at the center and then to decrease away from it.  
An expanding shell or a wide-angle outflow cannot be completely ruled out, however.  An interpretation of this pattern will be given in \S~\ref{discussion}. 
%

\subsubsection{The W51e8 core}
\begin{figure*}
\centering
\includegraphics[angle=-90,width=\textwidth]{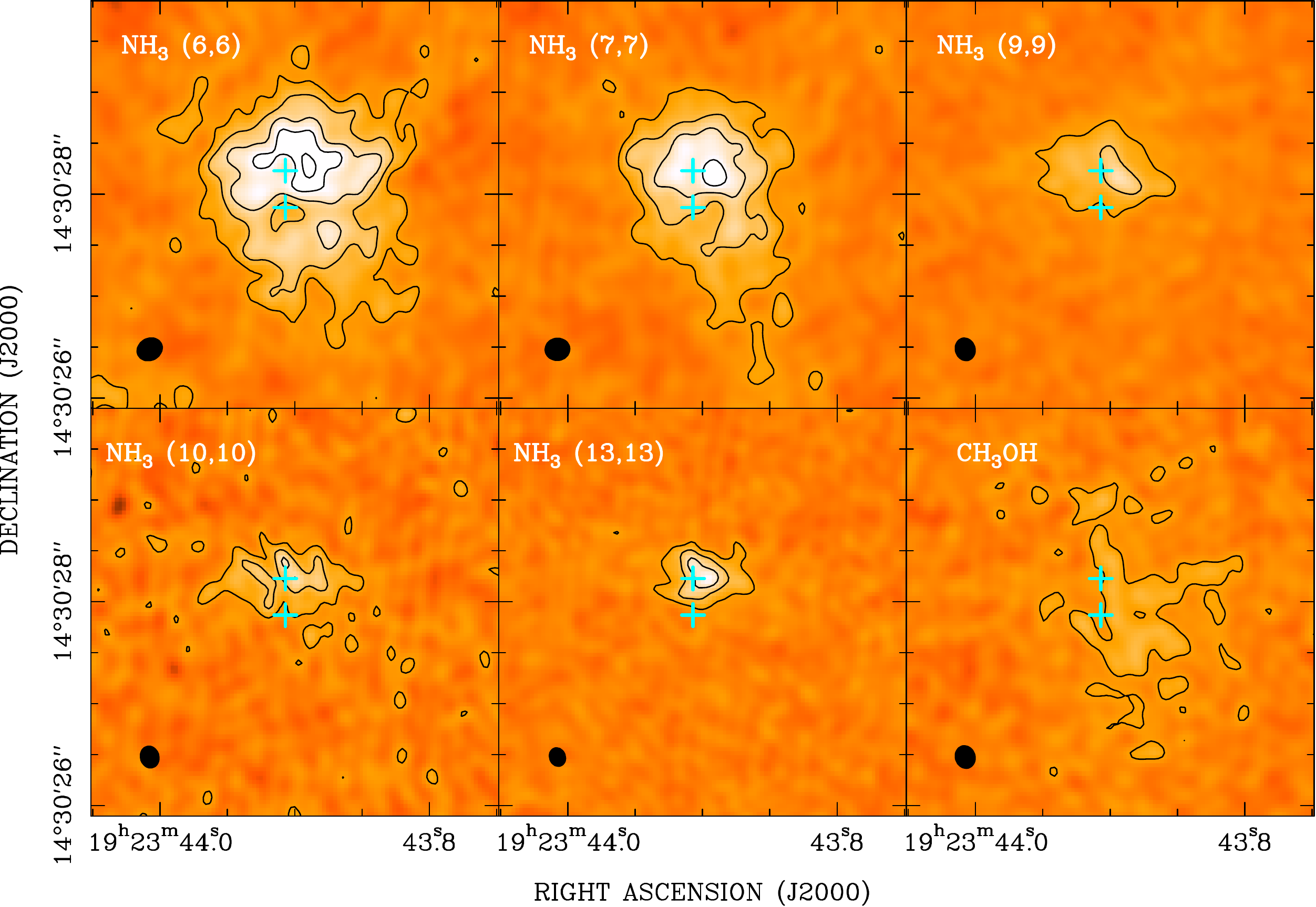}
\caption{Total intensity images of 5 inversion transitions of  NH$_3$ as well as the \met\,line towards W51e8. 
Contours of the \amm\,total intensity are also overplotted in the image, representing factors 1, 2, 3, 4,.... of 20~mJy~beam$^{-1}$~\kms. 
The images were integrated over the velocity range covering  the main hyperfine component for each transition (from 48~\kms~to 70~\kms).  
The velocity resolution was smoothed to 0.4~\kms\,for all transitions. 
The synthesized beams (0\pas19--0\pas26)  are shown in the lower left corner of each panel (see Table~\ref{obs}). 
The images were constructed with 0\pas04 pixels for all transitions. 
Note that ammonia emission  is detected towards W51e8  up to the (13,13) doublet. 
The cyan crosses mark the positions of the radio continuum peaks identified at 25 GHz (see Fig.~\ref{w51main}). 
}
\label{e8-mom0}
\end{figure*}

Figure~\ref{e8-mom0} shows the total intensity images of various NH$_3$ inversion transitions and the \met\,line towards W51e8. 
The emission shows a central stronger component which is elongated E-W across $\sim$2\arcsec\, and is detected in all transitions and a more diffuse, nearly-spherical core surrounding the central component, which is detected only at lower excitation. 
This indicates temperature variations within the core, with the lower-JK doublets tracing the cooler parts of the \amm\,core. 

\begin{figure}
\centering
\includegraphics[width=0.5\textwidth]{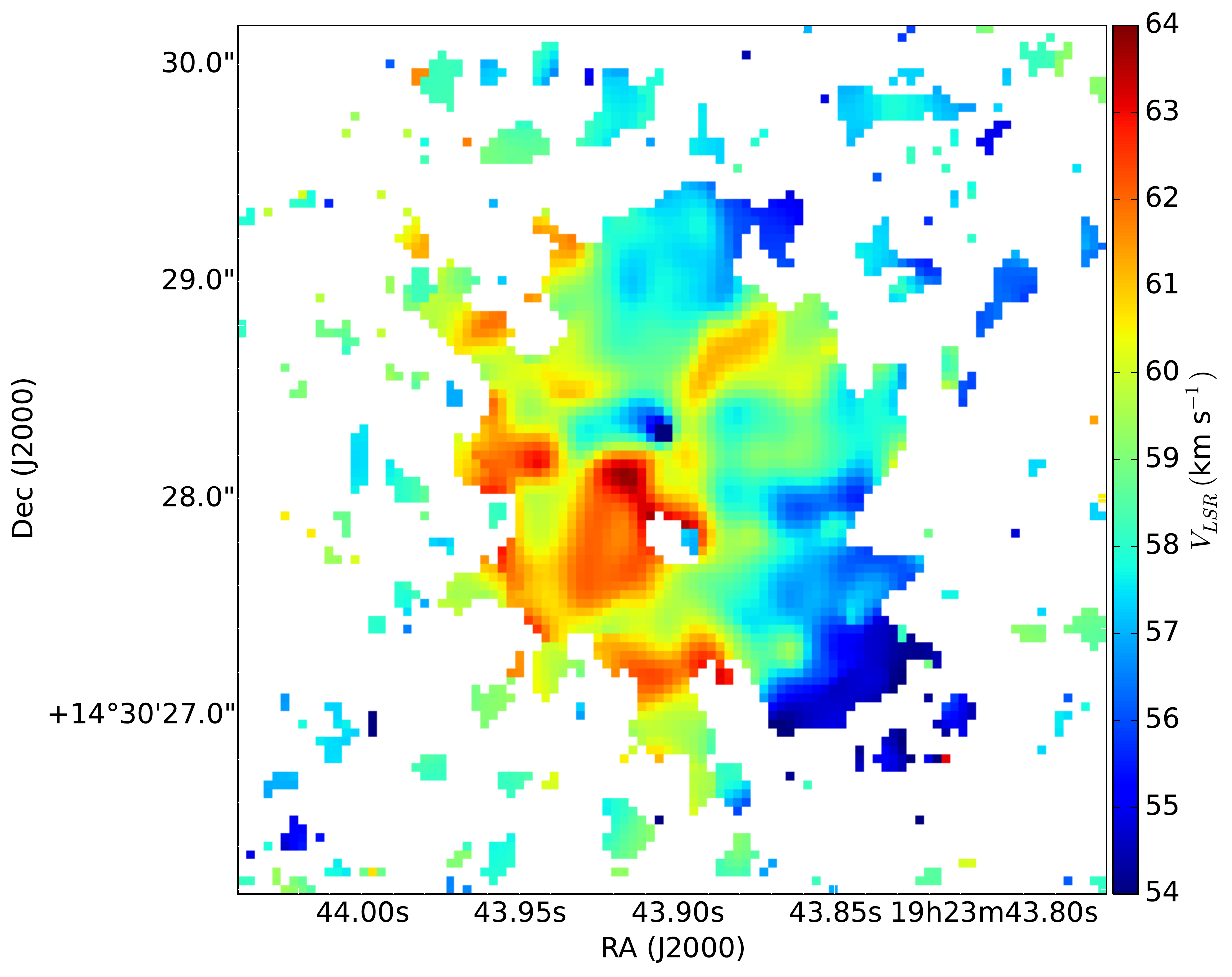}
\includegraphics[width=0.5\textwidth]{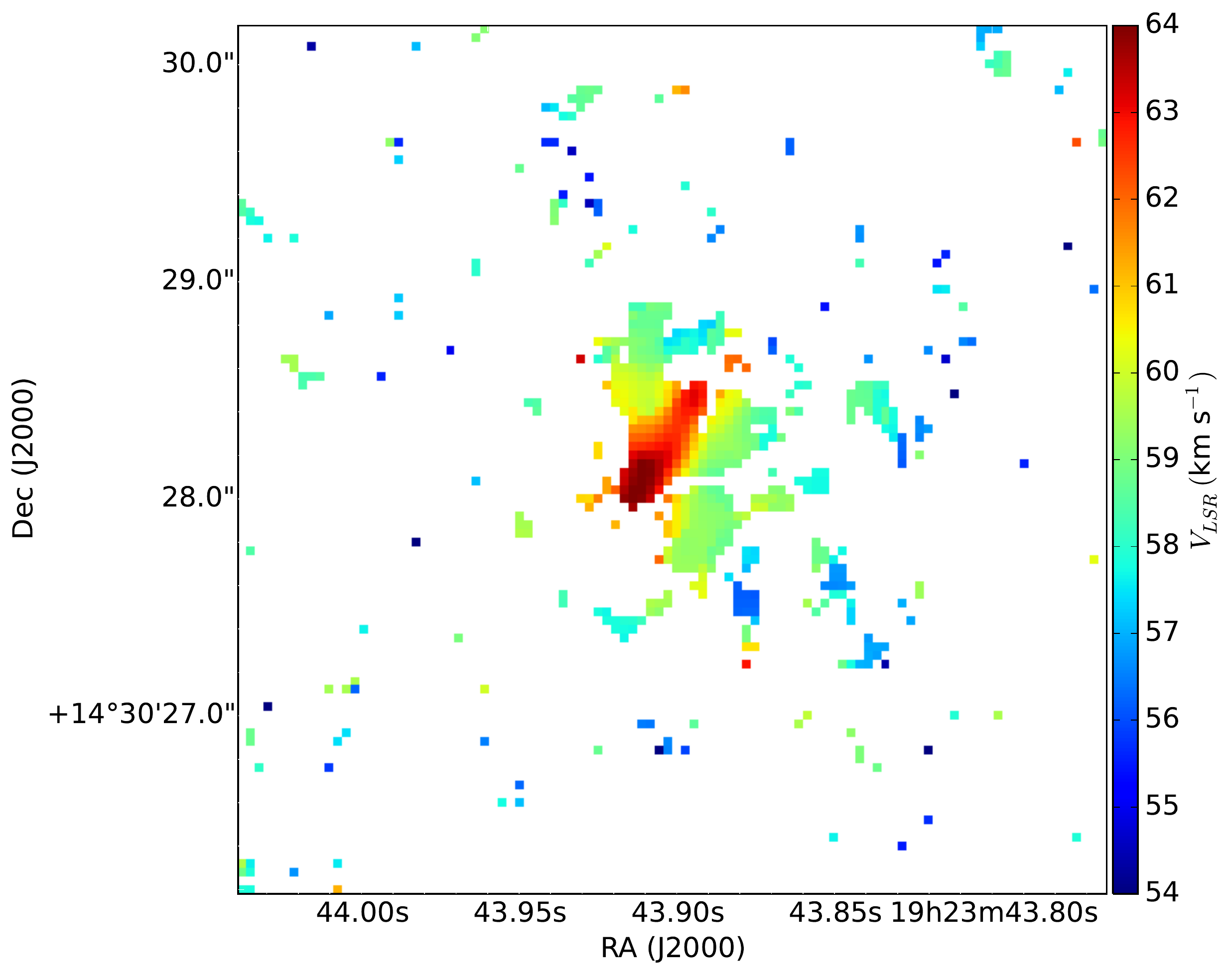}
\caption{Velocity field of the  NH$_3$ (6,6) inversion line for both the main line ({\it upper panel}) and the satellite components ({\it lower panel}), as measured in emission towards W51e8.   
Colors indicate V$_{LSR}$ in \kms\,(color bar). The images were constructed with  0\pas04 pixels. 
}
\label{e8-mom1}
\end{figure}

\begin{figure}
\centering
\includegraphics[angle=-90,width=0.5\textwidth]{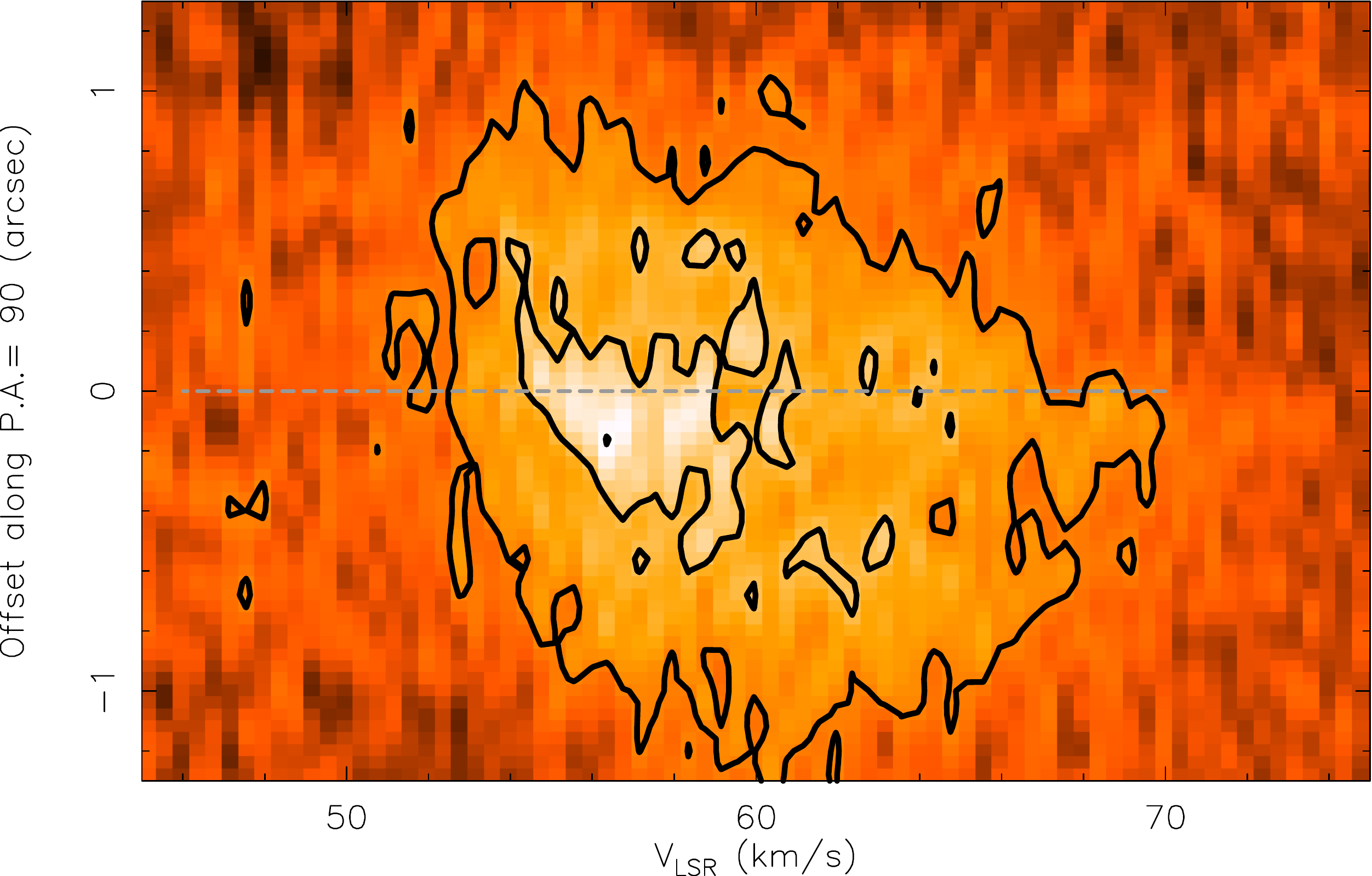}
\includegraphics[angle=-90,width=0.5\textwidth]{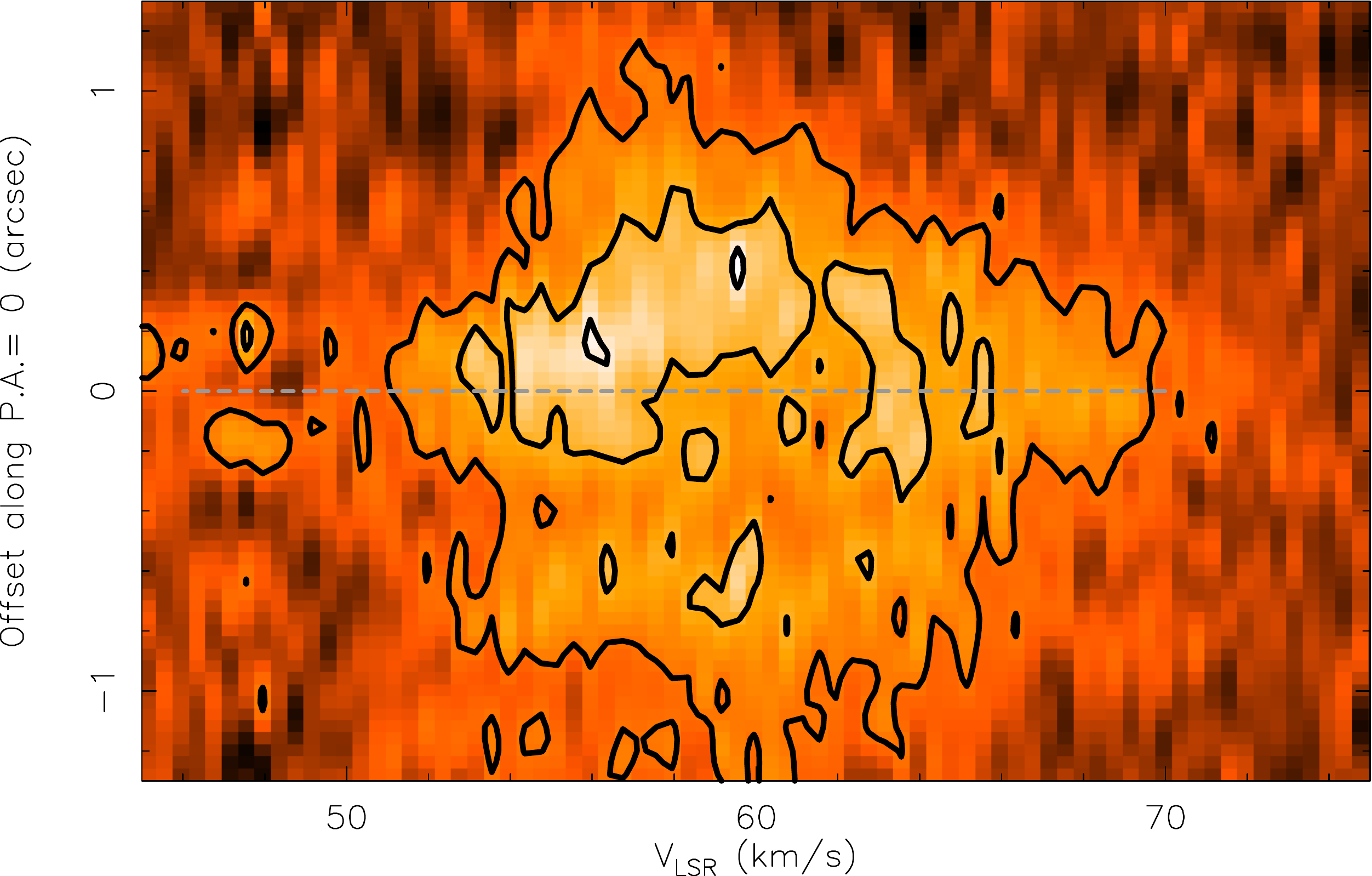}
\caption{Position-velocity diagram of the (J,K) = (6,6) line observed in emission around the protostar W51e8. 
The cut is taken at the presumed location of the protostar (see text) with a P.A. = 90\degree\,(top panel) and P.A. = 0\degree\,(bottom panel). The P.A. is measured from north to east.
The contours are 0.002 and 0.004 Jy beam$^{-1}$.   
}
\label{pv_e8}
\end{figure}

We also created velocity field maps for  both the main hyperfine line and one of the hyperfine satellites from the (6,6) doublet (see Figure~\ref{e8-mom1}). 
Although the velocity field looks quite complex, it is still possible to discern some regular patterns. 
For example, redshifted emission is almost exclusively located in the SE, while the emission to the west and north are mildly blueshifted; the lowest blue-shifted velocity is located to the SW. 
Therefore, the velocity field map suggests two directions with potential velocity gradients in the molecular gas surrounding W51e8: E-W and N-S.  
In order to confirm more confidently their presence, 
we made pv-diagrams of the (6,6) doublet at  P.A.=0\degree\,and  P.A. = 90\degree\,(lower and upper panels of Figure~\ref{pv_e8}, respectively). 
The cuts were taken at the peak of the 25~GHZ continuum emission, which also corresponds to the peak of the (13,13) \amm\,line  (this is the presumed location of the exciting protostar - see \S~\ref{discussion}). 
Although the structure is not symmetric with respect to the zero-position, these pv-plots support the presence of two velocity gradients in perpendicular directions. 

%

\begin{table*}
\caption{Parameters of the hyperfine components of \amm\, inversion lines observed around W51~Main.}             
\label{hf_lines}      
\centering                        
\begin{tabular}{ccccccccccc} 
\hline\hline                 
\noalign{\smallskip}
\multicolumn{1}{c}{Line} & \multicolumn{2}{c}{$\Delta \nu_{HF}$}& \multicolumn{2}{c}{$\Delta V_{HF}$} &   \multicolumn{1}{c}{$a_{ms}$}&  \multicolumn{1}{c}{F$_{\rm peak}$} &  \multicolumn{1}{c}{$\Delta V_{1/2}$}&\multicolumn{1}{c}{F$_{\rm int}$} &\multicolumn{1}{c}{$V_{c}$}
& Peak Opacity 
\\ 
\multicolumn{1}{c}{(J,K)} & \multicolumn{2}{c}{(MHz)}  & \multicolumn{2}{c}{(\kms)}  & & \multicolumn{1}{c}{(Jy)} & \multicolumn{1}{c}{(km/s)}      &  \multicolumn{1}{c}{(Jy~km/s)} & \multicolumn{1}{c}{(km/s)}
& ($\tau$)
\\
& Inner & Outer & Inner & Outer&&&&&\\
\noalign{\smallskip}
\hline
\noalign{\bigskip}
 \multicolumn{11}{c}{W51e2-W (HC HII)} \\
\noalign{\smallskip}
(6,6)   &  $\pm$2.24 & $\pm$2.62   &$\pm$26.9 & $\pm$31.4  & 0.0081 & -0.162 $\pm$0.001   & 4.29$\pm$0.04 & -0.74 $\pm$0.01 & 57.55$\pm$0.01 \ \  & $81\pm9$ \\
(7,7)   &  $\pm$2.34 & $\pm$2.68   &$\pm$27.3 & $\pm$31.2  & 0.0060 & -0.0642$\pm$0.0009 & 4.52  $\pm$0.12   & -0.31$\pm$0.01 & 57.22$\pm$0.03 \ \  & $39\pm8$ \\
(9,9)   &  $\pm$2.48 & $\pm$2.75   &$\pm$27.0 & $\pm$30.1  & 0.0037 & -0.0097$\pm$0.0003 & 3.60$\pm$0.20$^a$   & -0.067$\pm$0.003 & 57.36$\pm$0.16 \ \  & $12\pm9$ \\
\noalign{\bigskip}  
 \multicolumn{11}{c}{W51e2-E (Protostar)} \\
\noalign{\smallskip} 
(6,6)   &  $\pm$2.24 & $\pm$2.62   &$\pm$26.9 & $\pm$31.4  & 0.0081 & 0.0225 $\pm$ 0.0003 & 9.02 $\pm$ 0.12 & 0.217 $\pm$ 0.004 & 56.40 $\pm$ 0.09 \ \  & $89\pm86$ \\
(7,7)   &  $\pm$2.34 & $\pm$2.68   &$\pm$27.3 & $\pm$31.2  & 0.0060 & 0.0104 $\pm$ 0.0002 & 9.56 $\pm$ 0.24 & 0.106 $\pm$ 0.003 & 56.39 $\pm$ 0.18 \ \  & $43\pm59$ \\
(9,9)   &  $\pm$2.48 & $\pm$2.75   &$\pm$27.0 & $\pm$30.1  & 0.0037 & 0.0034 $\pm$ 0.0002 & 6.93$\pm$ 0.59 & 0.025 $\pm$ 0.003 & 56.47 $\pm$ 0.43 \ \  & $43\pm124$ \\
\noalign{\bigskip}  
 \multicolumn{11}{c}{W51e2-NW (Protostar)} \\ 
\noalign{\smallskip} 
(6,6)   &  $\pm$2.24 & $\pm$2.62   &$\pm$26.9 & $\pm$31.4  & 0.0081 & 0.0150$\pm$0.0005 & 8.54$\pm$0.32 & 0.136$\pm$0.007  \ \  & 55.27$\pm$0.22  & $20\pm15$\\
(7,7)   &  $\pm$2.34 & $\pm$2.68   &$\pm$27.3 & $\pm$31.2  & 0.0060 & 0.0043$\pm$0.0004 & 8.83$\pm$1.03 & 0.040$\pm$0.006  \ \  & 54.13$\pm$0.76  & $10\pm16$\\ 
(9,9)   &  $\pm$2.48 & $\pm$2.75   &$\pm$27.0 & $\pm$30.1  & 0.0037  & --  & --    & -- & -- \ \  & -- \\
\noalign{\bigskip}
 \multicolumn{11}{c}{W51e2-E+NW (Entire core)} \\ 
\noalign{\smallskip}
(6,6)   &  $\pm$2.24 & $\pm$2.62   &$\pm$26.9 & $\pm$31.4  & 0.0081 & 0.0675 $\pm$ 0.0009 & 9.6$\pm$0.16  & 0.692 $\pm$ 0.014  & 55.28 $\pm$0.11 \ \  & $21\pm2$\\
(7,7)   &  $\pm$2.34 & $\pm$2.68   &$\pm$27.3 & $\pm$31.2  & 0.0060 & 0.0186$\pm$0.0008 & 9.1$\pm$0.5 & 0.180$\pm$0.012  \ \  & 56.39 $\pm$0.35 & $10\pm16$ \\ 
(9,9)   &  $\pm$2.48 & $\pm$2.75   &$\pm$27.0 & $\pm$30.1  & 0.0037  & --  & --   --   & -- & -- \ \  & -- \\
\noalign{\bigskip}
 \multicolumn{11}{c}{W51e8} \\
\noalign{\smallskip}
(6,6)   &  $\pm$2.24 & $\pm$2.62   &$\pm$26.9 & $\pm$31.4  & 0.0081 & 0.0469$\pm$0.0004 &  14.1$\pm$0.4 & 0.705$\pm$0.007 &59.85$\pm$0.18 & $23\pm 3$ \\
(7,7)   &  $\pm$2.34 & $\pm$2.68   &$\pm$27.3 & $\pm$31.2  & 0.0060 & 0.0267$\pm$0.0009 &   9.2$\pm$0.4 & 0.260$\pm$0.01 &60.03$\pm$0.30 & $25\pm3$ \\
(9,9)   &  $\pm$2.48 & $\pm$2.75   &$\pm$27.0 & $\pm$30.1  & 0.0037 & 0.0083$\pm$0.0006 &   14.1$\pm$0.4 & 0.129$\pm$0.009 &63.5$\pm$1.3 & $28\pm6$ \\
\noalign{\smallskip}
\hline   
\end{tabular}
\tablefoot{
  The  frequency separations ($\Delta \nu_{HF}$, cols.~2 and 3) of the four satellite components were calculated  using Equation~\ref{freq_hf} in Appendix~\ref{app_b}. 
The corresponding velocity separations ($\Delta V_{HF}$) are also reported in cols.~4 and 5. 
$a_{ms}$ (col.~6) is the theoretical ratio of the satellite line  to the main line strengths  calculated  using Equations~\ref{int_m} in Appendix~\ref{app_b}. 
F$_{\rm peak}$, F$_{\rm int}$, and $\Delta V_{1/2}$ were fitted simultaneously for all the  hyperfine quadrupole satellite components assuming a Gaussian shape for each of the hyperfine lines.
The values reported in the table are average values of the four Gaussians fitted to the HFS satellites. 
$\tau$ (col. 10) is the line opacity estimated numerically with Equation A.1 from Paper I. \\
a : The best-fit parameters for the (9,9) line come from a fit that allowed the velocity offset of the two hyperfine lines to vary, since for the (9,9) line only, the fixed-offset hyperfine fit was of poor quality.\\
}
\label{nh3_hf}
\end{table*}

\subsection{Spectral profiles}
\label{spec}
We extracted spectral profiles towards individual sources by mapping each spectral channel and summing the flux density in each channel map  for each transition.  Figures~\ref{spec-e2w}, \ref{spec-e2e+nw}, and \ref{spec-e8} show such spectral profiles extracted towards the W51e2-W HC HII region, the sources seen in \amm\,emission in the  W51e2 core and the W51e8 core, respectively.  
As displayed in these plots, the  \amm\,profiles are not simply composed of a single spectral component. 
In fact, owing to the interaction with the quadrupole moment of the nitrogen nucleus, 
each \amm\,inversion line is actually split into five components,  
a ``main component'' and four symmetrically spaced ``satellites'', 
which make up the \amm\, hyperfine structure (HFS). 
The frequency separations and relative intensities of the four satellite components can be calculated 
using quantum mechanics formalism for a symmetric molecular rotor (see Appendix~\ref{app_b} for the relevant equations): 
these parameters are reported in Table~\ref{nh3_hf} (at least for the lines observed in this study).  
The satellite lines are spaced $\sim$27-31 \kms\,from their main components, so they are well resolved from each other. 

The entire  HFS spectra for the observed \amm\,inversion lines are shown in the left panels of Figures~\ref{spec-e2w}, \ref{spec-e2e+nw}, and \ref{spec-e8}, whereas the middle and right panels  show in more detail the profiles of the \amm\,main central hyperfine components and the \met\,line, respectively.  
Individual spectral profiles  do not show the presence of multiple velocity components, so they can be reasonably well fitted by single Gaussian profiles. 

In order to derive the line parameters, we used five-component Gaussian models to fit the HFS in each of those inversion transitions, 
where we fixed their velocity separations according to the calculated values, assuming the  same line-widths for all four hyperfine components, to reduce the number of free parameters.  In order to derive the kinematics in optically thin lines, we also fit the hyperfines only with a 4-component gaussian model again with fixed offsets and linewidths. 
The fitted line parameters are reported in Table~\ref{nh3_lines} for the main components and in Table~\ref{nh3_hf} for the hyperfine components\footnote{Since we did not manage to obtain good fits to the absorption in the most optically thick lines (6,6) and (7,7) with the multi-gaussian models, we also attempted a fit using
models that account for the optical depth of the main and hyperfine components (see \S~\ref{model_fit}).}.  

In the following, we will describe properties of spectral profiles in individual sources. 

\subsubsection{The HC HII region W51e2-W}

\begin{figure*}
\centering
\includegraphics[width = 170pt, keepaspectratio = true]{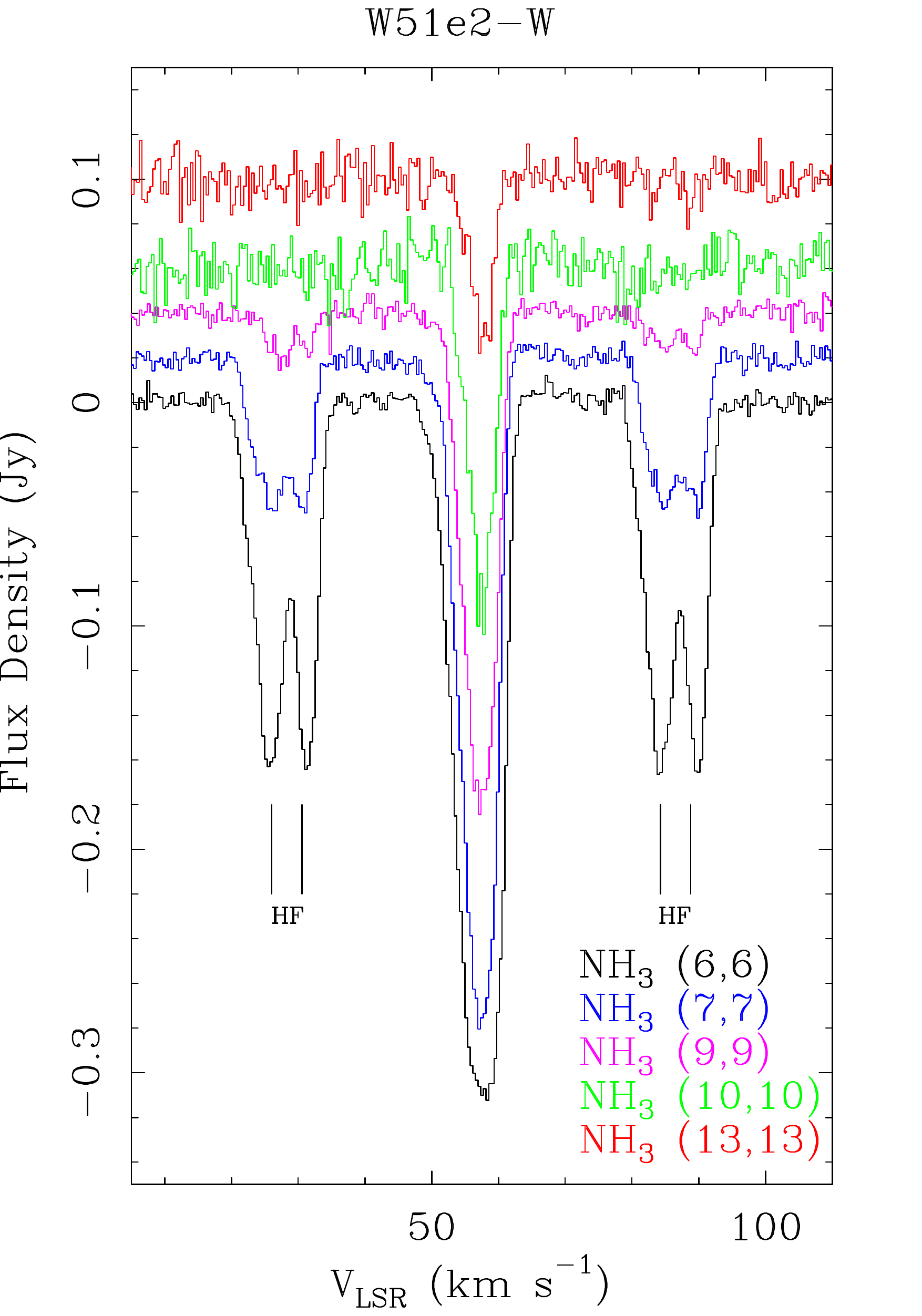}
\includegraphics[width = 170pt, keepaspectratio = true]{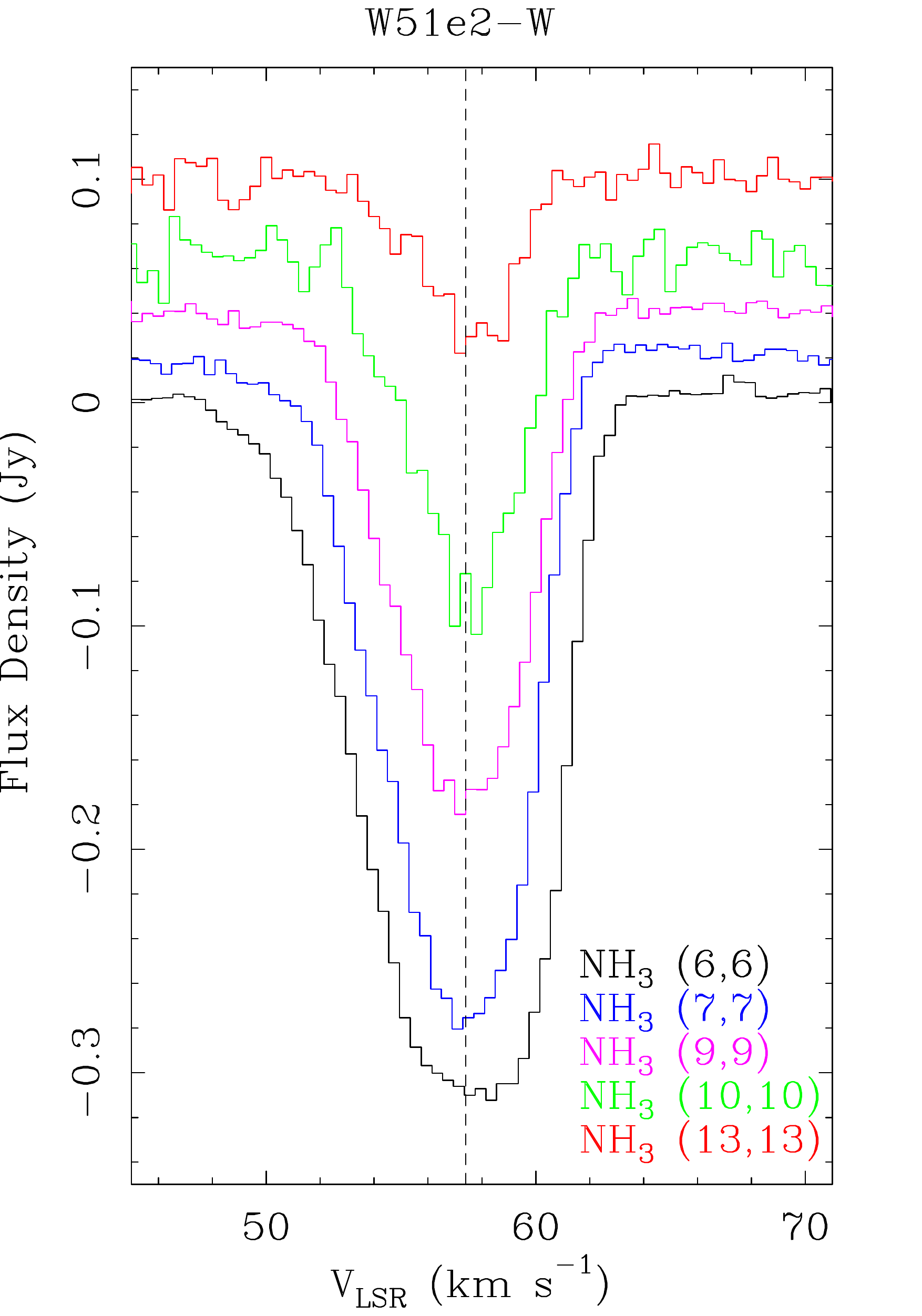}
\includegraphics[width = 170pt, keepaspectratio = true]{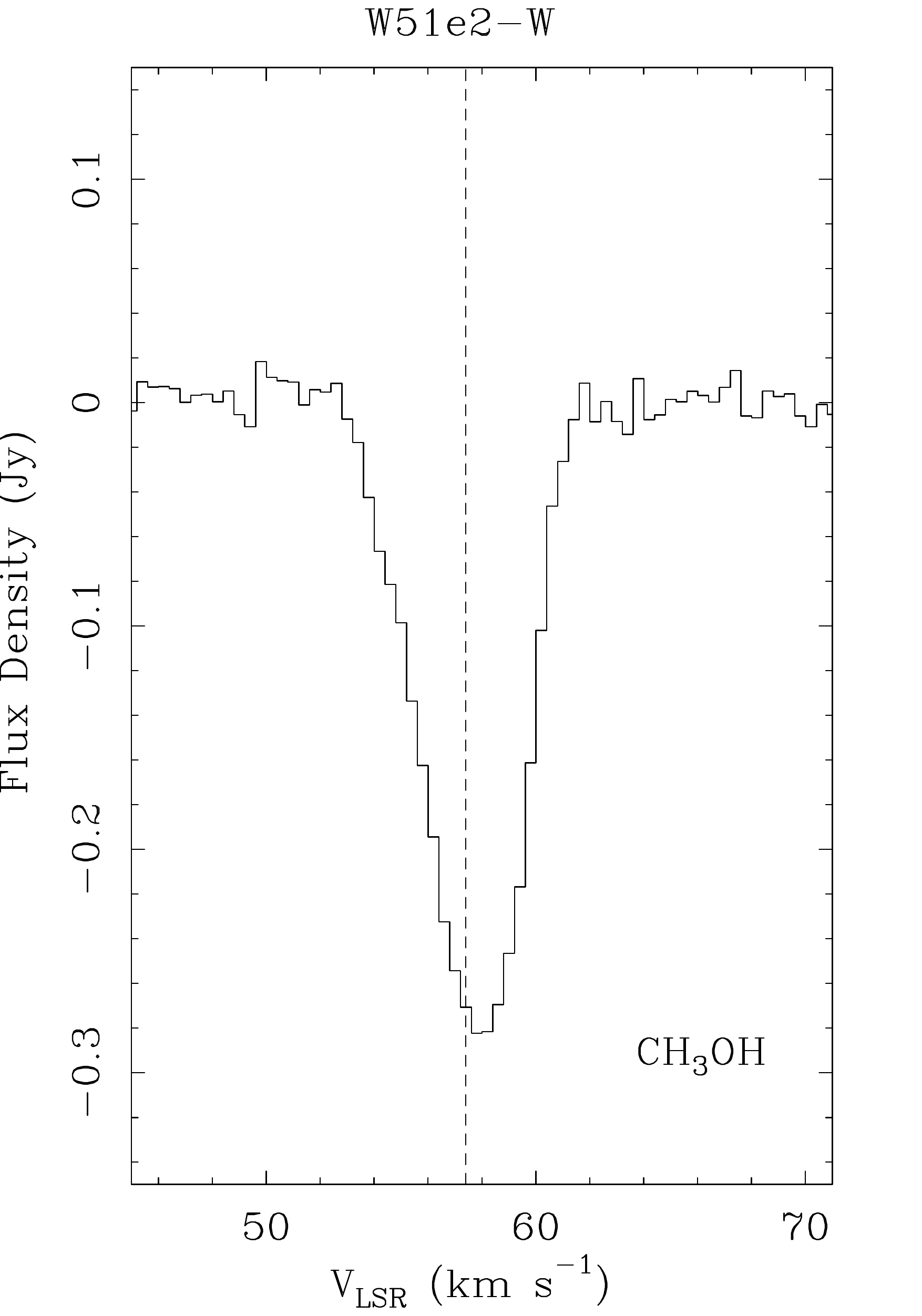}
\caption{Spectral profiles of  the absorption lines integrated over the radio continuum emission in W51e2-W.  
({\it Left and central panels}) Spectral profiles of \amm\, for transitions (6,6), (7,7), (9,9), (10,10), and (13,13). 
An offset in flux density is applied to transitions adjacent in energy, to better display individual profiles.  
The lower state energy levels of transitions shown here are $\sim 408-1691$\,K (see Table~\ref{obs}).
The hyperfine satellite lines, separated by $\sim \pm24-31$ \kms\,(see Table~\ref{nh3_hf}), are clearly detected for the (6,6), (7,7), and (9,9) lines ({\it left panel}).
A narrower velocity range is displayed in the {\it central panel}, in order to show more clearly the line profiles 
of the main hyperfine component of each inversion transition. 
({\it Right panel})  Spectral profile of the $J_K$= 13$_2$-13$_1$ line of CH$_3$OH ($\nu_{rest}=27.473$ GHz). 
The velocity resolution is 0.4~\kms\, ({\it all  panels}), and the vertical dashed line indicates a velocity of  57.4~\kms\,({\it central and right panels}). 
}
\label{spec-e2w}
\end{figure*}

The absorption spectral profiles for both the \amm\,and the \met\,transitions are shown in Figure~\ref{spec-e2w}. 
Gaussian fitting provides similar central velocities, V$_c$=57.1--57.5~\kms, for multiple transitions of \amm, from (6,6) to (13,13), but quite different FWHM line-widths, $\Delta V_{1/2}$=3.9--7.2~\kms\,(see Table~\ref{nh3_lines}). 
We do not believe this is a physical effect, and we ascribe the larger  line-widths (as well as the slightly lower central velocities) inferred for the lower-excitation lines to their higher opacities. 
In fact, perhaps the most striking feature of the \amm\,spectral profiles shown in the left panel of Figure~\ref{spec-e2w} is the prominence of the two pairs of hyperfine satellites, which reach a relative intensity with respect to the main component of nearly 50\%, indicating extreme optical depth values (see last column of Table~\ref{nh3_hf}). 

When we fit simultaneously line-widths and opacities (see \S~\ref{model_fit}), we derive similar line-widths  for the main hyperfine components at different excitation, of about 4~\kms. 
This value is consistent with the parameters fitted for the  hyperfine satellites in the (6,6) and  (7,7) doublets, which are expected to be more optically thin than the main line.  
Likewise, we consider the central velocities estimated from the hyperfine satellites more reliable than the main component. 
In conclusion, for the \amm\,core associated with the HC HII region W51e2-W, we infer a systemic velocity of 57.4~\kms\,and a line-width of  4.4~\kms\,(see Table~\ref{nh3_hf}). 


%
%
%
%
\subsubsection{The W51e2-E and NW sources and surrounding core}
\begin{figure*}
\centering
\includegraphics[width=0.3\textwidth]{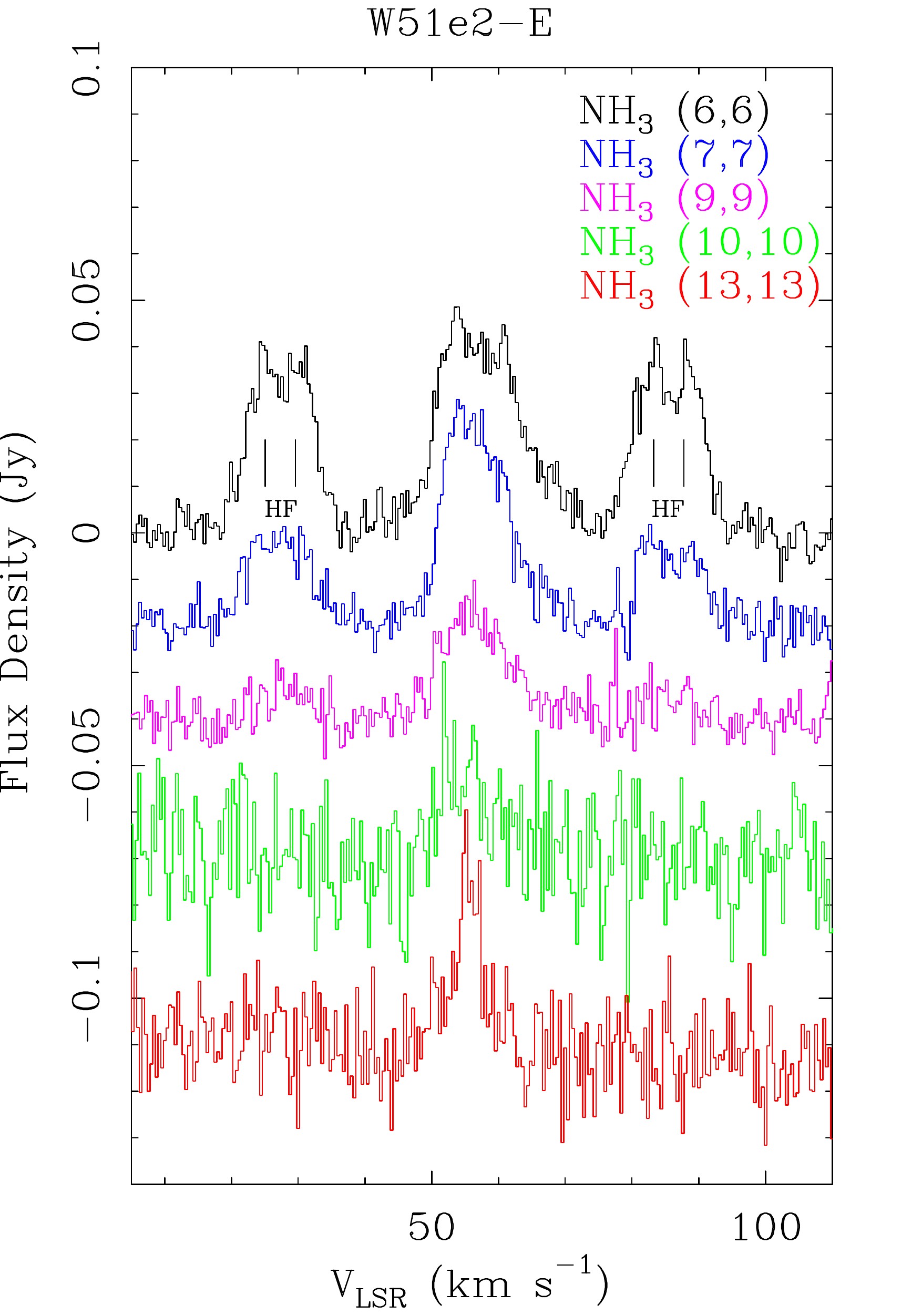}
\includegraphics[width=0.3\textwidth]{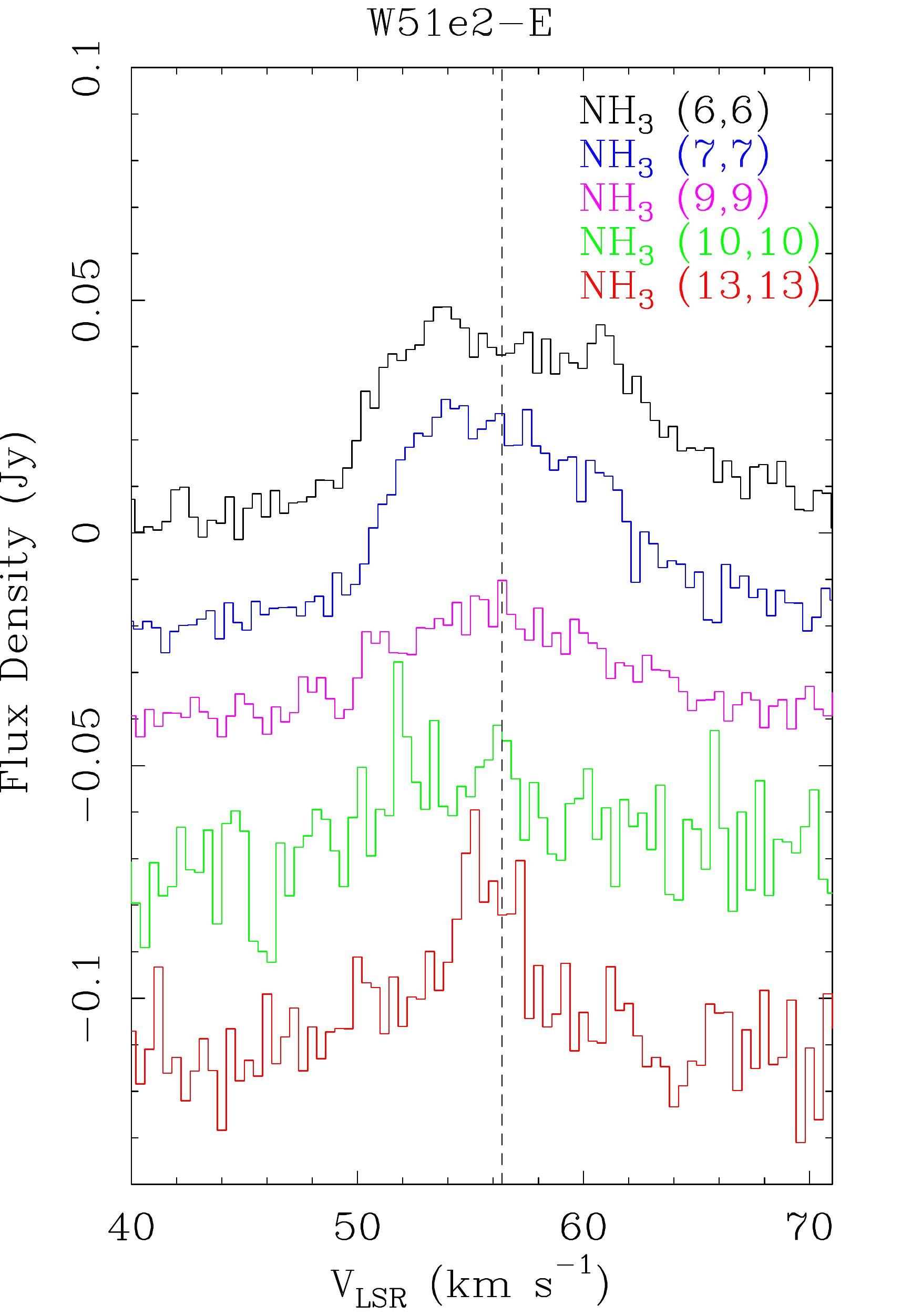}
\includegraphics[width=0.3\textwidth]{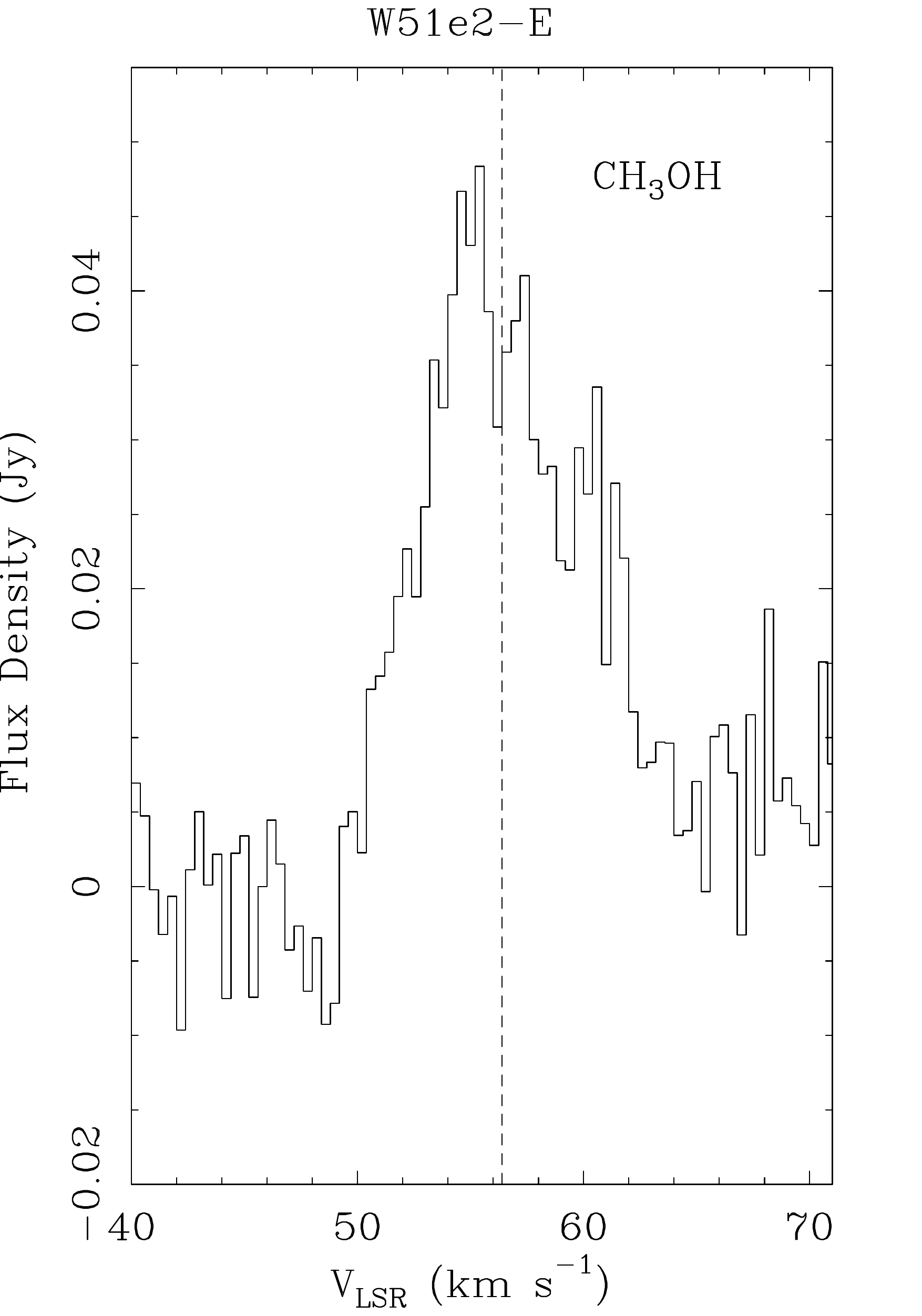}
\includegraphics[width=0.3\textwidth]{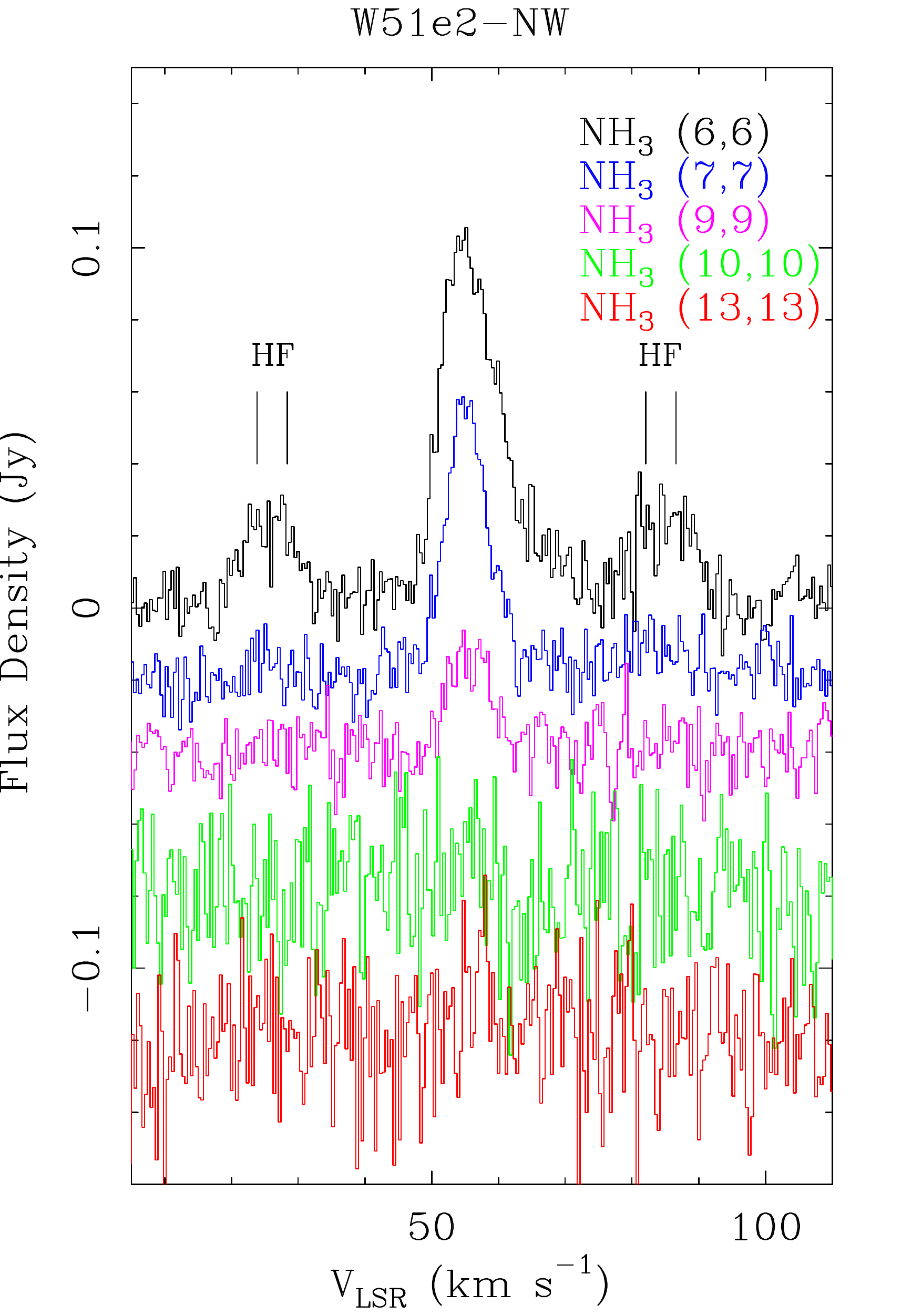}
\includegraphics[width=0.3\textwidth]{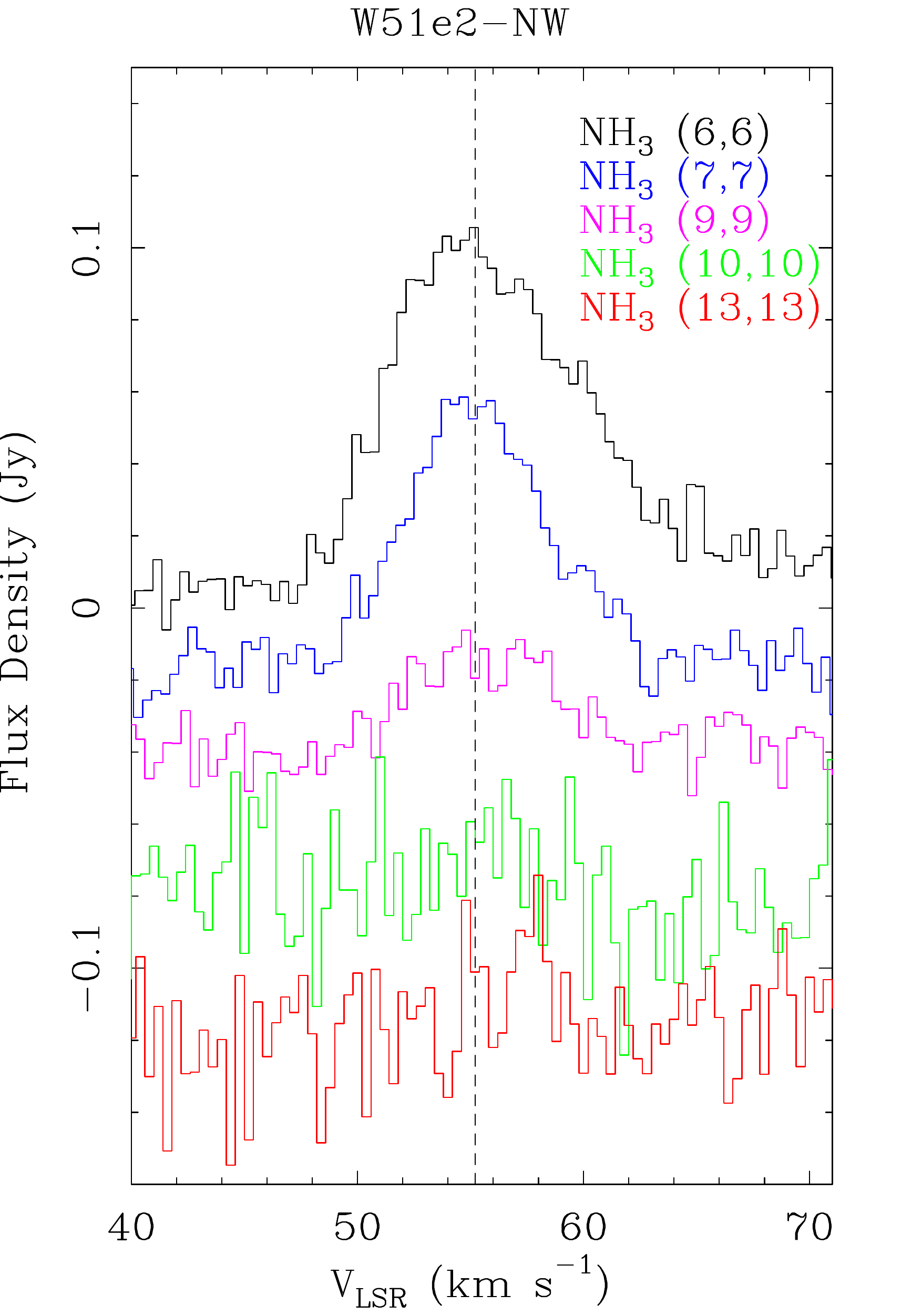}
\includegraphics[width=0.3\textwidth]{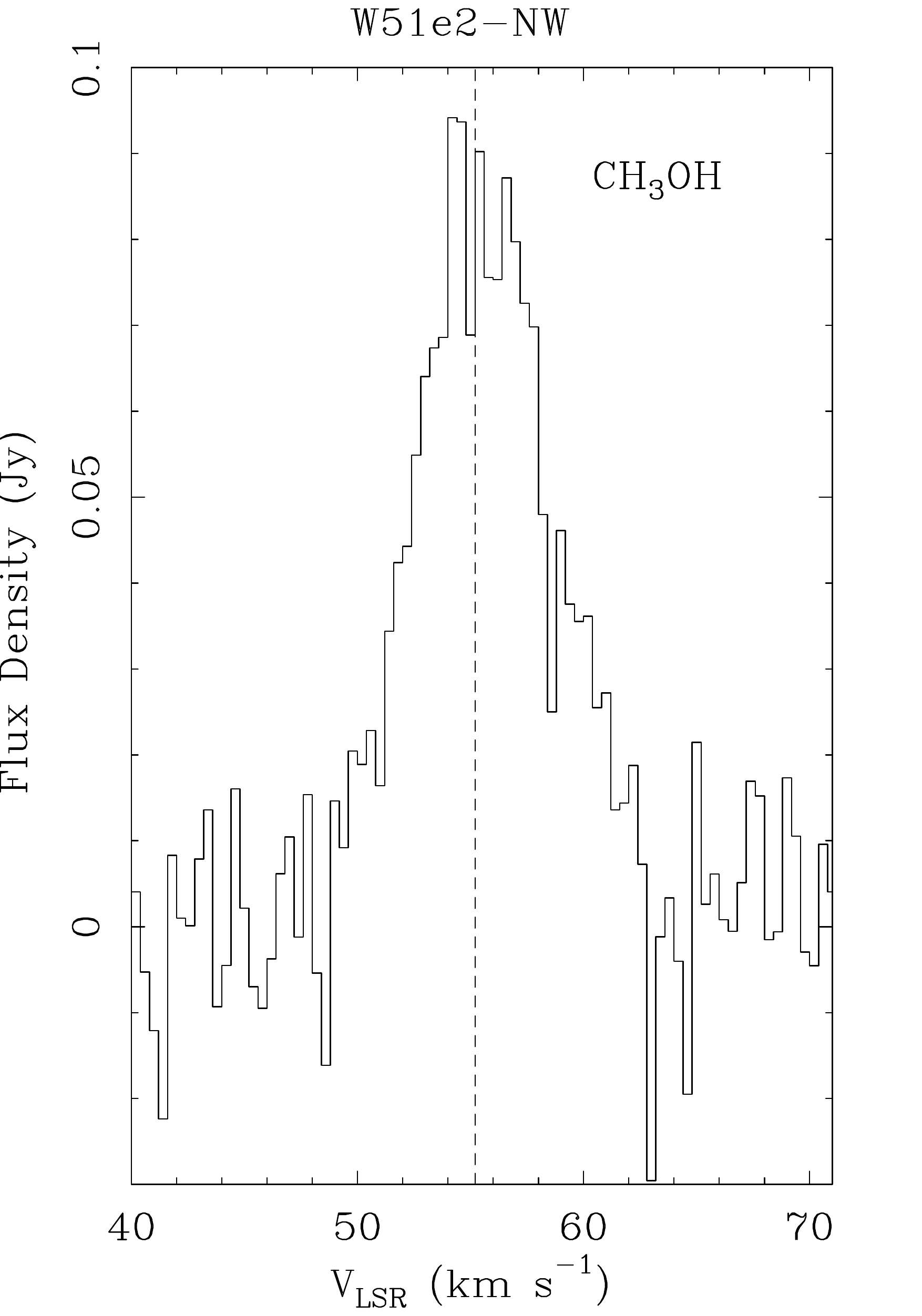}
\includegraphics[width=0.3\textwidth]{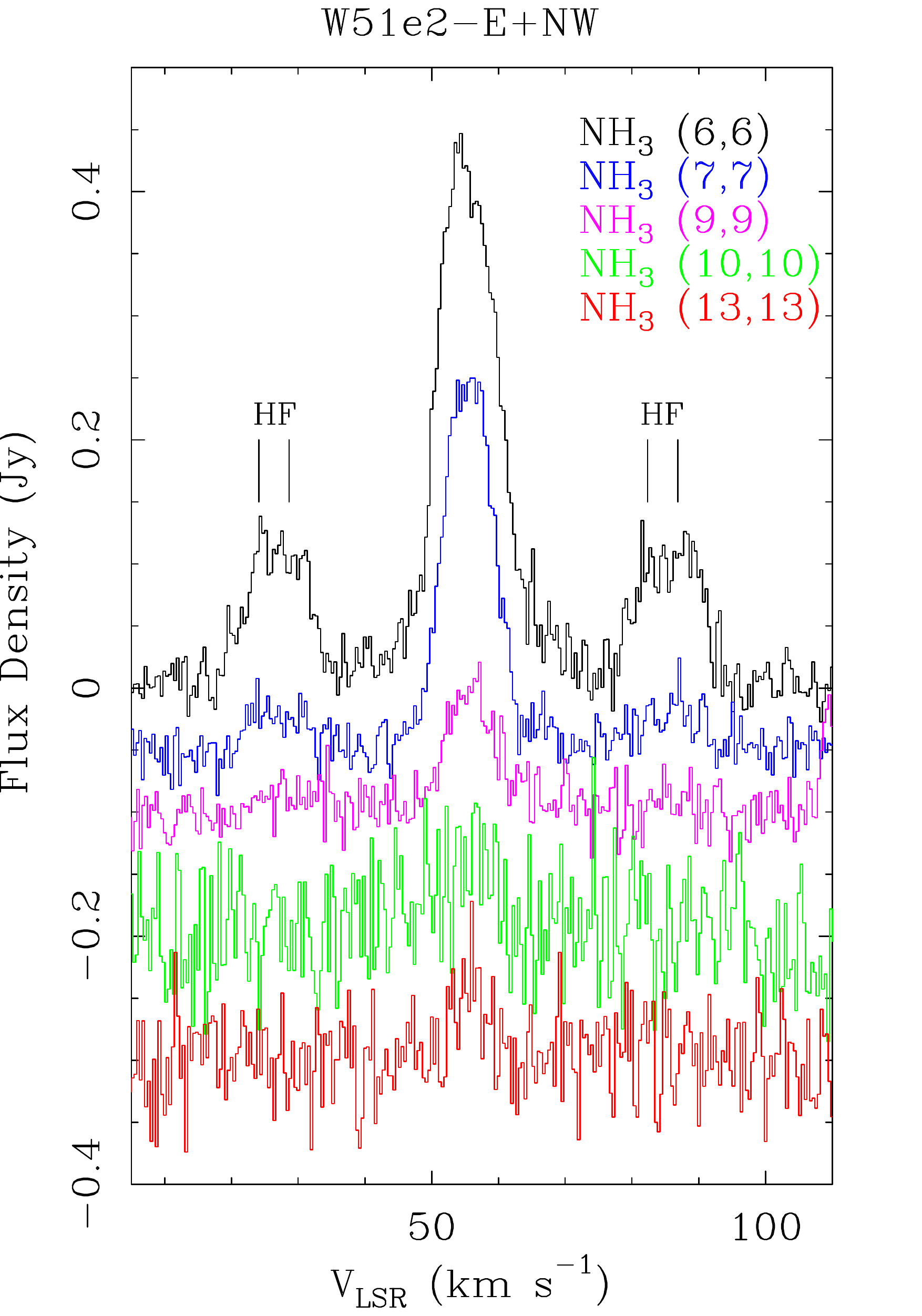}
\includegraphics[width=0.3\textwidth]{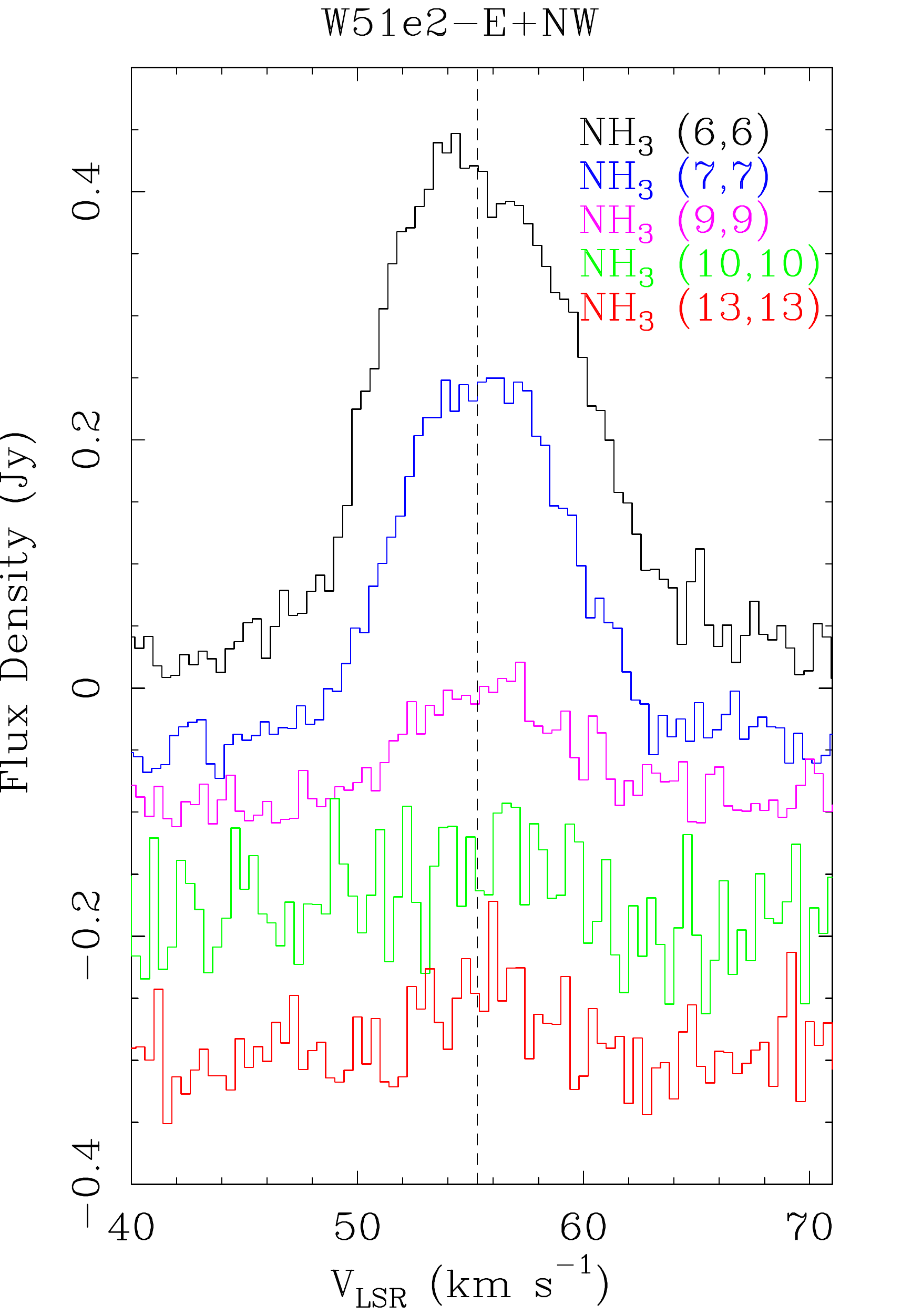}
\includegraphics[width=0.3\textwidth]{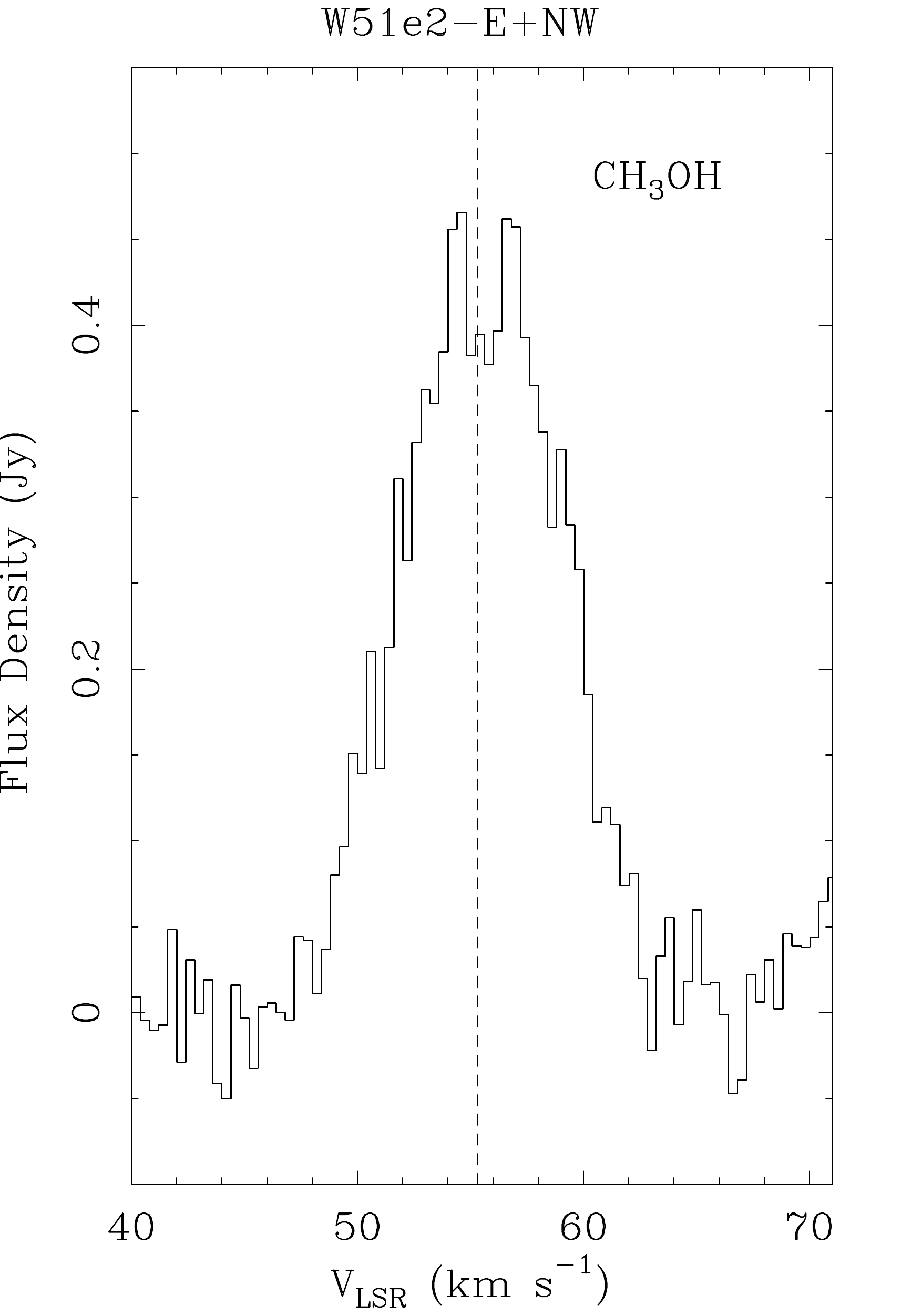}
\caption{Spectral profiles of  emission lines arising from hot molecular gas in the surroundings of the HC HII region. 
The spectra displayed here are integrated towards the dust continuum sources W51e2-E ({\it top row}), W51e2-NW  ({\it middle row}), and toward the entire molecular core seen in emission ({\it bottom row}), respectively. 
({\it Left and central panels}) Spectral profiles of \amm\, for transitions (6,6), (7,7), (9,9), (10,10), and (13,13). 
An offset in flux density is applied to transitions adjacent in energy, to better display individual profiles.  
The lower state energy levels of transitions shown here are $\sim 408-1691$\,K (see Table~\ref{obs}).
The hyperfine satellite lines, separated by $\sim \pm24-31$ \kms\,(see Table~\ref{nh3_hf}), are clearly detected for the (6,6), and (7,7) doublets in the three cases; the (9,9) HF lines are detected only towards W51e2-E ({\it top row, left panel}).
A narrower velocity range is displayed in the {\it central panel}, in order to show more clearly the profiles 
of the main hyperfine component of each inversion transition. 
({\it Right panel})  Spectral profile of the $J_K$= 13$_2$-13$_1$ line of CH$_3$OH ($\nu_{rest}=27.473$ GHz). 
The velocity resolution is 0.4~\kms\, ({\it all  panels}), and the vertical dashed line in the central and right panels indicate velocities of 56.4~\kms\,({\it top row}), 55.2~\kms\,({\it middle row}), and 55.3~\kms\,({\it bottom row}), respectively. 
}
\label{spec-e2e+nw}
\end{figure*}

We extracted spectra towards  W51e2-E 
and W51e2-NW\footnote{The positions and radii defining the areas where the  spectra were integrated are reported in Table~\ref{nh3_lines}.},
 as well as across the whole core, defined by the region where the \amm\,(6,6) inversion line is observed in emission (Figure~\ref{w51e2em-mom0}). 
 These spectra are shown in Figure~\ref{spec-e2e+nw} (top, middle, and bottom panels, respectively). 
All spectra show prominent satellites in the lower excitation lines, indicating high optical depths. 
This is reflected in the quite different values estimated for the central velocities 
and line-widths for different excitation lines (see Table~\ref{nh3_lines}). 

Owing to their symmetry, the optically thin lines unambiguously define the systemic velocity of the core.
Therefore, to derive  central velocities and line-widths, we  again rely on the more optically-thin hyperfine satellites:
 56.4, 55.2, 55.3 ~\kms\,and  9.3, 8.7, 9.3,~\kms\,for the W51e2-E and W51e2-NW sources and the entire W51e2 core, respectively. 
%
%
\subsubsection{W51e8}
\begin{figure*}
\centering
\includegraphics[width=0.3\textwidth]{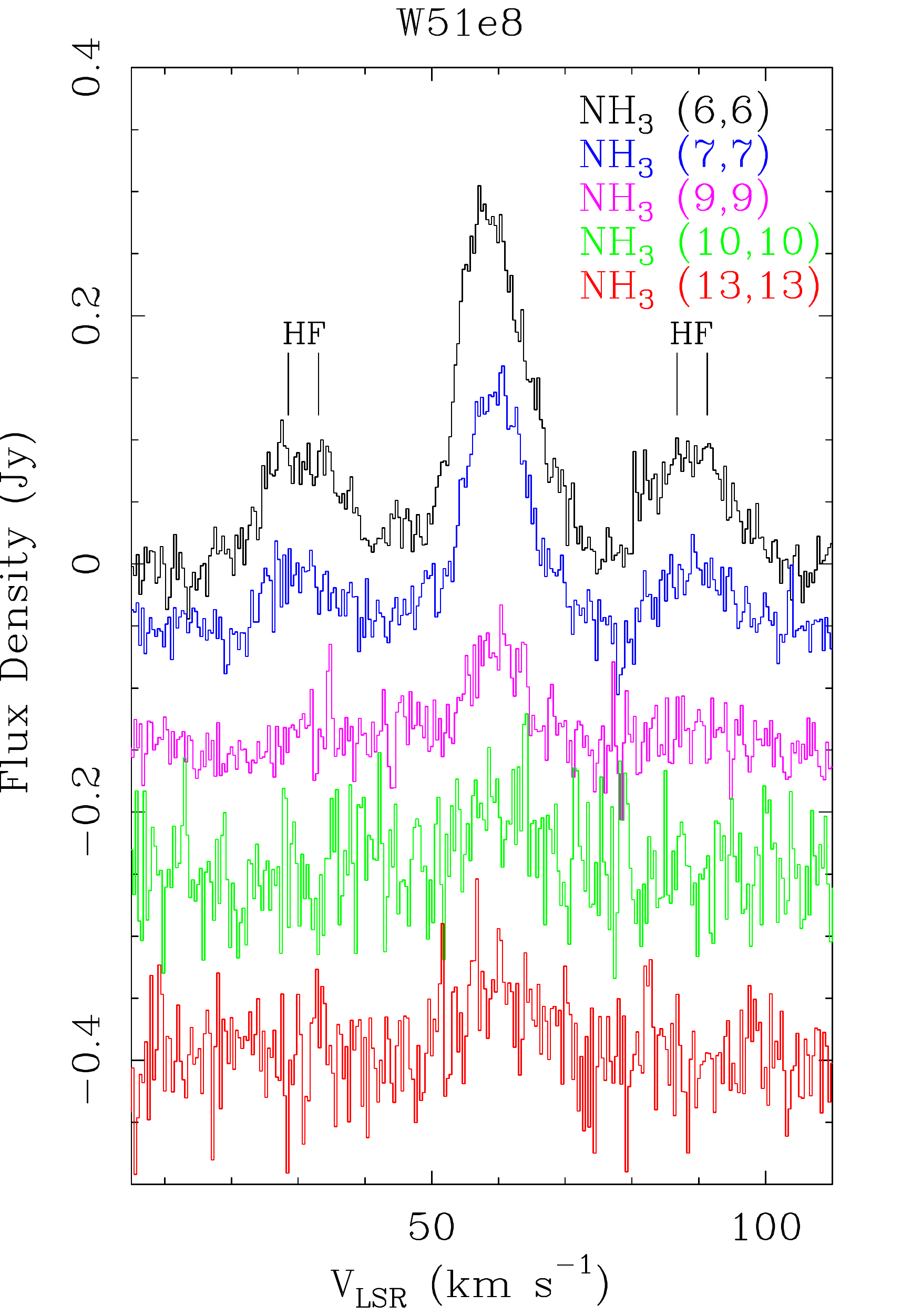}
\includegraphics[width=0.3\textwidth]{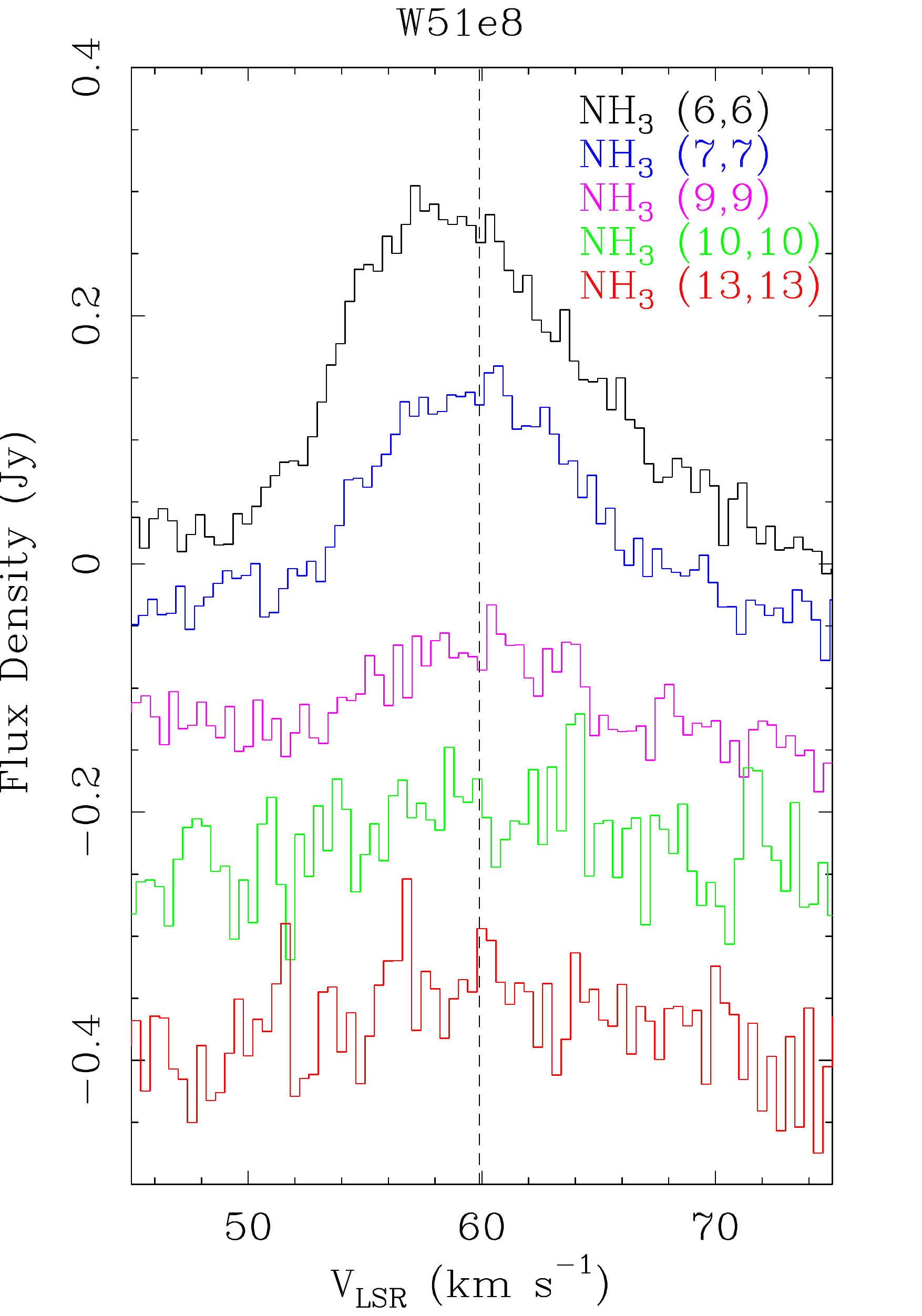}
\includegraphics[width=0.3\textwidth]{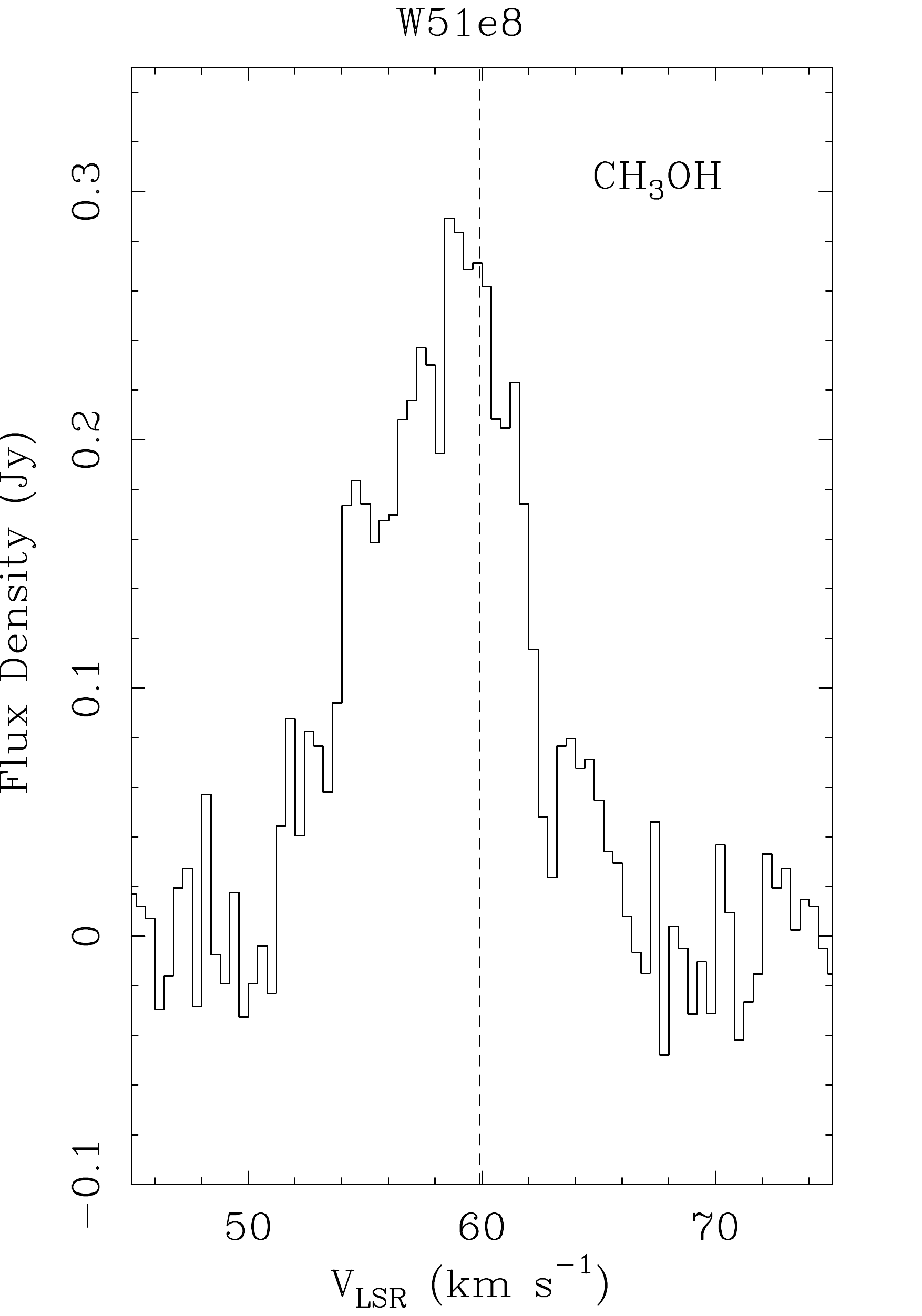}
\caption{Spectral profiles of  emission lines arising from hot molecular gas around W51e8.  
({\it Left and central panels}) Spectral profiles of \amm\, for transitions (6,6), (7,7), (9,9), (10,10), and (13,13). 
An offset in flux density is applied to transitions adjacent in energy, to better display individual profiles.  
The lower state energy levels of transitions shown here are $\sim 408-1691$\,K (see Table~\ref{obs}).
The hyperfine satellite lines, separated by $\sim \pm24-31$ \kms\,(see Table~\ref{nh3_hf}), are detected for the (6,6), (7,7), and (9,9) lines ({\it left panel}).
A narrower velocity range is displayed in the {\it central panel}, in order to show more clearly the line profiles 
of the main hyperfine component of each inversion transition. 
({\it Right panel})  Spectral profile of the $J_K$= 13$_2$-13$_1$ line of CH$_3$OH ($\nu_{rest}=27.473$ GHz. 
The velocity resolution is 0.4~\kms\, ({\it all  panels}), and the vertical dashed line indicates a velocity of  59.9~\kms\,({\it central and right panels}). 
}
\label{spec-e8}
\end{figure*}

For W51e8, we integrated across an approximately spherical  region (with a radius of 3\pas7) where the \amm\,(6,6) inversion line is observed in emission (Figure~\ref{e8-mom0}). 
The spectral profiles are shown in Figure~\ref{spec-e8}. 
We derive central velocities varying in the range 58.5 to 60.6~\kms\,and line-widths in the range 7.5--13~\kms, from multiple transitions of \amm,  as well as the \met\,line. 
From the hyperfine satellites detected from the (6,6) and (7,7) doublets, we infer a consistent systemic velocity of $\sim$60~\kms, but their line-widths remain fairly uncertain, in the range 8--14~\kms. 
We conclude that W51e8 is significantly redshifted with respect to the W51e2 core, and has a large internal velocity dispersion (probably $>$10~\kms), possibly indicating very turbulent and/or outflowing gas (see discussion in \S~\ref{discussion}).

\section{Analysis: determining physical conditions}
\label{phycon}

\amm\,is a high-density gas tracer, and its inversion lines provide an excellent probe of the gas kinetic temperature \citep{HoTownes83}. 
 Using the  parameters measured for the HFS of the five \amm~inversion transitions observed in W51~Main, and in the assumption of LTE (i.e., the metastable inversion lines are thermalized), 
we can estimate the physical conditions of the \amm\,gas, such as optical depth ($\tau$), rotational temperature (\trot), and column density (\ncol).  
The methodology as well as the formalism adopted to analyze the \amm\,data are described in Appendix~\ref{app}. 

 We detected  hyperfine pairs of satellite lines 
for the (6,6), (7,7), and (9,9) transitions towards  cores e2-W, e2-E, e2-NW\footnote{The (9,9) doublet was not detected in W51e2-NW.}, and e8 (Table~\ref{hf_lines}). 
In particular, the hyperfine satellites are remarkably prominent in the (6,6) doublet, where they show relative intensities (with respect to the main line) from roughly 50\% to nearly 100\% towards cores e2-W and e2-E, respectively (see Figures~\ref{spec-e2w} and \ref{spec-e2e+nw}). 
The expected theoretical value of the ratio of the satellite   to the main component strengths for the lines targeted here is $<$1\% assuming LTE (see column 6 in Table~\ref{hf_lines}), indicating very large optical depths even for these highly excited transitions. 
Indeed, the values of optical depths that we derive vary in the range 10-100 from the (9,9) down to the (6,6) line  (see Table~\ref{hf_lines}), indicating exceptionally high opacity in the gas seen both in absorption and emission.
The W51e2-E core is the most optically thick and the W51e8 is the least optically thick core.  

After estimating the optical depths, we used rotational temperature diagrams (RTDs) to derive rotational temperature  and column densities of the \amm\,gas towards all the cores (see description  in Appendix~\ref{app} and Figures~\ref{fig_rtd_abs}, \ref{fig_rtd_em}). 
Since we do not have direct estimates of the opacity for all transitions, and we cannot exclude that even the most highly-excited transitions are optically-thick, we used only the transitions (6,6), (7,7), and (9,9) in our  analysis of the physical conditions. 
We estimated  an average rotational temperature of 140~K for the W51e2 complex, which rises to 174~K, 173~K, and 144~K for the individual cores e2-W, e2-E, and e2-NW, respectively. For W51e8, we estimated  a slightly higher temperature of  about 200~K. 
Using an RTD analysis with CH$_3$CN lines at 2mm, \citet{Zhang98} found  rotation temperatures of 140 K and 130 K, while \citet{Remijan04} estimated 153 K and 123 K by using CH$_3$CN transitions at 3 mm and 1 mm, for W51e2 and W51e8, respectively. While these previous estimates for W51e2 are  consistent with ours, the lower temperatures inferred for W51e8 could be ascribable to a temperature gradient and to the coarser angular resolution ($1-3$\arcsec) of previous observations\footnote{In W51e2 we find that the gas temperature in different cores is the same within 30~K, so the impact of  different angular resolutions on the temperature estimates should less important than for W51e8.}. 

For the column density, we can make a reliable estimate only for the absorption against the HII region, where the good SNR in both the main and the satellite spectral components enabled accurate estimates of the gas opacity.  
For the gas seen in emission, where our opacity estimates   have large error bars,  we provide  only lower limits to the true column density (assuming optically thin gas). 
After estimating the \amm\,column density, we can calculate the volume density  and the total mass of the molecular gas in each individual  core in W51~Main. 
In this calculation, we assume an ammonia fractional abundance of [\amm]/[H$_2$]=10$^{-7}$ \citep[e.g.][]{Mauersberger86}\footnote{We explicitly note that the value of [\amm]/[H$_2$] is known to no better than an order of magnitude, therefore the values of mass  and molecular density are just order-of-magnitude estimates. In addition, this ratio is fixed, therefore our analysis neglects the effect of chemistry on the location and abundance of ammonia molecules within W51~Main. For a model with variable fractional abundance, please see \citet{Osorio09}.}  and a spherical gas distribution within the core radius\footnote{This corresponds to approximately  the average between the semi-major and semi-minor axes of the ellipse used to integrate the spectra in each core (see Table~\ref{nh3_lines}), except for the absorption in the HC HII region.}. 

Since the HC HII region is not resolved in our maps, we take as an upper limit to its radius 320 AU \citep[0\pas06][]{Shi10a}.  According to this analysis, for the molecular gas surrounding the HC HII region we derive a volume density of $\sim 6.9 \times 10^{9}$~\cmc, and an upper limit to  the molecular mass of about 5 M$_\odot$. 

For the entire core e2, which surrounds the HC HII region and is seen in \amm\,emission, we derive lower limits to both the volume density ($\sim 3 \times 10^{6}$~\cmc) and the gas mass ($\sim$80~\ms). 
The density increases towards the positions of the dusty sources W51e2-E and W51e2-NW, where we  estimate (lower limits to the) gas masses of 18~\ms\,and 32~\ms\,(within about half an arcsecond and one arcsecond from the two sources, respectively). 
For W51e8, we estimate a gas density of $\sim 3 \times 10^{6}$~\cmc\,and a  gas mass of about 70~\ms, similar to the entire W51e2 core. Unlike the latter, however, we do not see evidence of multiple sources in W51e8 (besides central condensation).
Our estimates are slightly lower than but overall consistent with the values for the gas mass derived by \citet{Hernandez14} from 1.3 mm continuum dust emission imaged with the SMA at $\gtrsim$1\arcsec\,resolution (96 and 86~\ms, for W51e2 and W51e8, respectively).

\begin{table*}
\caption{High-mass YSOs hosting hot cores in W51~Main.}             
\label{ysos}      
\centering                        
\begin{tabular}{lcllccccccccc} 
\hline\hline                 
\noalign{\smallskip}
\multicolumn{1}{c}{Source} & &RA (J2000) & DEC (J2000) & & \vlsr & \trot$^{(a)}$ & N & R$_{\rm core}^{(b)}$ & M$_{\rm gas}^{(c)}$  & Outflow & Disk \\
\multicolumn{1}{c}{Name} &  & \multicolumn{1}{c}{(h:m:s)} &  \multicolumn{1}{c}{(\degree:':")} & & (\kms) & (K) & (\cmc) & (AU) & (\ms) & (Y/N) & (Y/N) \\
\noalign{\smallskip}
\hline
\noalign{\smallskip}  
W51e2-W  & & 19:23:43.9096 & 14:30:34.551 & & 57.4 & 174 & $6.9 \times 10^{13}$ & 320 &  5  & N& Y \\ 
W51e2-E  & & 19:23:43.9618 & 14:30:34.558 & & 56.4 & 173 & $5.1 \times 10^7$ & 2430 &  18 & Y& N \\
W51e2-NW  & & 19:23:43.8900 & 14:30:35.630 & & 55.2 & 144 & $1.6 \times 10^7$  & 4460 &  32 & Y& N \\
W51e8 & & 19:23:43.9073 & 14:30:28.197 & & 59.9 & 204 & $2.7 \times 10^6$ & 10200 &  69  & Y& N \\ 
\noalign{\smallskip}
\hline   
\end{tabular}
\tablefoot{\\
(a) In the case of emission lines, the rotational temperatures  are calculated in the optically-thin assumption, therefore the quoted values are lower limits to true kinetic temperature  of the gas.
(b) The radius is defined by the area where we integrated the spectra (see Table~\ref{nh3_lines} and Sect.~\ref{spec}), except for W51e2-W, for which we used the deconvolved size of the compact continuum emission quoted by \citet{Shi10a}.  \\
(c) The gas mass is calculated in a sphere of radius R$_{\rm core}^{(a)}$ for a given volume density $\rho$. For the emission, with poor estimates of the opacity, this is calculated in the optically-thin assumption. Therefore the quoted values are lower limits to the gas mass.The mass for absorption should be instead regarded as an upper limit (since we do not know the actual size of the HC HII region).   These mass estimates assume [\amm]/[H$_2$]=10$^{-7}$. 
}
\end{table*}

\section{Discussion}
\label{discussion}
We have identified two main centers of HMSF activity in W51-Main, which are exciting hot cores and are presumably hosting one or multiple high-mass YSOs at their centers: 
the W51e2 complex (forming a multiple system) and the W51e8 core ($\sim$6\arcsec\,southward of W51e2). 
In order to characterize the nature of these hot cores and their exciting sources, and more generally assess the star formation activity in W51~Main, we  analyze here our \amm\,maps in the context of previously published high-resolution data. 

The main physical and kinematic  properties  of the identified high-mass YSOs and/or hot cores in W51~Main are summarized in Table~\ref{ysos}. 

\subsection{The W51e2 complex}

\begin{figure}
\includegraphics[angle=-90,width=0.5\textwidth]{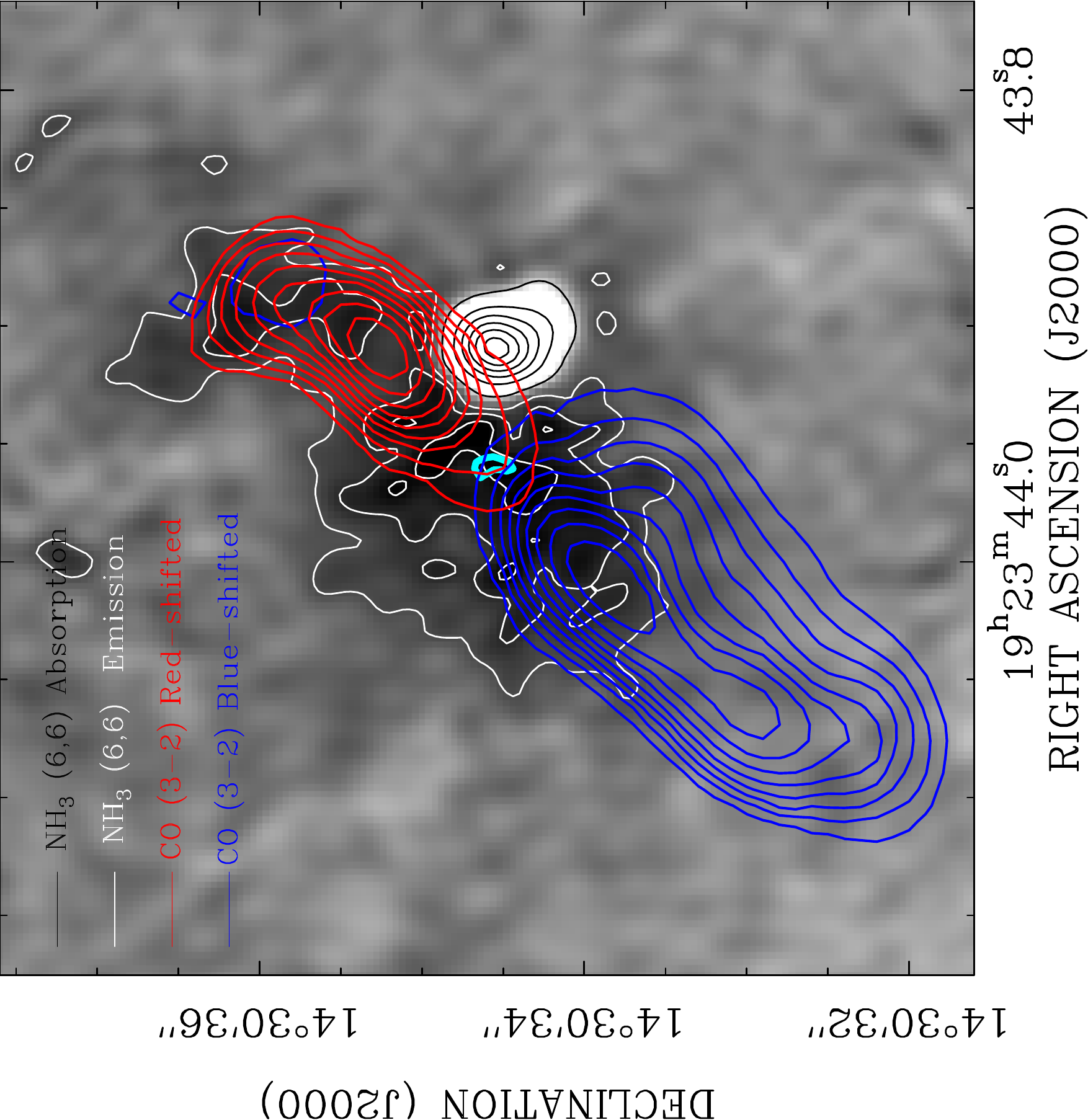}
\caption{
Overlay of the CO J=3-2 $0^{th}$ moment image from the SMA (blue and red contours; \citealt{Shi10b}) onto the total intensity (0th moment) map of the (6,6) inversion transition of  NH$_3$ 
 (gray scale and white+black contours). 
The \amm\,emission is displayed with white contours, representing 30\% to 90\% with steps of 20\% of the line peak for the (6,6) line, 107 mJy beam$^{-1}$ km s$^{-1}$).  
 The \amm\,absorption is displayed with black contours, representing factors 1, 5, 9,.... of --50~mJy~beam$^{-1}$, for all transitions.  
The CO 3-2 line contours are from 10\% to 90\% with steps of 10\% of the line peak  
(67 Jy beam$^{-1}$ km s$^{-1}$ for the red contours and 
82 Jy beam$^{-1}$ km s$^{-1}$ for the blue contours). 
Blue and red contours correspond  to blue- and redshifted gas, 
with integrated velocity ranging from --124 to --12 km s$^{-1}$ 
and from +10 to +116 km s$^{-1}$,  respectively. 
The cyan contour locates the peak of the total intensity map of the most highly-excited \amm\,transitions (13,13): we assume that this peak pinpoints the high-mass YSO driving the CO outflow. 
}
\label{nh3+co}
\end{figure}

\begin{figure}
\includegraphics[width=0.5\textwidth]{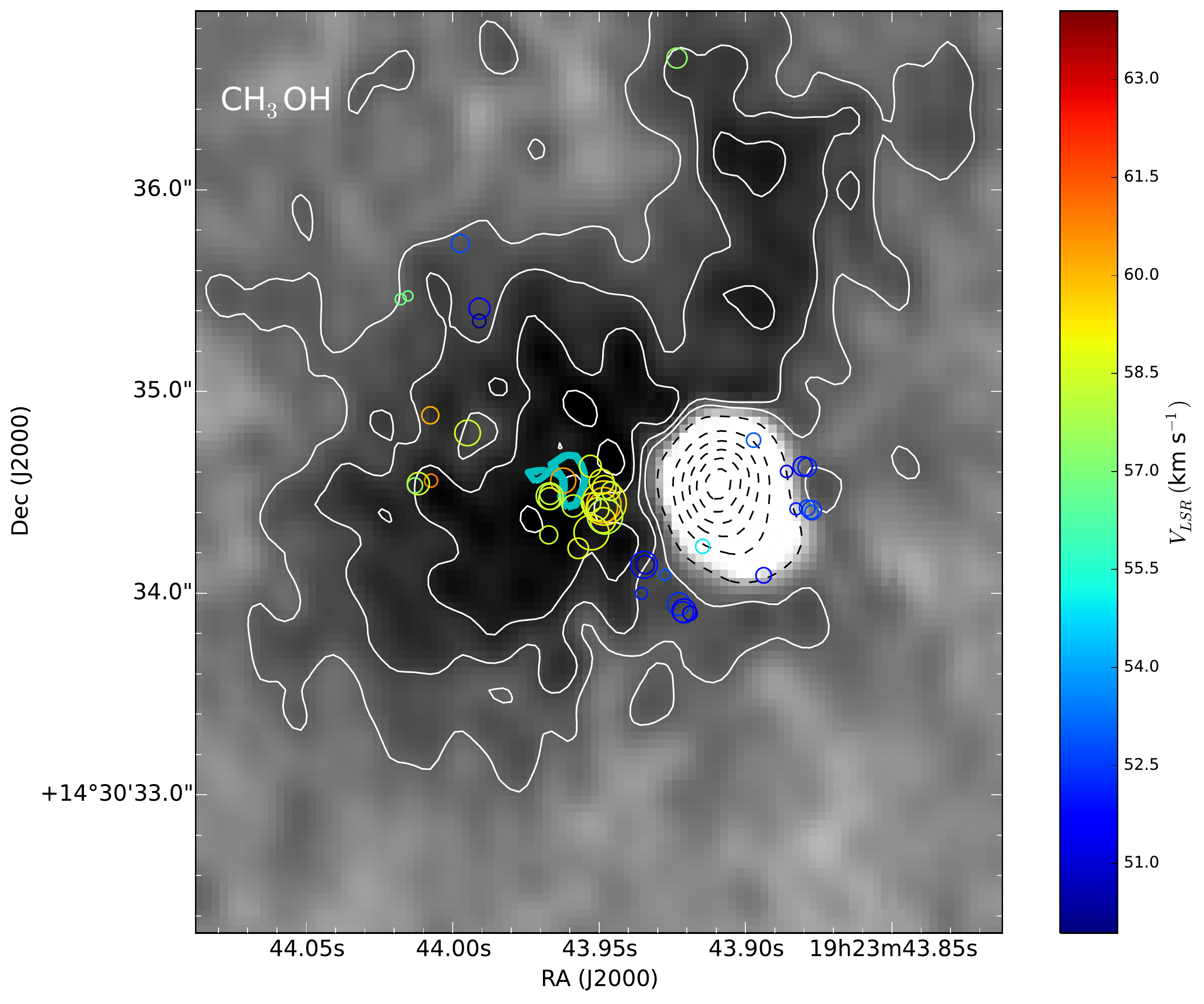}
\includegraphics[width=0.5\textwidth]{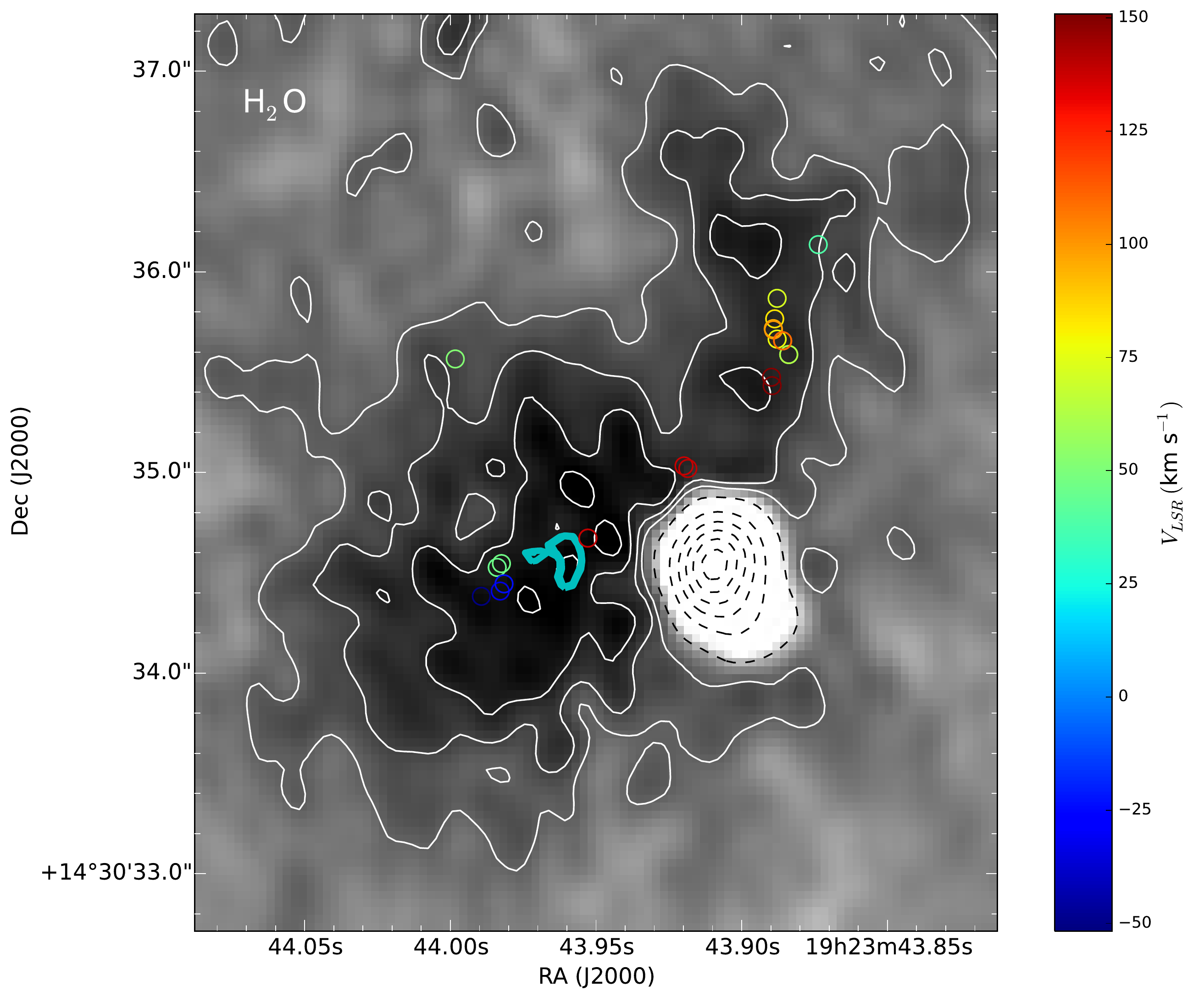}
\includegraphics[width=0.5\textwidth]{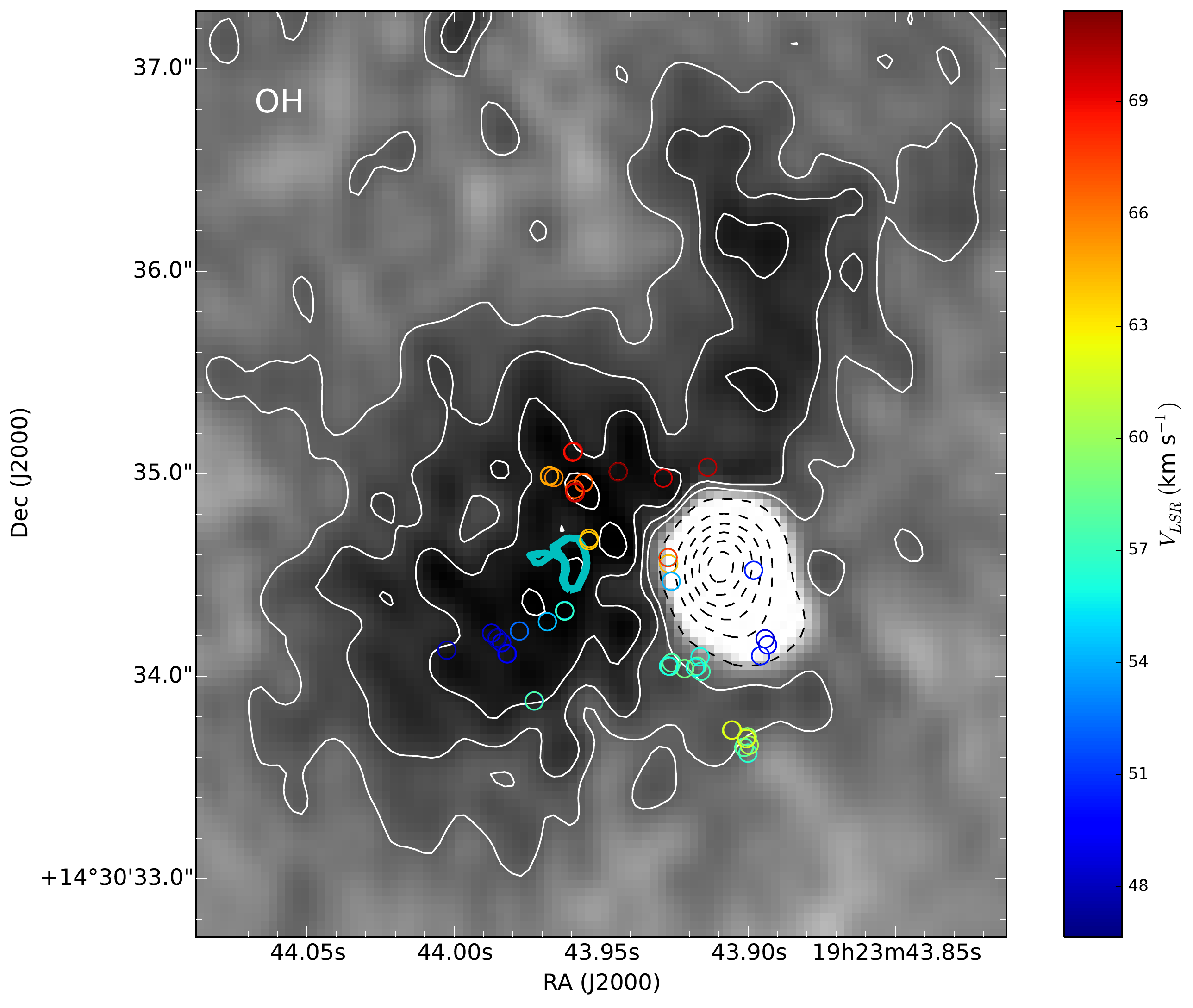}
\caption{\scriptsize
Overlay of molecular masers detected around W51e2 onto the total intensity (0th moment) map of the (6,6) inversion transition of  NH$_3$ 
 (gray scale and white+black contours). 
 The circles show positions of \met\,masers  detected with MeRLIN \citep[][\it top panel]{Etoka12}, 
 \wat\,masers detected with the VLBA \citep[][\it middle panel]{Sato10}, and
 OH masers detected with the VLBA \citep[][\it bottom panel]{FishReid07}, respectively. 
   Colors denote  l.o.s.  velocity in \kms\,(color scale on the right-hand side in each panel). 
The \amm\,emission is displayed with white contours, representing 20\% to 100\% with steps of 20\% of the line peak for the (6,6) line, 107 mJy beam$^{-1}$ km s$^{-1}$).  
 The \amm\,absorption is displayed with black contours, representing factors 1, 5, 9,.... of --50~mJy~beam$^{-1}$.  
The cyan contour locates the peak of the total intensity map of the most highly-excited \amm\,transitions (13,13); the bulk of \met\,maser emission concentrates around the same peak: we assume that its position pinpoints the high-mass YSO W51e2-E. 
}
\label{e2-nh3+mas}
\end{figure}

\begin{figure}
\includegraphics[width=0.5\textwidth]{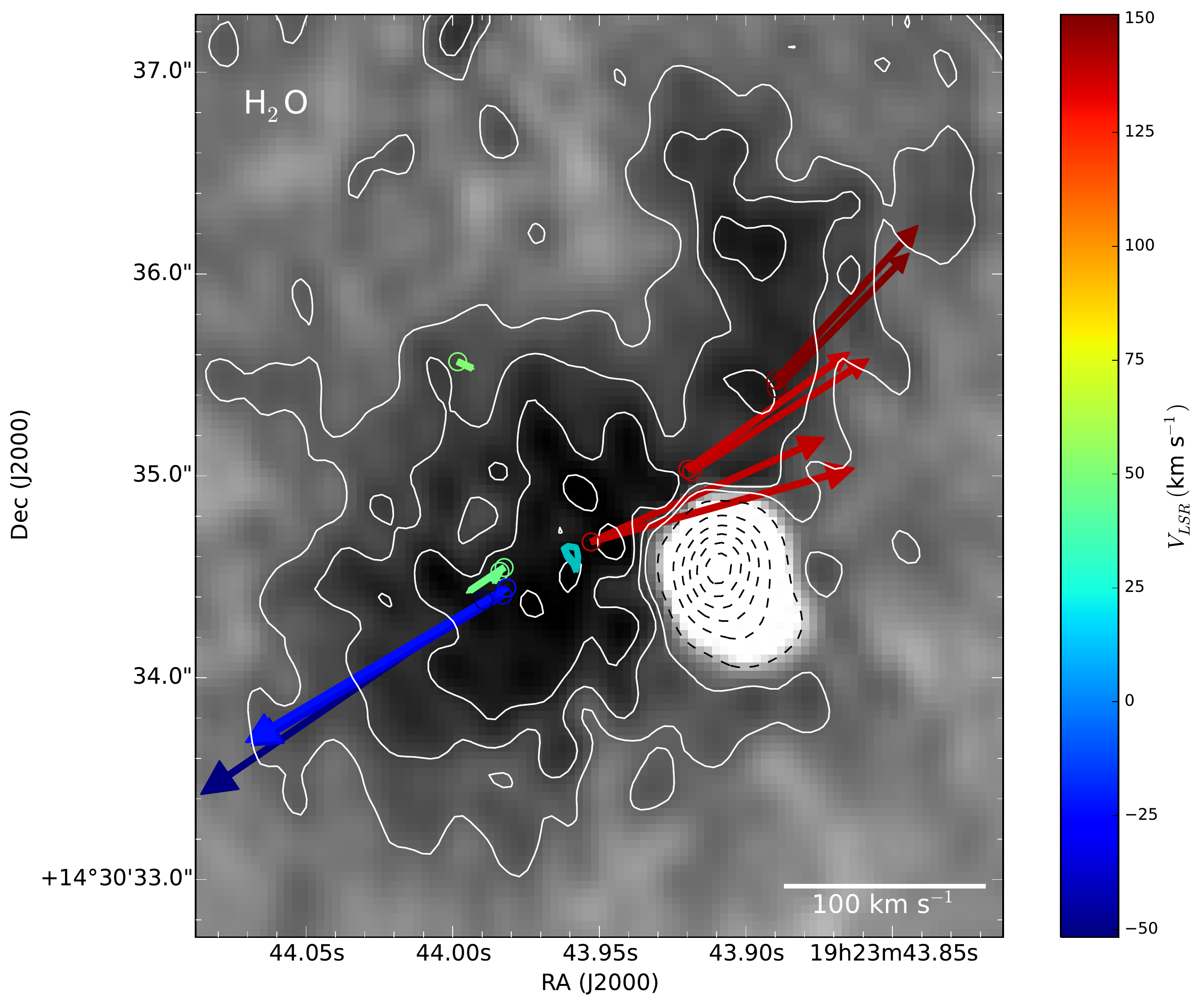}
\includegraphics[width=0.5\textwidth]{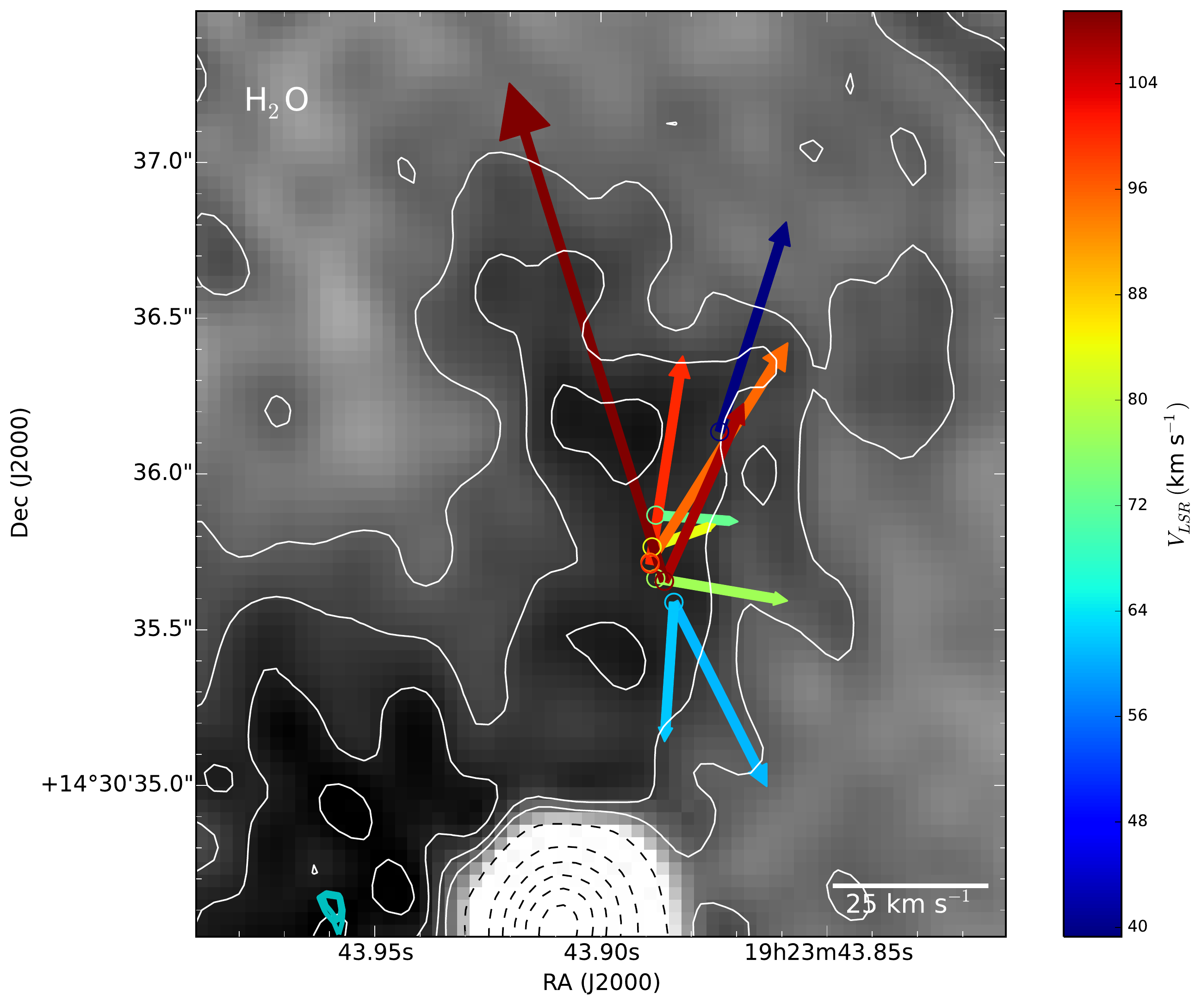}
\caption{
Outflows identified by \wat\,proper motions towards the YSOs W51e2-E (upper panel) and W51e2-NW (lower panel) \citep{Sato10}. 
The position (circles) and proper motions (arrows) of water masers are overlaid onto the total intensity (0th moment) map of the (6,6) inversion transition of  NH$_3$  (same as Figure~\ref{e2-nh3+mas}). 
   Colors denote  l.o.s.  velocity (color scale on the right-hand side in each panel) and   
 the scale for the proper motion amplitude is given in the lower right corner of each panel (both in  \kms). 
The diverging proper motions identify the  origin of the fast  outflows, i.e. the protostellar position.  
Note that the peak of the total intensity map of the most highly-excited \amm\,transitions (13,13) (cyan contours) lies at the origin of  the \wat\, maser outflow: we assume that its position pinpoints the high-mass YSOs W51e2-E. 
}
\label{e2-nh3+h2o}
\end{figure}

\begin{figure}
\includegraphics[width=0.5\textwidth]{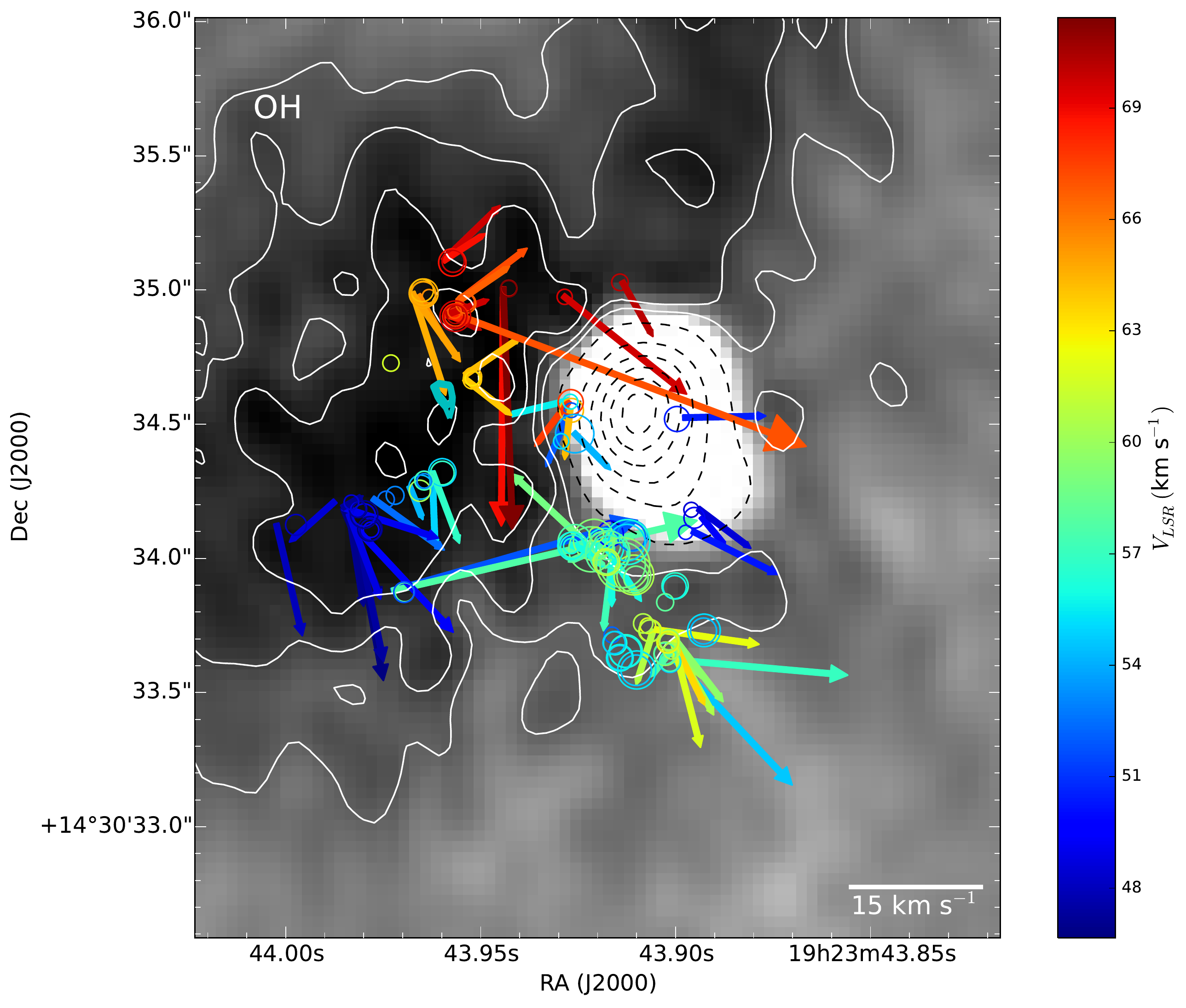}
\caption{
OH maser proper motions measured with the VLBA towards W51e2-E \citep{FishReid07}. 
The position (circles) and proper motions (arrows) of OH masers are overlaid onto the total intensity (0th moment) map of the (6,6) inversion transition of  NH$_3$  (same as Figure~\ref{e2-nh3+mas}). 
  Colors denote  l.o.s.  velocity (color scale on the right-hand side) and   
 the scale for the proper motion amplitude is given in the lower right corner  (both in  \kms). 
The  proper motions seems to trace a slow wide-angle expansion around W51e2-E, as well as a filamentary structure south of the HC HII region.  
}
\label{e2-nh3+oh}
\end{figure}

We present here a series of overlays of our \amm\,images with  different star formation tracers in the W51e2 complex.
In particular, Figure~\ref{nh3+co} shows an  overlay of the CO $J=3-2 \ 0^{th}$ moment image from the SMA (blue and red contours; \citealt{Shi10b}) onto the total intensity (0th moment) map of the (6,6) inversion transition of  NH$_3$ (with white contours displaying emission and black contours displaying absorption). 
Figures~\ref{e2-nh3+mas}, \ref{e2-nh3+h2o}, and \ref{e2-nh3+oh},   show the same (6,6) \amm\,total intensity map with overlaid different molecular maser species (\met, \wat, OH) detected around W51e2\footnote{ We choose the (6,6) ortho-transition as a "standard" for the \amm\,dense gas distribution because it is the strongest among the observed transitions. We do not believe this introduces a bias due to e.g.  differences between ortho and para-species or the possibility of maser emission. Indeed, the (6,6) emission displays a morphology similar to the strongest  observed para-transition (7,7) (see Figure~\ref{w51e2em-mom0}).
We also believe that the (6,6) emission is unlikely to be masing given that it is apparently consistent with the other lines (assuming a single LTE temperature) and given that there are no narrow, extremely bright features in the spectral profiles (see for ex., Figures~3 and 4 of Paper II for an example of \amm\,(6,6) maser line in  W51 North). }.  
In the following, we will  discuss separately physical properties of the three cores: 
W51e2-E, W51e2-W,  and W51e2-NW.

{\bf W51e2-E.} 
Figure~\ref{nh3+co}  reveals a core of dense hot molecular gas at the center of the CO outflow and clearly shows that  the driver of the CO outflow is not the HC HII region, but the dusty source W51e2-E (as first noticed by \citealt{Shi10b}).  
Interestingly, we find that the peak of the total intensity map of the most highly-excited (13,13) \amm\,transition (shown with the cyan contour), presumably locating the hottest gas, falls  at the center of the blue- and red-shifted lobes of the outflow. 
Therefore, we assume that this peak locates the position of the embedded protostar driving the CO outflow, i.e. W51e2-E. 
  \citet{Shi10b} derived  a mass-loss rate of  $10^{-3}$~\msyr\, and a mechanical power of 120~\ls, which is  an order of magnitude larger than expected for an early B-type star \citep[e.g.][]{Arce07}. This finding suggests that the protostellar core W51e2-E is forming an O type star.  In alternative, a cluster of B type stars could in principle explain the high mechanical power but would not be expected to drive a collimated outflow, and therefore can be excluded in this case. 
Based on dust emission, \citet{Shi10a} estimated 140~\ms\,available in the whole core. 
We estimate nearly 20~\ms\ of gas (assuming [\amm]/[H$_2$]=10$^{-7}$) within about half an arcsecond from the presumed location of the protostar, indicating a significant amount of material in the immediate vicinity of the protostar available for accretion. 

Besides the large mass and the powerful outflow, another indication of the presence of a high-mass protostar is provided by the excitation of \met, \wat, and OH masers around W51e2-E (Figure~\ref{e2-nh3+mas}). 
In particular, Class II \met\,masers are interesting because they are a typical signpost for HMSF.  
\citet{Etoka12} used  MERLIN  
  to show that the bulk of methanol maser emission comes from a compact ($\sim$0\pas5) ring-like structure centered approximately at the (13,13) \amm\,emission peak  (Figure~\ref{e2-nh3+mas}, top panel). 
Besides the ring, masers are distributed also to the NE and SW of the ring, across 1\pas5, and show a clear velocity coherence,  with blueshifted emission to the SW and redshifted emission to the NE, for a total  velocity extent of about 10~\kms.  
Since this structure is roughly perpendicular to the  CO outflow (P.A. $\sim$ 150\degree), 
a natural explanation could be that it traces an accretion flow: the structure across 1\pas5 could be an infalling envelope and the central ring (within 0\pas5) may probe a compact and dense disc or torus around the central protostar. 
Accretion however is not the only possibility. 
An alternative interpretation would be that the red- and blueshifted maser components are diverging from  W51e2-E with an expansion velocity of $\sim$5~\kms, tracing a slow and episodic wide-angle  outflow along NE-SW  (in this scenario, the central ring would indicate a younger outflow event). 

\wat\,masers show a bipolar structure along NW-SE, with redshifted velocities to the NW and blueshifted velocities to the SE  (Figure~\ref{e2-nh3+mas}, middle panel), in agreement with the CO emission.
\citet{Sato10} measured their proper motions,  clearly identifying a fast outflow (V$_{\rm exp} = 120 \pm 12$~\kms) arising from  W51e2-E along NW-SE (Figure~\ref{e2-nh3+h2o}, top panel), i.e. perpendicular to the \met\,maser distribution and along the more extended molecular outflow  seen in the CO (3-2) line \citep{Shi10b}. 

Finally, \citet{FishReid07} observed several transitions of OH masers at 1.7 GHz with the VLBA and measured their positions, l.o.s. velocities, and proper motions (bottom panel of Figure~\ref{e2-nh3+mas} and Figure~\ref{e2-nh3+oh}). 
The OH masers appear to be distributed in two main clusters. The first cluster is associated with W51e2-E and is distributed along the CO outflow, with accordingly redshifted spots to the NW and blueshifted to the SE: these may be tracing the innermost portion of the outflow, along with the \wat\,masers, at least based on positions and l.o.s. velocities. Proper motions seem to show a more complex kinematic structure, although they seem to globally indicate a wide-angle expansion around W51e2-E (at much lower velocity than the \wat\,masers though). There are however quite a few redshifted spots which have velocity vectors pointing inward towards W51e2-E: these may potentially probe infalling gas. 
The second cluster is excited south of the HC HII region, where OH masers (along with few CH$_3$OH masers) arise in two groups, located in correspondence of the contours of hot \amm\,gas engulfing  the HII region to the south. 
It is not clear if these OH masers are associated with W51e2-E or with the HC HII region, or are excited by a third (undetected) source. Likewise, their kinematics is rather unclear, although they seem to  have  proper motions diverging from W51e2-E. 

In summary, these high-angular resolution measurements of molecular masers  provide convincing evidence of a fast outflow ($\sim$100 \kms), a slower expanding wide-angle shell ($\sim$10 \kms), and potentially an infalling envelope associated with W51e2-E.   
 Interestingly, our hot \amm\,measurements provide supporting evidence for the latter.  

Although deriving the velocity field for the emission is more problematic than for absorption (owing to lower SNR), we have four lines of evidence indicating  accretion/infall of  the hot thermal  gas around W51e2-E.
First, the spectral profiles 
show that the lower-excitation (more optically thick) lines are double peaked, with the blueshifted component stronger than the redshifted one, while the higher-excitation (more optically thin) lines are more symmetric (\S~\ref{spec}). 
This feature is a well-understood signature of infall in a centrally condensed core. In fact, in a collapsing core, the blue side and the red side of the line arise from the rear and the front side of the core, respectively. The redshifted portion of the emission comes mainly from the outer (and cooler) region in the front side, whereas the blueshifted portion comes from the inner (and hotter) region in the rear side. This geometrical asymmetry produces a stronger blue shoulder in line profiles relative to the red shoulder. This asymmetry decreases with the decrease of optical depth, and the line becomes symmetric when it is optically thin because radiation from different parts of the core is not absorbed. In exceptional cases, like the strong hot-core G31.41+0.31, it has been possible to display this effect with spatially resolved maps of the intensity profiles as a function of distance to the center \citep[e.g.,][]{Mayen14}. 
In general, however, the SNR as the well as the spatial resolution is too poor to attempt such an analysis, as it is also the case for W51e2. 
Therefore, the signature of infall simply from spectral profiles remains ambiguous; we cannot exclude for example that an asymmetric spectral profile (skewed to the blue side) could arise from an outflow with a stronger blueshifted lobe.  
Nevertheless, a second line of evidence supporting the hypothesis of infall is provided by the pv-diagram, which shows a C-shaped structure (in the blueshifted side) and, although less clear, an o-shaped structure (when the redshifted side is also included; see Figure~\ref{pv_e2e}). This feature is expected for a radially infalling core where the l.o.s. velocity displacement is expected to be maximum at the center and then to decrease away from it.
In addition, the velocity field map shows some weak redshifted emission in the vicinity of W51e2-E (Figure~\ref{e2em-vel}), which further supports the hypothesis of infalling gas. 
Finally, towards W51e2-E we estimate the highest \amm\,column density in the W51e2 complex (with the exception of the HII region), which is inconsistent with outflowing gas. 

We conclude that   W51e2-E is a high-mass protostar, driving a powerful outflow and potentially associated with an infalling massive envelope.


{\bf W51e2-W.} 
In the majority of previous studies, the  HC HII region was thought to be the center of star formation activity in the W51e2 complex. Recent studies with increasing angular resolution, including the present one, are however showing that this is not the case. 
The overlays in Figures~\ref{nh3+co} and \ref{e2-nh3+h2o}, show that there is   no molecular outflow arising from the HC HII region, at least based on CO emission and/or \wat\,masers.  The presence of an ionized outflow has been postulated in order to interpret  the SW extension   \citep{Gaume93}, but our new measurements do not bring in any additional  evidence.  
The bulk of \met\,maser emission is concentrated onto the nearby companion W51e2-E, and only a few maser spots are observed  in the western edge of the HC HII region, having blue-shifted emission (Figure~\ref{e2-nh3+mas}, upper panel), in agreement with the velocity structure observed in \amm\,(see Fig.~\ref{mom1}). 
Several OH masers are still observed at the edges of the HII region (Figure~\ref{e2-nh3+mas}, lower panel). Their proper motions indicate an isotropic slow expansion from the HII region with a velocity of $\lesssim$10~\kms\, (Figure~\ref{e2-nh3+oh}),  consistent with the generally accepted scenario where OH masers are excited during the late stages of expanding UC HII regions \citep[e.g.,][]{FishReid07}.

Although these findings point to a much lower degree of star formation activity with respect to the nearby W51e2-E,  
 the absorption spectrum of the \amm\,inversion lines (detected up to the most excited levels) 
indicates the presence of a hot core still surrounding the HC HII region. 
We estimate an upper limit of about 5~\ms\ of molecular material   (assuming [\amm]/[H$_2$]=10$^{-7}$)  still present in the immediate environs of the HC HII region (within 0\pas06, 320 AU).  
This material could be potentially accreting onto the central YSO, either in the form of an accretion disk or an infalling envelope, provided that it has not been stopped by the intense radiation pressure from the central O-type star and the thermal pressure from the ionized gas in the HII region.
Indeed, the observed velocity gradient provides evidence for rotation  in the molecular core surrounding the HC HII (\S~\ref{w51e2-res}).   

\citet{ZhangHo97} noticed an increase of the gradient seen in  the \amm\,(3,3) line moving inward in radius (with a slope $r^{-1.2}$), and suggested that the rotating  material was spinning-up during collapse. We do not spatially resolve the gradient and therefore we cannot measure the rotation curve. Nevertheless,  our \amm\,velocity field maps do not display a clear  steepening of the gradient with  increasing  excitation energy.  If the molecular gas were in differential (e.g., Keplerian) rotation, one would  expect warmer gas closer to the central YSO to move faster. We cannot exclude however that this effect is hidden by optical depth effects and/or insufficient angular resolution.

Finally, since the absorption profiles appear quite symmetric and do not display a redshifted component, we do not see evidence for infall in the HII region. 
Gravitational collapse of the W51e2 core was suggested based on lower resolution images of  lower-excitation transitions of \amm\,\citep{ZhangHo97} and other high-density tracers like CS \citep{Zhang98}. The key evidence supporting infall was provided by both inverse P-Cygni spectral profiles (with blueshifted emission and redshifted absorption) and a C-shaped emission in  pv-diagrams.
Since these previous studies were conducted at $\ge$1\arcsec\,resolution, 
 both effects could be the consequence not of infall but of insufficient angular resolution, resulting in spatial blending of (redshifted) absorption and (blueshifted) emission from the two distinct YSOs, separated by less than 1 arcsecond: W51e2-W (having a more redshifted velocity) and W51e2-E (having a more blueshifted velocity).
Additionally, the molecular absorption has a central velocity consistent with the Hn$\alpha$ radio RLs  \citep[][Ginsburg et al, in prep]{KetoKlaassen08}, and not offset as it would be expected if the molecular gas were infalling\footnote{Even in the assumption of infall of the ionized gas, we would still expect the (optically-thin) RLs to be at the systemic velocity.}.
We conclude that based on our new \amm\,measurements, the evidence of infall towards the W51e2-W HII region is lacking.

{\bf W51e2-NW.} 
W51e2-NW is a massive core first detected in dust emission 1\arcsec\,north of W51e2-W,  
and could be an additional YSO in the W51e2 clump \citep{Shi10a}.  
Since no continuum emission has been detected at $\lambda \ge 7$ mm, this YSO could be at an earlier phase of star formation. 
Since we detect hot \amm\, up to the (9,9) doublet, W51e2-NW must be also exciting a hot molecular core.   
We estimate 32~\ms\,for its mass   (assuming [\amm]/[H$_2$]=10$^{-7}$), consistent with the estimate from dust emission \citep[40~\ms;][]{Shi10a}.
Given its location in the  path of the molecular outflow, however,  we cannot conclusively establish if W51e2-NW is an  independent protostar  or  just a component of the same outflow driven by W51e2-E.
The presence of a compact outflow that arises from the center of the W51e2-NW core (Figure~\ref{e2-nh3+h2o}, lower panel), identified by \citet{Sato10} using \wat\,maser proper motions, supports the protostellar scenario.    
This H$_2$O maser outflow has lower velocity  (V$_{\rm exp} \sim 20$~\kms) than the outflow driven  by W51e2-E (V$_{\rm exp} \sim 100$~\kms) and it  is elongated N-S, consistent with the distribution of the hot ammonia total intensity  (Figure~\ref{e2-nh3+h2o}, lower panel). 
The simultaneous presence of hot dust and molecular gas, along with a bipolar molecular outflow, provides evidence that W51e2-NW is indeed an additional (high-mass) YSO forming in the same W51e2 core.

\subsection{W51e8}

  Our \amm\,measurements demonstrate that W51e8 is  a massive hot core (T$\sim$200~K, M~$\sim$70~\ms)\footnote{ The value quoted for the temperature is a lower limit and the one for the mass assumes [\amm]/[H$_2$]=10$^{-7}$.}. 
  \citet{ZhangHo97} detected W51e8 at 22~GHz but not at 8.4~GHz, and speculated that the continuum flux is dominated by dust emission heated from a high-mass YSO.  
We detected a two-component structure in the 25~GHz continuum emission (e.g. see Fig.~\ref{w51main}) and measured a declining flux density for the stronger northern component from 25~GHz to 36~GHz, inconsistent with emission by dust (the southern component is not detected above 27 GHz). 
Ginsburg et al. (2016) conducted a sensitive continuum study of the W51 complex from 3.5 to 22.5~GHz, and measured spectral indices for both components of the radio continuum in W51e8, showing that it is due to free-free emission. 

\begin{figure}
\centering
\includegraphics[width=0.5\textwidth]{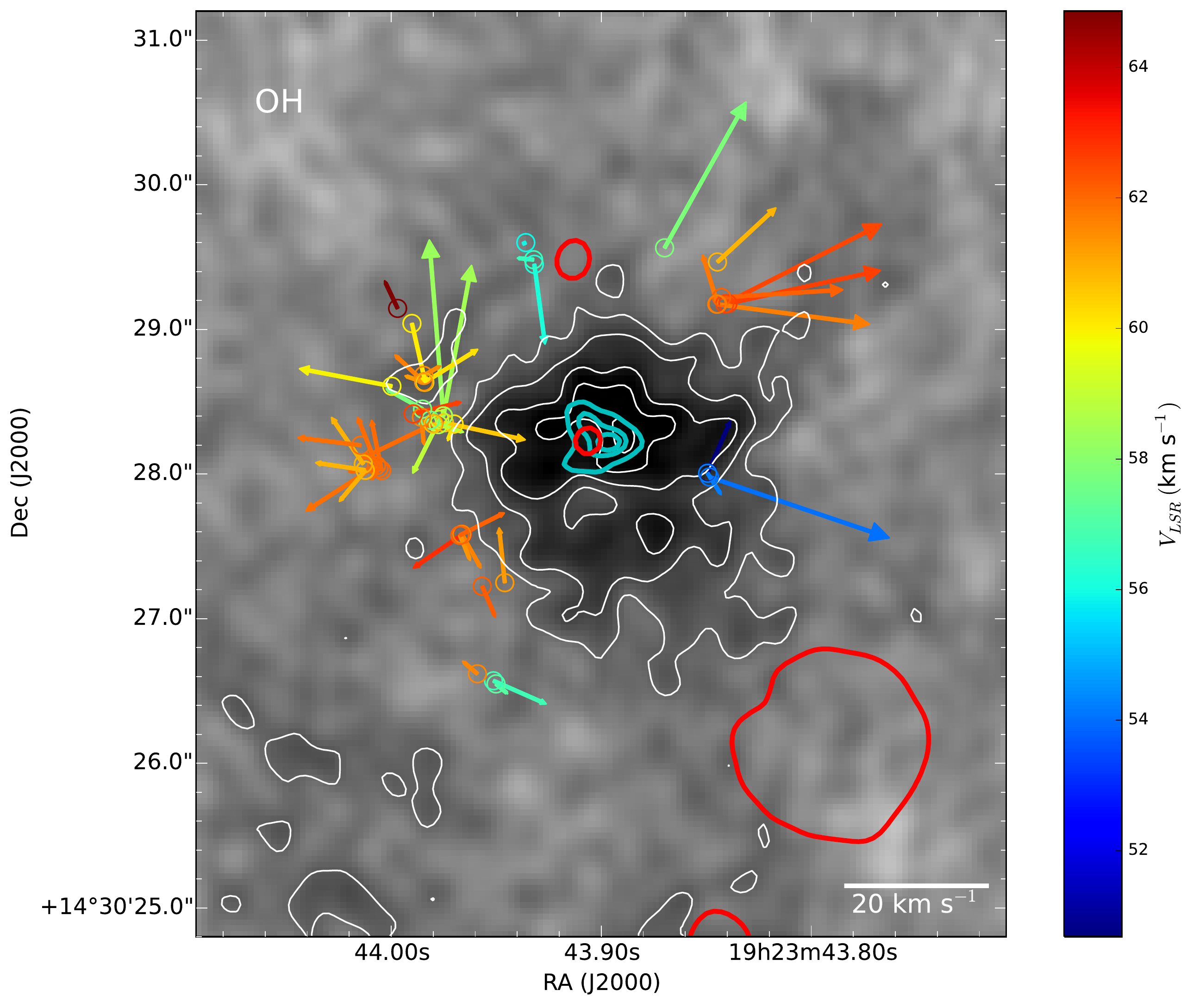}
\includegraphics[width=0.5\textwidth]{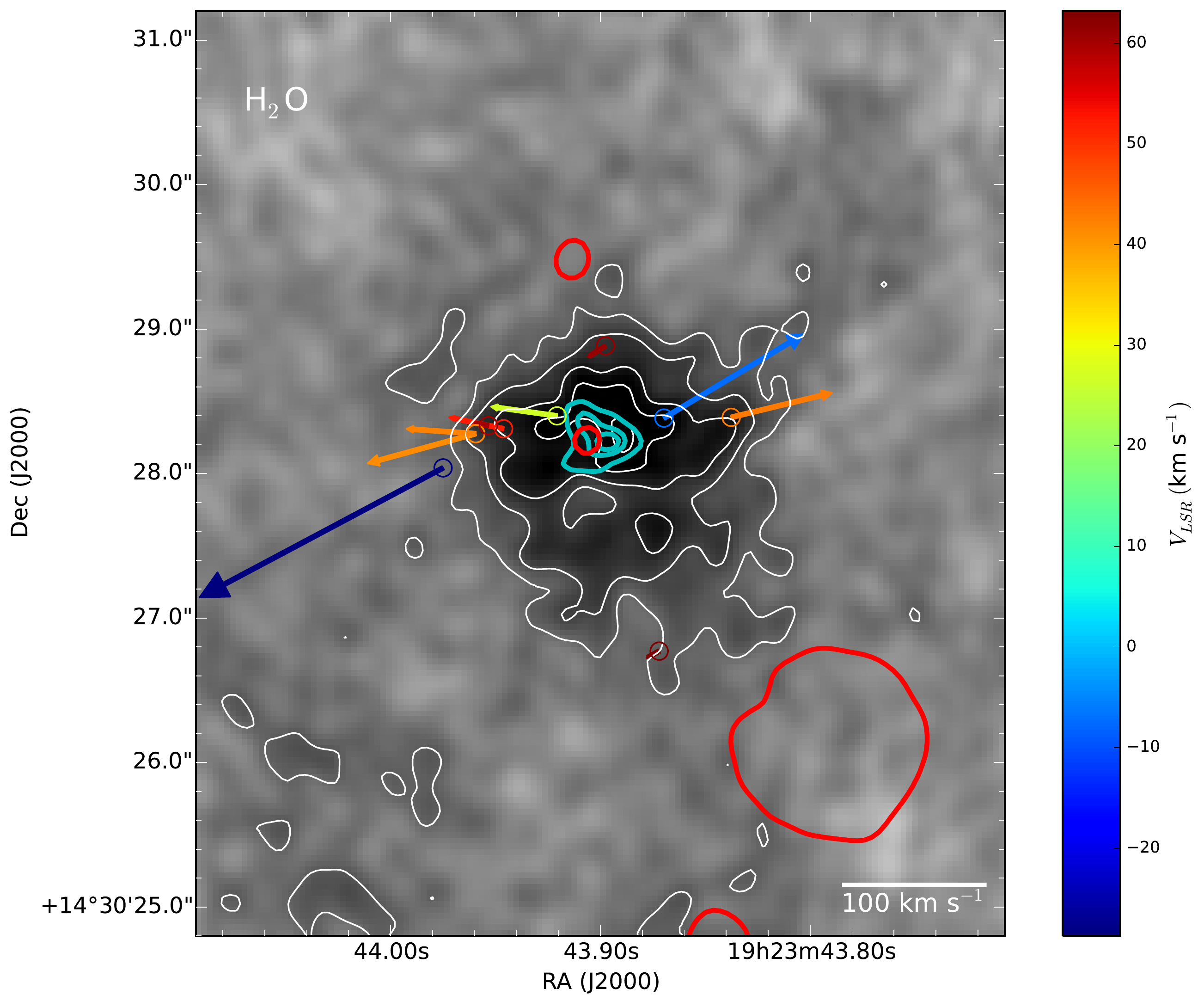}
\caption{
Overlay of molecular masers detected around W51e8 onto the total intensity (0th moment) map of the (6,6) inversion transition of  NH$_3$ 
 (gray scale and white+black contours). 
 The circles show positions and the arrows proper motions of OH masers \citep{FishReid07}   ({\it top panel}), and 
 \wat\,masers \citep{Sato10}  ({\it bottom panel}), respectively, both measured with the VLBA. 
 {  Colors denote  l.o.s.  velocity (color scale on the right-hand side in each panel) and   
 the scale for the proper motion amplitude is given in the lower right corner of each panel  (both  in \kms).
}
The \amm\,emission is displayed with white contours, representing 20\% to 100\% with steps of 20\% of the line peak for the (6,6) line, 107 mJy beam$^{-1}$ km s$^{-1}$).  
 The \amm\,absorption is displayed with black contours, representing factors 1, 5, 9,.... of --50~mJy~beam$^{-1}$~km s$^{-1}$), for all transitions.  
 The red contour indicates the 3 mJy flux level per beam   
 (corresponding to $\sim 4\sigma$ where $\sigma \sim0.7$  mJy beam$^{-1}$) for the 25 GHz continuum emission. Besides W51e1 to the SW, a peak of the radio continuum falls at the peak of the total intensity map of the most highly-excited \amm\,transitions (13,13), indicated with the cyan contours. 
 We assume that its position pinpoints a high-mass YSO at the center of W51e8, driving both a fast \wat\,maser outflow and a more slowly expanding OH maser shell. 
}
\label{e8-nh3+mas}
\end{figure}

Figure~\ref{e8-nh3+mas} shows the OH and  \wat\,masers detected  with the VLBA around W51e8 overlaid onto the   total intensity map of the (6,6) inversion transition. 
The latter shows a central stronger component which is elongated E-W, consistent with the \wat\,maser spatial distribution,
and a more diffuse, nearly-spherical core containing the central component, which is surrounded by an OH maser shell. 
The overlay in the top panel displays proper motions of  OH masers, which are globally suggestive of isotropic expansion of a molecular shell around W51e8 \citep{FishReid07}. A closer inspection reveals that expansion is the dominant motion in the maser clusters to the NW and NE, while masers in other locations show more complex proper motion patterns, including inflow. 
On the other hand, the lower panel shows that H$_2$O masers identify a bipolar outflow along approximately the E-W direction and with an expansion velocity of about 20~\kms\,\citep{Sato10}. 
The overlays show also the 25 GHz radio continuum peak and the total intensity of the (13,13) \amm\,line, are spatially coincident and fall at the center of the OH maser shell and the \wat\,maser bipolar outflow: we assume that this position locates the exciting high-mass YSO (this is the position reported in Table~\ref{ysos}). 

What about the velocity field of the \amm\,gas?
As in the case of W51e2-E, the \amm\,emission lines display  asymmetric spectral profiles which may in principle indicate infall:  
the lower-excitation (more optically thick) lines are double peaked with the blueshifted component stronger than the redshifted one, while the higher-excitation (more optically thin lines) are more symmetric. 
 \citet{Zhang98} claimed infall of the molecular core (infall speed of 3.5 \kms)  based on line asymmetries in CS and CH$_3$CN. 
However, in the case of W51e8 we do not have supporting evidence of infall from the spatially-resolved maps of the velocity field so the infall signature  from spectral profiles remains  ambiguous. 
In fact, alternative explanations are possible. 
For example, Figure~\ref{e2em-vel} shows predominantly redshifted emission towards the east and blueshifted emission towards the west, which may be related to the outflow. Consistently,   a pv-diagram at P.A. = 90 shows a velocity gradient in the central component (Figure~\ref{pv_e8}, upper panel). A pv-diagram in the perpendicular direction shows also a hint of velocity gradient which may indicate rotation in the  direction perpendicular to the outflow, although the signature is less clear than along the outflow axis (Figure~\ref{pv_e8}, lower panel). 
The \wat\,masers  show that the fastest components of the outflow have blueshifted velocities (Figure~\ref{e8-nh3+mas}, lower panel). 
Therefore, the asymmetric spectral profiles (skewed to the blue side) could  be due to an  outflow with a stronger blueshifted lobe. 
We conclude that the core surrounding W51e8 may be not contracting after all.

\section{Summary}

We imaged the  W51~Main complex with the JVLA at $\sim$0\pas2 resolution in five metastable inversion transitions of ammonia, which are emitted from doublet levels from about 400~K up to 1700~K above the ground state. 
We also imaged the radio continuum emission from 25 GHz to 36 GHz in the region. 
 Our \amm\, maps reveal the presence of four individual hot cores in W51-Main, presumably hosting high-mass YSOs at their centers which seem to be in different evolutionary stages: W51e2-W, W51e2-E, W51e2-NW, and W51e8. 
Our \amm\,(and continuum) measurements, along with previously published data, enabled us to derive  physical and dynamical properties of the identified YSOs, which we describe 
in detail below.

\begin{enumerate}

\item 
 The  HC HII region W51e2-W has been generally considered the center of star formation activity in the W51~Main complex. 
 The flux of Lyman continuum photons inferred from  radio recombination lines and the radio continuum flux, requires as an exciting source an O8-type YSO to maintain ionisation \citep[][; Ginsburg et al. in prep.]{Shi10a}. 
 Despite the fact that the exciting YSO has already reached the ZAMS, our \amm\,maps identify a hot molecular core ($T_{\rm gas} \sim 170$~K)  surrounding the HC HII region. The core has a clear velocity gradient ($\Delta V \sim$4.4~\kms) that we interpret as rotation in the E-W direction. 
 While this finding may suggest the presence of a rotating disk and that mass accretion maybe be still ongoing, 
the velocity field maps of \amm\,lines at different excitation energy do not show a clear velocity gradient increase toward the interior, 
as would be expected if material was spinning-up during accretion. 
Likewise, we do not discern a Keplerian velocity profile (at  spatial resolution of order of 1000 AU). 
Higher-resolution data of optically-thin lines from the same molecular species may be required to  confirm both effects. 
 While the lower limit to the mass of the central YSO is 20~\ms, we estimate a total molecular gas mass of only 5~\ms\, from the \amm\,lines (assuming $X_{NH_3}=10^{-7}$), suggesting that most of the mass has been already accreted onto the central star. 
The lack of outflow and/or \wat/\met\,maser activity  supports the idea  that the main accretion phase may be (nearly) over for this O-type YSO.

\vspace{0.5cm}
\item
W51e2-E (0\pas8 eastward from the HC HII region) possesses the typical features of an embedded high-mass protostar at an early stage of HMSF: bright continuum dust emission  but lack of free-free or RL  emission, hot core activity ($T_{\rm gas} \sim170$~K from \amm), strong \met\,masers, powerful molecular outflow,  organized magnetic-field structure  \citep{Tang09, Zhang14}. 
While we do not find clear indication of rotation in the circumstellar gas surrounding W51e2-E, there are some indications of collapse/infall. 
In fact, the \amm\,line profiles from lower-excitation (i.e. thicker) lines  show suppressed redshifted emission as compared with the blueshifted emission component, while the line asymmetry disappears in higher-excitation (i.e. thinner) lines. 
This is a well-understood signature of infall in a dense core. 
Supporting evidence is also provided  by an O-shaped pv-diagram and redshifted emission in the vicinity of W51e2-E. 
The indication of infalling gas, coupled with powerful outflow and maser activity, suggests that the center of accretion in the W51e2 complex is the embedded W51e2-E protostar rather than the neighboring HC HII region. 
We do not know the actual mass of the protostar, but we estimate about 20~\ms\,worth of molecular material  (assuming [\amm]/[H$_2$]=10$^{-7}$)  within half an arcsecond or about 2500 AU from the putative protostellar position (\citealt{Shi10a} estimated 140~\ms\,for the total core mass). 
This indicates that the natal core has enough material available for accretion onto the central protostar to become  an O-type star.

\vspace{0.2 cm} 
\item 
  W51e2-NW is an additional hot molecular core ($T_{\rm gas} \sim 140$~K from \amm)  in the W51e2 complex (1\arcsec\,north of the HC HII region), initially detected  in dust emission and now in hot \amm\, up to the (9,9) doublet. We estimate 32~\ms\,of molecular gas for this hot core   (assuming [\amm]/[H$_2$]=10$^{-7}$). W51e2-NW does not exhibit continuum emission at $\lambda>$7~mm nor \met/OH masers \citep{Etoka12,FishReid07}, but it drives an H$_2$O maser outflow: this may indicate a protostellar core, either at an earlier evolutionary stage or with a lower mass than W51e2-E. 

\vspace{0.2 cm} 
\item 
 W51e8 contains a large amount  ($\sim 70$~\ms,   assuming [\amm]/[H$_2$]=10$^{-7}$) of warm ($T\sim 200$~K) molecular gas, as derived from our \amm\,measurements. 
 It drives a bipolar outflow (traced by \wat\,masers) indicating the presence of a high-mass protostar at its center, whose position is pinpointed by the peaks in both the (13,13) \amm\,total intensity and the 25 GHz radio continuum emissions. 
The latter can be interpreted in terms of an ionized jet, pointing to a hot core stage prior to the onset of a HC HII region. 
The presence of an OH maser shell slowly expanding ($\sim$10~\kms) around the \amm\,core is however inconsistent with this hypothesis, since OH maser shells are a typical signpost of  expanding UC HII regions. 
Likewise, the lack of methanol maser emission, a typical signpost of HMSF, is puzzling, although that could be explained with the low  molecular density estimated for the e8 core,  $\sim 10^6$~\cmc, insufficient to efficiently pump  \met\,masers \citep[which require $N_{H_2} \ge 10^7$~\cmc;][]{Cragg05}.  
Moreover, it is not clear which is the dominant  motion in the hot molecular gas (although expansion is more probable). 
Further observations are needed to clarify the evolutionary stage and the gas kinematics of W51e8 within W51~Main. 
\end{enumerate}

This study of W51~Main demonstrates that high-angular resolution imaging (i.e., a few tenths of arcseconds) of high-excitation lines of \amm\,
at $\sim$1~cm wavelengths are well suited to study the kinematics and physical conditions of the hottest and densest molecular gas in HMSF complexes where  O-type stars are forming. 
 W51~Main offers  the rare opportunity to compare physical and dynamical properties of high-mass YSOs at different evolutionary stages forming in the same clump. 


\appendix
\chapter{}

\section{Methods to estimate the physical conditions of the \amm\,gas}
\label{app}

\subsection{Absorption}
\label{app_abs}

We follow the approach for determining column density in a given
rotational state from \citet{Mangum2015a}.  
For each transition, we compute  the column density of molecules in the lower transition state ($N_l$) 
 starting with Equation 30 in \citet{Mangum2015a}: 
 
$$
N_l = \frac{3h}{8\pi^3|{\mathbf \mu_{ul}}|^2}
\left[\exp\left(\frac{h\nu}{kT}\right) - 1 \right]^{-1} \int \tau_\nu dv.
$$

Using: 

$$|{\mathbf  \mu_{jk}}|^2 \equiv S\mu^2$$
$$ S = \frac{K^2}{J(J+1)} \ {\rm (for \ (J, K) \rightarrow \ (J, K)  \ transitions) }, $$

we find that the metastable states of \amm\,($\rm N_u$ = N(J, K)): 

\begin{equation}
\ \ \ \ \
N(J, K) = \frac{3h}{8\pi^3|{\mathbf \mu}|^2}  \frac{J(J+1)}{K^2}
\left[\exp\left(\frac{h\nu}{kT}\right) - 1 \right]^{-1} \int \tau_\nu dv.
\end{equation}

If we further assume that the Rayleigh-Jeans approximation applies ($h\nu \ll kT$):

$$
\exp\left(\frac{h\nu}{kT}\right) - 1 = \frac{h\nu}{kT},
$$
we get the following:

\begin{eqnarray}
  \ \ \ \  \ \ \ \ \ \ \ \ \  N(J, K) = \frac{3h}{8\pi^3|{\mathbf \mu}|^2}  \frac{J(J+1)}{K^2} \frac{kT_{ex}}{h\nu} \int \tau_\nu dv =  \nonumber \\ 
 = \frac{3k}{8\pi^3|{\mathbf \mu}|^2} \frac{J(J+1)}{K^2} \frac{T_{ex}}{\nu} \int \tau_\nu dv.   \ \ \ \ \ \  
\end{eqnarray}

For \amm, $\mu = 1.468 \times 10^{18}$~esu~cm. Inserting this value for $\mu$ and the other physical constants results in the following:

\begin{equation}
  \ \ \ \ \ \ \ \ \ \ \ \ \ \  
  N(J,K)  = 7.7 \times 10^{13} \ \frac{J(J+1)}{K^2} \ \frac{T_{ex}}{\nu}  \int \tau_\nu dv.
\end{equation} 

We use peak optical depth and FWHM line widths to approximate the integration over the line profile, therefore Equation A7 in  \citet{Mangum2015a} applies, implying  that a factor of 0.94 should be included in the prefactor of the equation for $N(J,K)$ above, resulting in:

\begin{equation}
\label{N_tau}
 \ \ \  \ \ \ \ \ \ \ \ \ \ \ \ \ \  N(J,K)  = 7.3 \times 10^{13} \ \frac{J(J+1)}{K^2} \ \frac{\Delta v}{\nu} \ \tau \ T_{ex},
\end{equation} 
where 
$N(J,K)$ is in cm$^{-2}$,
$\Delta v$ is the linewidth (in \kms), 
$\nu$ is the transition frequency (in GHz), 
\tex\,is the excitation temperature (in K), 
and $\tau$ is the line opacity (derived numerically solving  Equation~A.1 from Paper I). 

In order to find the total column density, we assume that all the energy levels  are populated according to a Boltzmann distribution, 
characterized by a single \trot: 
\begin{equation}
\label{eqboltz1}
\ \ \ \ \ \ \  \ \ \ \ \ \ \ \ \ \ \ \ \ \  \ \ \ N_{tot} = \frac{N(J,K)}{g(J,K)} \ Q(T_{rot}) \ e^{E_{l}/T_{rot}}  
\end{equation} 
where 
$g(J,K)=g_{op}\ (2J+1)$ are the statistical weights ($g_{op}=1$ for para- and  $g_{op}=2$ for ortho-transitions), 
$Q(T_{rot})$ is the partition function (under the high-temperature assumption), and $E_l$ is the lower state energy of the (J,K) transition (in K). 

Since we observed more than two transitions, we can use RTDs to fit simultaneously the rotational temperature and the column density. 
Rearranging Equation~\ref{eqboltz1}, using Equation~\ref{N_tau} for the column density $N(J,K)$, and taking the natural logarithm of both sides, 
we have:  
\begin{eqnarray}
\label{eq_rtd_abs}
\ \  \ \ \ \ \ \ \ \log{\left[ 7.3 \times 10^{13} \frac{J(J+1)}{K^2}\frac{1}{g_{op}(2J+1)} \frac{\Delta v}{\nu} \tau \right] }  \nonumber \\ 
= \log{\frac{N_{tot}}{T_{ex} Q(T_{rot})}} - \frac{E_l}{T_{rot}}
\end{eqnarray}

\begin{figure}
\centering
\includegraphics[width=0.5\textwidth]{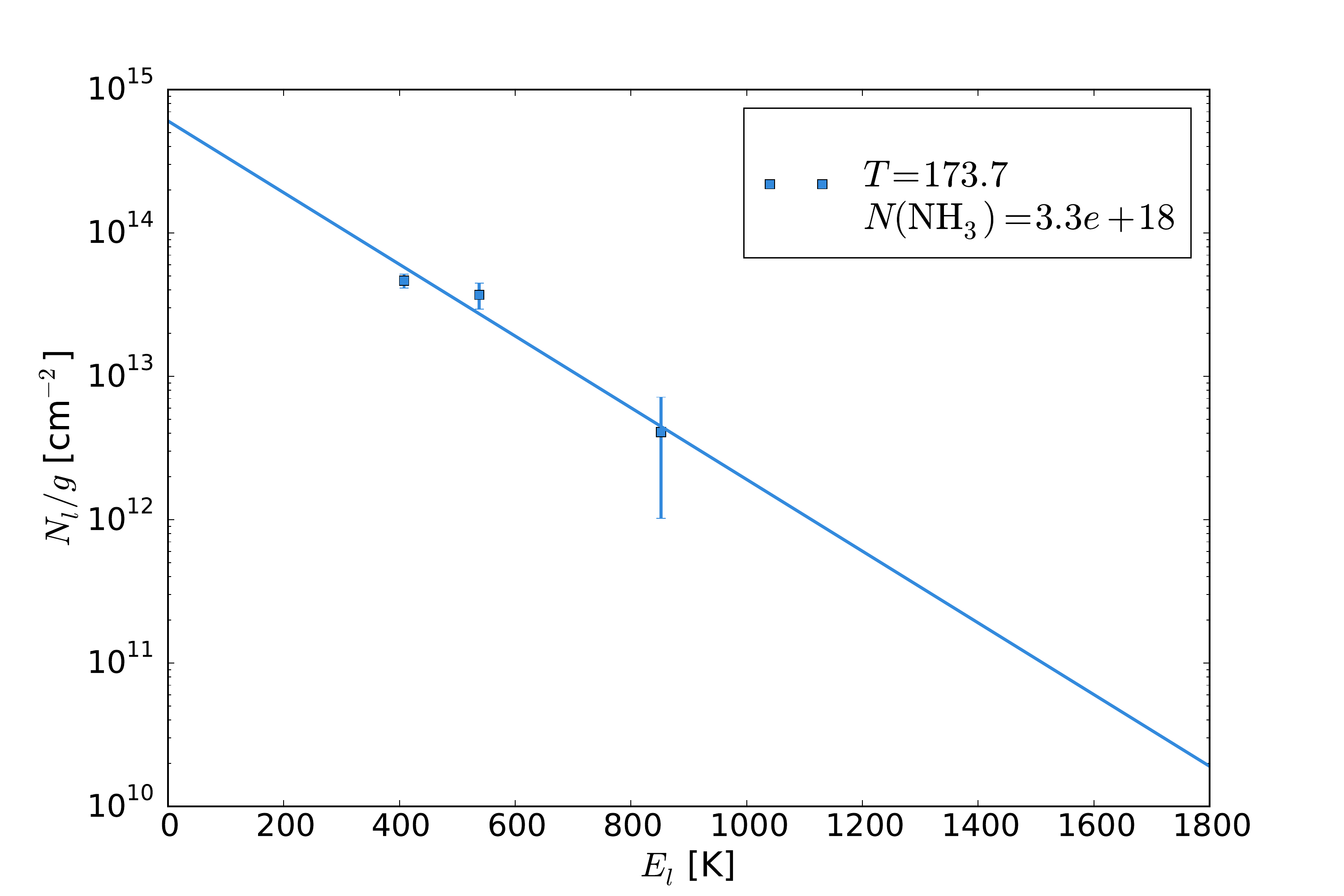}
\caption{Rotational \amm\,temperatures determined for the core around the HC HII region W51e2-W. The temperatures are fit to the slope of the normalized \amm\,column density $N_l$ (in units of \cmc) plotted against the energy of each transition $E_l$ (in units of Kelvins),  using Eq.~\ref{eq_rtd_abs}. 
Only transitions with measured optical depths were used in the fits: (6,6), (7,7), and (9,9).  
}
\label{fig_rtd_abs}
\end{figure}

Equation~\ref{eq_rtd_abs} shows that the logarithm of the left member is a linear function of $E_l$ 
(if all transitions have the same \tex), 
with slope $-1/T_{rot}$ and intercept $log\frac{N_{tot}}{T_{ex} Q(T_{rot})} $ at  $E_l=0$. 
We can then determine \trot\, and $N_{tot}$ from a least squares fit of lower state energy  to the optical depth for different transitions in log space (Fig.~\ref{fig_rtd_abs}). In this calculation, we assume the \amm\,gas to be in LTE, and therefore \tex=\trot.

\subsection{Emission}
As in the case for absorption, we  follow the approach for determining column density in a given
rotational state from \citet{Mangum2015a}.  For each transition, we compute 
$N_u$ using the optically thin version of Equation 30 in \citet{Mangum2015a}:
\begin{equation}
\ \ \ \ \ \ \ \ \ \ \ \ \ \  \ \ \ \ \ \ \ \ \ \ \ \ \ \   N_u = \frac{3 k_B}{8\pi^3 \nu \mu_{lu}^2} S_\nu f^{-1} F_\tau,
\end{equation}
where $f$ is the filling fraction and $F_\tau = \frac{\tau}{1-e^{-\tau}}$ is the optical depth correction
if we have computed an optical depth or 1 otherwise\footnote{Please refer to Section 11 in \citet{Mangum2015a} for more details on the optically thin approximation.}.  
Line strengths $\mu_{lu}$ were retrieved from the JPL molecular database \citep{Pickett1998a} via
the \texttt{astroquery} (\url{http://www.astropy.org/astroquery/}) interface to the splatalogue (\url{splatalogue.net}) website.
We then fit a linear function:
\begin{equation}
\label{eq_rtd_em}
\ \ \ \ \ \ \ \ \ \ \ \ \ \   \log{ \left[N(J,K) \right] } = -E_u/T_{rot} + \log{ [N_{tot}  /Q(T_{rot})]}
\end{equation}
to the data, where $E_u$ is the upper energy level of a given state, $N_{tot}$ is
the total column density, and $Q(T_{rot})$ is the partition function derived using
equation 15.48 of \citet{Wilson2009a}   (under the high-temperature assumption).  

\begin{figure*}
\centering
\includegraphics[width=0.48\textwidth]{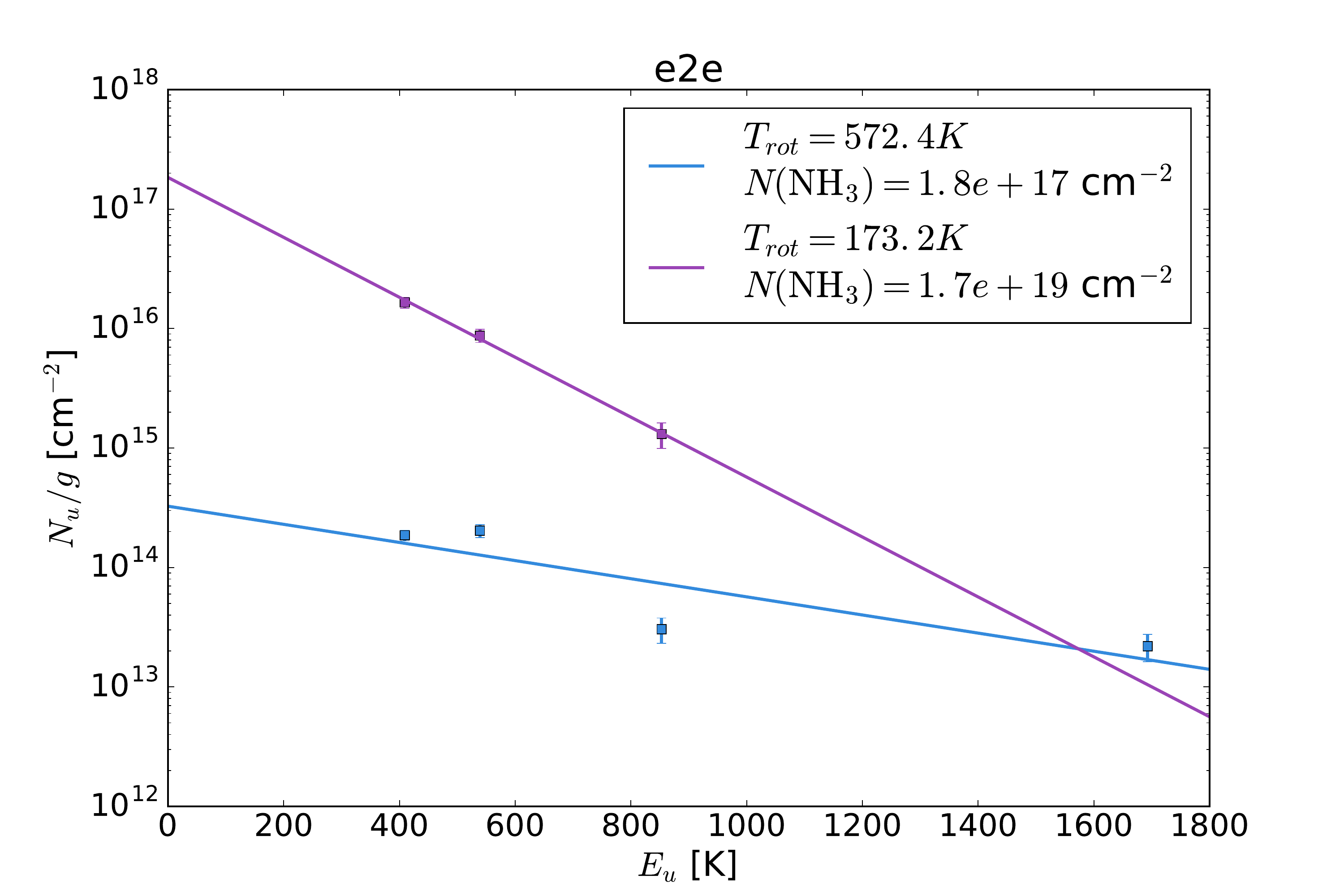}
\includegraphics[width=0.48\textwidth]{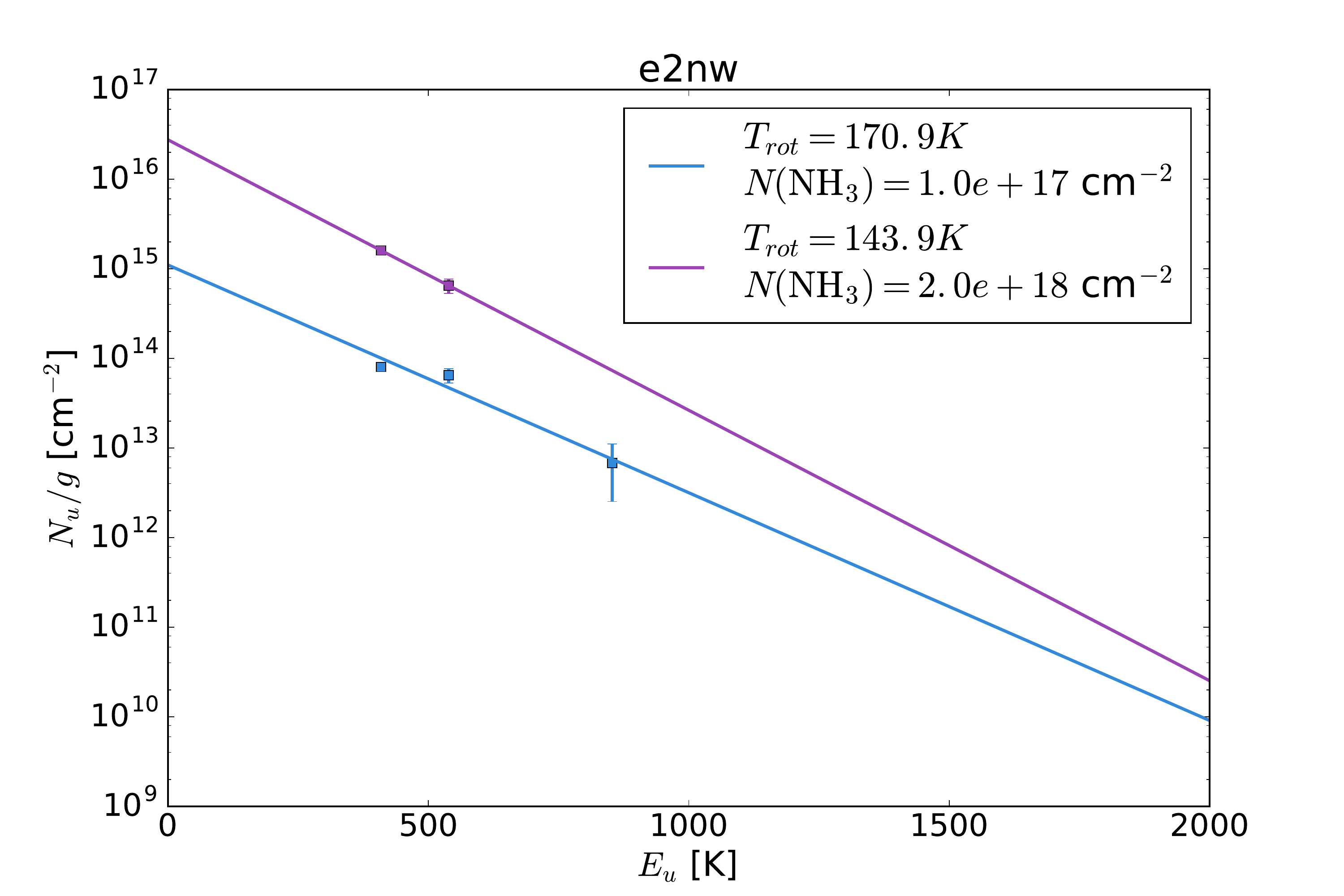}
\includegraphics[width=0.48\textwidth]{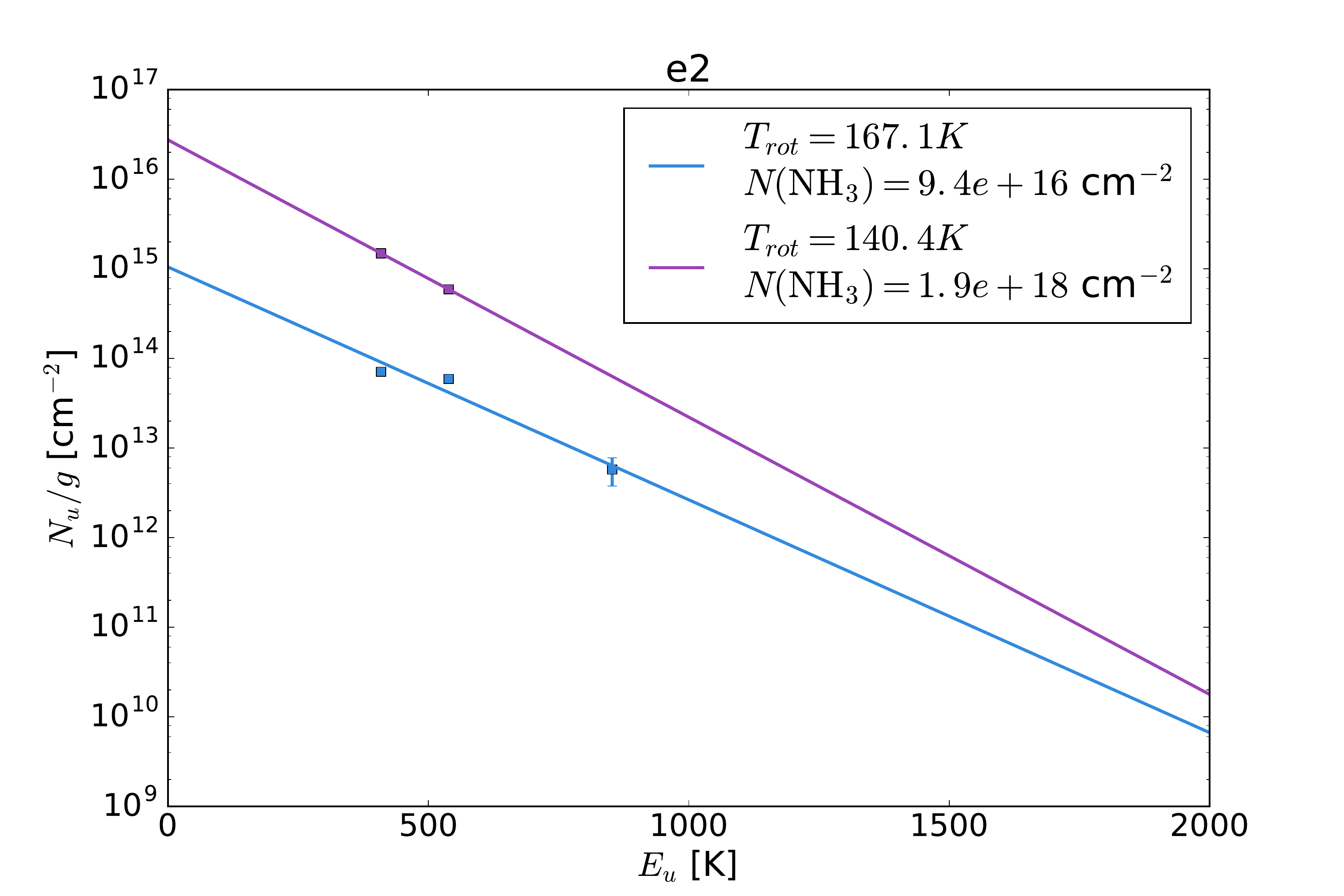}
\includegraphics[width=0.48\textwidth]{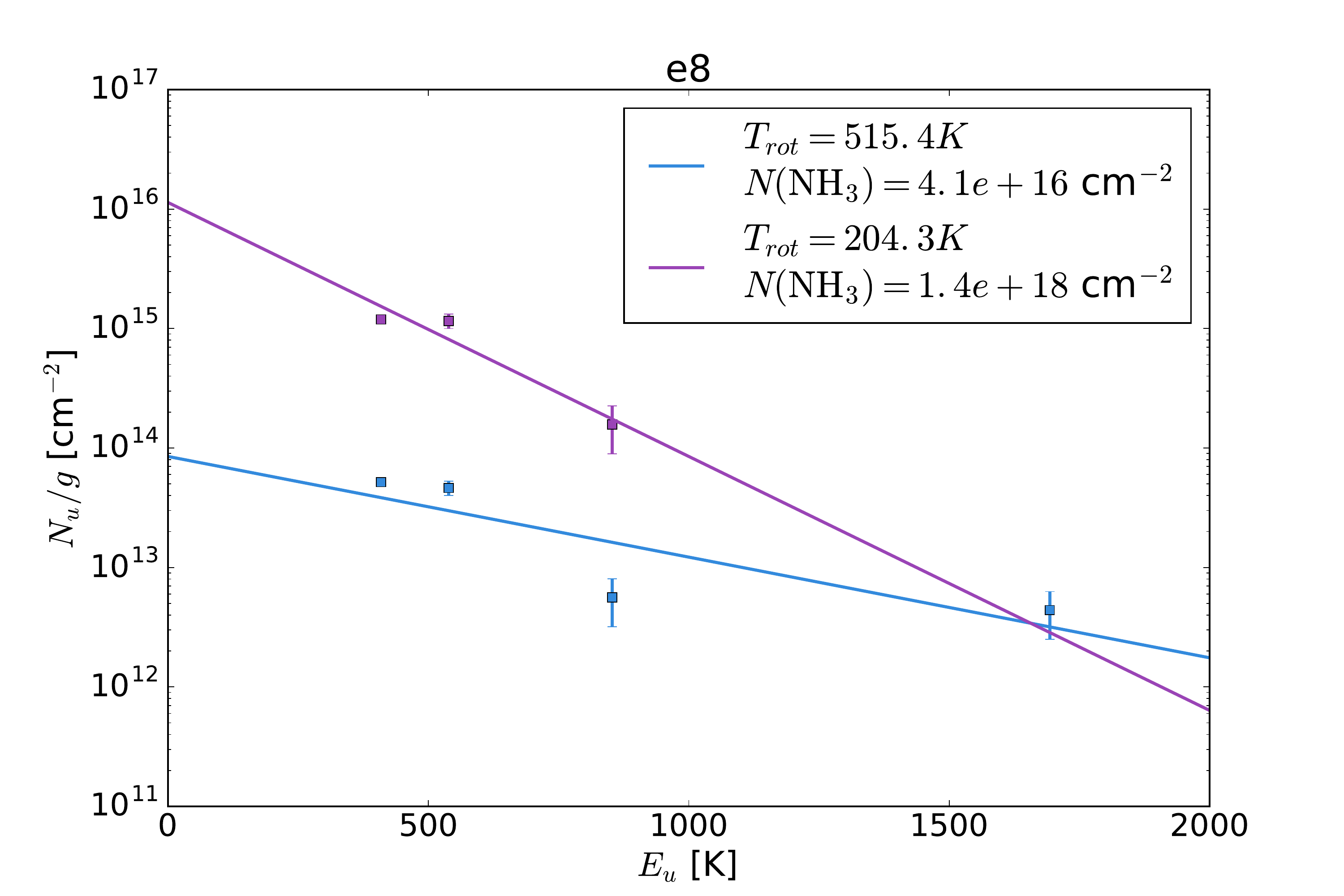}
\caption{Rotational \amm\,temperatures determined for the protostars W51e2-E and W51e2-NW, and cores W51e2 and W51e8 (from left to right, and from top to bottom). The temperatures are fit to the slope of the normalized \amm\,column density $N_u$ (in units of \cmc) plotted against the energy of each transition $E_u$ (in units of Kelvins), using Eq.~\ref{eq_rtd_em}. 
The blue line is a fit to the measured columns without opacity  correction. 
The magenta line is the fit after applying the opacity  corrections, using the optical depth values in Table~\ref{nh3_hf}. 
We excluded the (10,10) transition from the fit because of much higher rms.
}
\label{fig_rtd_em}
\end{figure*}

The results of the fit are shown in Figure~\ref{fig_rtd_em} for all the cores seen in \amm\,emission in W51~main.

%
\subsection{Hyperfine line model fitting}
\label{model_fit}
Owing to high optical depths, simple Gaussian fitting could not reproduce accurately the observed HFS spectral profiles of the \amm\,(6,6) and (7,7) inversion lines both in absorption  and emission.  
In particular, at an optical depth much larger than 10, the main hyperfine component of the \amm\, (6,6) and (7,7) inversion lines should be flat-topped. In contrast, the observed lines shown in Figures~\ref{spec-e2w} and ~\ref{spec-e2e+nw} exhibit a Gaussian-like profile. One explanation is that the gas in the dense core is clumpy and highly structured. If a dense core is filled with pockets of hot and dense gas with  a large optical depth and a small filling factor, the \amm\, spectra from these individual pockets are optically thick, hence exhibit strong satellite hyperfines. If these pockets move at different velocities (due to infall, rotation, or turbulence), the sum of the spectra within the telescope beam can appear Gaussian-like depending on the kinematics and relative intensities of the individual spectra.
The idea of a clumpy medium in relation to \amm\,spectra was discussed in the work by \citet{SollinsHo05} when analyzing high resolution \amm\,(3,3) absorption spectra in G10.6-0.4. However, a detailed and exact modeling the \amm\,spectra in W51 is beyond the scope of this paper. 
For a physically-motivated model to explain ammonia emission in a spatially-resolved hot-core, see for example \citet{Osorio09}.

Here we use a two-stage process to account for the optical depth effect when evaluating the line parameters in absorption\footnote{This issue affects the lines both in absorption  and emission, but we applied this model only to the absorption lines, which have sufficiently high SNR and narrow widths in the HF components to enable us to  fit their centroids precisely.}. 
\begin{enumerate}
\item Because our earlier gaussian component
fitting revealed that the line profiles and centroids of the
hyperfines are not always the same or even consistent with the central
component, we used a model that allowed the offset of the hyperfine
component to vary symmetrically around the central velocity. We
assumed all components had the same intrinsic width.  The central line was \emph{not} modeled, so that the velocities and widths come only from the optically thin hyperfine lines.  This fitting revealed that the hyperfine velocity offsets are only slightly
different from the theoretical values given in Table 3; there were
apparent differences based on the Gaussian fits, but these likely
resulted from a shift in the main line's centroid. 

\item We modeled the hyperfines and main line simultaneously accounting
for the optical depth, but keeping the offsets and widths fixed.  This
fit was not very good, as the hyperfines and main line cannot be explained
by the same optical depth and excitation temperature.  In particular, the
hyperfine and main lines cannot be simultaneously modeled with an
excitation temperature $T_{ex}=2.73$ K (i.e., the coldest excitation temperature expected to be observed in a line), which is assumed in the simple
approach used in point 1.  
In fact, both the (6,6) and (7,7) line profiles
can be approximately reproduced if the excitation temperature
$T_{ex}\sim1800$ K and the optical depths are somewhat higher. 

The shape in the main line profiles is likely due to excitation temperature and optical depth variations with velocity, suggesting that there is significant unresolved velocity structure within the absorption profile (i.e., a clumpy medium).
\end{enumerate}

\begin{figure}
\centering
\includegraphics[angle=0,width=0.5\textwidth]{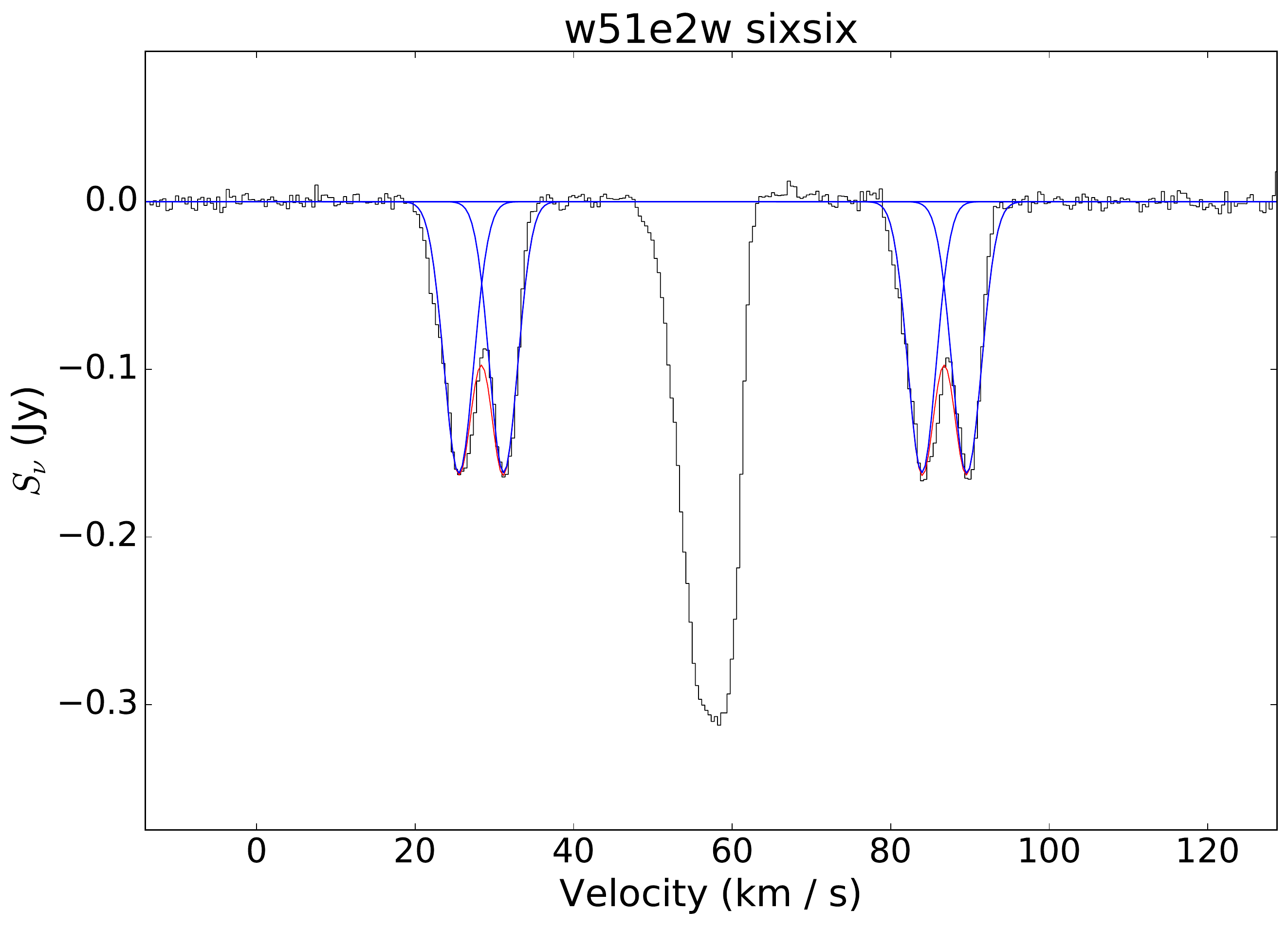}
\caption{Spectrum of W51e2w in the \amm\ (6,6) transition fitted as described in Appendix~\ref{model_fit}.}
\label{sp6e2w}
\end{figure}


\chapter{}
\section{ \amm\,inversion lines hyperfine structure}
\label{app_b}
Each \amm\,inversion transition displays a complex spectral profile due to hyperfine interactions. 
In particular, the interaction of the electric quadrupole moment of the nitrogen nucleus\footnote{The quadrupole moment results from the non-spherical distributions of charge within the nitrogen nucleus.}, with the electric field due to the electrons, 
splits each  (J,K) inversion energy level  into three levels characterised by the quantised nuclear spin of the nitrogen, $I_N=1$, and the total angular momentum, ${\bf F} = {\bf I_N} + {\bf J}$. 
The following selection rules apply:
$$F = (J-1, \ J, \ J+1) \ , \ \ \Delta F = (0, \pm1) $$
(and obviously $\Delta K= 0, \ \Delta J= 0$ for metastable inversion lines).
Therefore, each \amm\,inversion line is splitted into five components,  
a central ``main'' line ($\Delta F=0$) and two pairs of "satellite" lines  ($\Delta F=\pm1$),
symmetrically spaced with respect to the main line\footnote{Weaker magnetic interactions due to spin-spin interactions among the hydrogen nuclei and the nitrogen nucleus lead to further splitting, which produces more HF components separated by up to a few tens of KHz. This is much smaller than our velocity resolution, therefore we neglect the splitting due to magnetic interactions.}. 
The frequency separations  of the four satellite components  with respect to the main line can be calculated 
from the nuclear quadrupole energy, $E_Q$, for a symmetric molecular rotor \citep[e.g.,][ P. Ho, PhD Thesis,1972]{TownesSchawlow75}:

\begin{eqnarray}
\label{freq_hf}
E_Q = \frac{e q Q  \left( \frac{3 K^2}{J (J+1)} -1 \right)  \times \left[0.75 C (C+1)-I_N (I_N+1) J (J+1) \right]  }{2 I_N (2 I_N-1) (2J-1) (2J+3)}  
\end{eqnarray}
where 
$$
C = F (F+1) -I_N (I_N+1) - J (J+1). 
$$
The product $e q Q = 4.09$ is the quadrupole coupling constant, where $Q$ is the nuclear quadrupole moment, 
$q$ is the second derivative of the coulomb potential, 
and $e$ is the electronic charge. 
The frequency separation of the outer satellites from the main line is given by $ E_Q(F = J-1) - E_Q(F = J)$, 
while  the inner satellites are separated from the main line by $ E_Q(F = J+1) - E_Q(F = J)$ . 
These frequency separations of the satellites are of  the order of 2~MHz for the  lines observed in this study and are reported in  cols.~2 and 3 of Table~\ref{nh3_hf}. 

The intensities of the hyperfine components are given by \citep[e.g.,][ P. Ho, PhD Thesis, 1972]{TownesSchawlow75}: 

\begin{eqnarray}
\label{int_m}
\small
I_{Main} = \frac{ A \left[ J (J+1.) + F (F + 1) - I_N (I_N+1) \right]^2  (2F + 1)}{F (F+1 )}  \\ 
 (\Delta F= 0, F = J-1, \ J, \ J+1)   \ \ \ \ \ \ \ \ \ \ \ \ \  \nonumber 
\end{eqnarray}
\begin{eqnarray}
\label{int_i}
I_{Sat_1} = \resizebox{.9\hsize}{!}{$ \frac{ -A (J + F + I_N +2) (J + F - I_N +1) (J - F + I_N) (J-F-I_N-1) }{F +1} $ }\nonumber \\ 
(\Delta F= +1, F = J-1, \ J)   \ \ \ \ \ \ \ \ \ \ \ \ \ \ \ \ \ \ \ \ \ \ \ \ \  \nonumber 
\end{eqnarray}
\begin{eqnarray}
\label{int_o}
I_{Sat_2} = \resizebox{.9\hsize}{!}{$ \frac{ -A (J + F + I_N +1) (J + F - I_N) (J - F + I_N +1) (J - F - I_N) }{F} $} \nonumber \\ 
 (\Delta F= -1, F = J, \ J+1)  \ \ \ \ \ \ \ \ \ \ \ \ \ \ \ \ \ \ \ \ \ \ \ \ \  \nonumber 
\end{eqnarray}

for the main line, and the two pairs of (inner and outer) satellite components, respectively\footnote{Note that the main line is a combination of three components, corresponding to $F = (J-1, \ J, \ J+1)$, which along with the two pairs of satellites, lead effectively to seven components in the HFS of \amm\,inversion lines.}.  
$A$ is normalisation factor chosen so that the total intensity across all hyperfine components is equal to 1: 
\begin{eqnarray}
\resizebox{.95\hsize}{!}{$ A^{-1}=\left[I_{Main} (F = J-1)+I_{Main} (F = J)+I_{Main} (F = J+1) \right]+ $} \nonumber  \\
 + \left[ I_{Sat_1}(F = J-1)+I_{Sat_1}(F = J)\right] +   \ \ \ \ \ \ \ \ \ \ \ \ \ \ \ \ \ \ \ \ \ \  \ \ \ \ \ \ \ \   \\
+ \left[ I_{Sat_2}(F = J)+I_{Sat_2}(F = J+1) \right], \ \ \ \ \ \ \ \ \ \ \ \ \ \ \ \ \ \ \ \ \ \ \ \ \  \ \ \ \ \ \ \ \ \nonumber 
\end{eqnarray}
 (e.g., $A= 1.53 \times 10^{-4}$ for $(J,K)=(6,6)$).  \\
The satellites have approximately equal intensities in each pair, typically below 1\% of the main line (see col.~6  of Table~\ref{nh3_hf}).  
\\

\begin{acknowledgements}
We thank the anonymous referee for a thorough effort in reviewing the manuscript and a very constructive report. 
We are grateful to Dr Vincent Fish for providing the OH proper motion measurements. 
We are grateful to Riccardo Cesaroni for useful discussions.
These data were obtained under JVLA program 12A-274.
Spectral line fitting was performed using the \texttt{pyspeckit} package \citep{Ginsburg2011c}
\end{acknowledgements}


\bibliographystyle{aa}
\bibliography{biblio}  

\end{document}